\documentclass[PAPER, cernpreprint, atlasdraft=false, UKenglish, texmf, orcidlogo]{atlasdoc}
\usepackage{atlaspackage}
\usepackage{atlasbiblatex}

\usepackage{atlasphysics}
\graphicspath{{logos/}{figures/}}

\usepackage{ANA-HION-2024-02-PAPER-defs}

\AtlasTitle{Charged-hadron and identified-hadron ($K^\mathrm{0}_\mathrm{S}$, $\Lambda$, $\Xi^\mathrm{-}$) yield measurements in photo-nuclear Pb+Pb and $p$+Pb collisions at $\sqrt{s_\mathrm{NN}} = 5.02$~TeV with ATLAS}

\AtlasAbstract{%
This paper presents the measurement of charged-hadron and identified-hadron ($K^\mathrm{0}_\mathrm{S}$, $\Lambda$, $\Xi^\mathrm{-}$) yields in photo-nuclear collisions using 1.7 $\mathrm{nb^{-1}}$ of $\sqrt{s_\mathrm{NN}} = 5.02$~TeV Pb+Pb data collected in 2018 with the ATLAS detector at the Large Hadron Collider.
Candidate photo-nuclear events are selected using a combination of tracking and calorimeter information, including the zero-degree calorimeter.
The yields as a function of transverse momentum and rapidity are measured in these photo-nuclear collisions as a function of charged-particle multiplicity.
These photo-nuclear results are compared with 0.1 $\mathrm{nb^{-1}}$ of $\sqrt{s_\mathrm{NN}} = 5.02$~TeV $p$+Pb data collected in 2016 by ATLAS using similar charged-particle multiplicity selections.
These photo-nuclear measurements shed light on potential quark-gluon plasma formation in photo-nuclear collisions via observables sensitive to radial flow, enhanced baryon-to-meson ratios, and strangeness enhancement.
The results are also compared with the Monte Carlo \textsc{dpmjet-iii} generator and hydrodynamic calculations to test whether such photo-nuclear collisions may produce small droplets of quark-gluon plasma that flow collectively.}

\AtlasRefCode{HION-2024-02}

\PreprintIdNumber{CERN-EP-2025-040}

\AtlasJournalRef{Phys. Rev. C 111 (2015) 064908}
\AtlasDOI{10.1103/7lx4-x8rw}

\AtlasCoverEgroupAnalysisTeam{atlas-hion-2024-02-analysis-team@cern.ch}

\hypersetup{pdftitle={ATLAS document},pdfauthor={The ATLAS Collaboration}}

\begin{document}

\maketitle

\tableofcontents

\section{Introduction}
\newcommand{\AtlasCoordFootnote}{%
ATLAS uses a right-handed coordinate system with its origin at the nominal interaction point (IP)
in the center of the detector and the $z$-axis along the beam pipe.
The $x$-axis points from the IP to the center of the LHC ring,
and the $y$-axis points upwards.
Cylindrical coordinates $(r,\Phi)$ are used in the transverse plane,
$\Phi$ being the azimuthal angle around the $z$-axis.
The pseudorapidity is defined in terms of the polar angle $\theta$ as $\eta = -\ln \tan(\theta/2)$.
Angular distance is measured in units of $\Delta R \equiv \sqrt{(\Delta\eta)^{2} + (\Delta\Phi)^{2}}$. Both the rapidity and pseudorapidity are calculated in the center-of-mass frame per nucleon pair.
}

When ultra-relativistic beams of lead nuclei are brought into collision, the processes often studied are those for which the nuclei have an impact parameter smaller than twice the nuclear radius. Such collisions are now understood to create droplets of quark-gluon plasma (QGP) that flow as a nearly perfect fluid, i.e., hydrodynamically~\cite{Harris:2023tti}.
However, the strong electromagnetic (EM) fields of the fully ionized nuclei can also induce interactions when the nuclei have significantly larger impact parameters~\cite{Bertulani:2005ru,Klein:2019qfb}. In the equivalent photon approximation, these strong EM fields correspond to a flux of quasi-real, high-energy photons. Importantly, the nuclei can produce high-energy photons coherently from the entire nucleus, resulting in an enhancement to the photon spectrum over a broad energy range which is proportional to $Z^2$ (e.g., atomic number Z = 82 for Pb).

As a result, the rates for EM interactions (which include photon--photon and photon--nucleus scatterings) are large enough to be measurable in \PbPb collisions at the Large Hadron Collider (LHC). Such collisions are commonly referred to as ``ultra-peripheral collisions'' (UPCs) because they can occur when the impact parameters between the incoming nuclei are large enough such that there is no hadronic interaction between the nuclei.  The ATLAS Collaboration has measured UPC events where the basic interactions are photon--photon collisions~\cite{HION-2016-02,HION-2016-05,HION-2018-19,STDM-2019-19},  including light-by-light scattering and scattering where two leptons in the final state are produced.  The ATLAS Collaboration has also measured UPC photo-nuclear collisions, for example in the case of dijet production~\cite{HION-2022-15}. In photo-nuclear reactions, the photon could act as a point-like particle interacting with a parton in the nucleus (the ``direct'' case). However, the vector-meson dominance picture suggests that the photon could fluctuate to a vector meson, for example a $\rho$ meson, which then interacts with the Pb nucleus (the ``resolved'' case)~\cite{Bertulani:2005ru,Baltz:2007kq}. Therefore, some subset of these collisions could be considered as $\rho$--nucleus collisions, albeit at a lower center-of-mass collision energy than the nucleon-nucleon $\sqrt{s_{NN}}$, depending on the $\rho$ energy.   Hence, such events will have an overall rapidity boost of the center-of-mass frame in the direction of the nucleus.

Two-particle azimuthal correlations have been measured in photo-nuclear ($\gamma+$Pb) events by ATLAS~\cite{HION-2018-27}. These results indicate significant non-zero elliptic ($v_2$) and triangular ($v_3$) flow coefficients. These coefficients have been interpreted in terms of a hydrodynamically flowing medium~\cite{Zhao:2022ayk}, and alternatively in terms of scattering diagrams in the glasma framework~\cite{Shi:2020djm}. The $v_2$ values are significantly smaller in photo-nuclear events compared with \pPb events at the same charged particle multiplicity \Nch.  The lower elliptic flow in photo-nuclear events may be explained by the stronger longitudinal decorrelations in the rapidity-shifted photo-nuclear events in hydrodynamic calculations~\cite{Zhao:2022ayk}. These authors make the specific prediction that the radial flow~\cite{Heinz:2013th} is essentially the same in photo-nuclear and \pPb collisions, as measured via the mean transverse momentum, \pT, of charged and identified particles. The formation of a small QGP droplet may also lead to other manifestations of QGP seen in heavy-ion collisions such as a baryon/meson enhancement~\cite{Fries:2003vb} and strangeness enhancement~\cite{ALICE:2016fzo,STDM-2018-60}. A measurement by CMS of two-particle correlations in the lower-multiplicity $\gamma+p$ system did not find evidence for collective effects when compared to the expectation from event generators~\cite{CMS-HIN-18-008}.

This paper presents the yields of charged hadrons and identified strange hadrons
(\ks, \lmda, \csd) in photo-nuclear collisions using 1.7 $\mathrm{nb^{-1}}$ of $\sqrt{s_\mathrm{NN}} = 5.02$~\TeV \PbPb data collected in 2018 with the ATLAS detector at the LHC. The results are compared with measurements in \pPb collisions specifically to test the predictions of Ref.~\cite{Zhao:2022ayk}.    Additionally, these photo-nuclear data are compared with the photo-nuclear Monte Carlo (MC) \dpmjet~\cite{Engel:1994vs,Roesler:2000he} generator and with hydrodynamic calculations~\cite{Zhao:2022ayk} to test the hypothesis of a small QGP droplet formation.

\section{The ATLAS detector}

The ATLAS detector~\cite{PERF-2007-01} at the LHC~\cite{Evans:2008zzb} covers nearly the entire solid angle around the collision point~\footnote{\AtlasCoordFootnote}.  It consists of an inner tracking detector surrounded by a thin superconducting solenoid, electromagnetic and hadronic calorimeters, and a muon spectrometer incorporating three large superconducting magnets. The inner-detector system (ID) is immersed in a \SI{2}{\tesla} axial magnetic field and provides charged-particle tracking in the range $|\eta| < 2.5$.

The high-granularity silicon pixel detector covers the vertex region and typically provides three measurements per track. An innermost insertable B-layer~\cite{ATLAS-TDR-19} has been operating as a part of the pixel detector since 2015. It is followed by the silicon microstrip tracker (SCT) which usually provides four two-dimensional measurement points per track. These silicon detectors are complemented by the transition radiation tracker (TRT), which enables radially extended track reconstruction up to $|\eta| = 2.0$.

The calorimeter system covers the pseudorapidity range $|\eta| < 4.9$. Within the region $|\eta|< 3.2$, electromagnetic calorimetry is provided by barrel and endcap high-granularity lead/liquid-argon (LAr) electromagnetic calorimeters, with an additional thin LAr presampler covering $|\eta| < 1.8$, to correct for energy loss in material upstream of the calorimeters. Hadronic calorimetry is provided by the steel/scintillating-tile calorimeter, segmented into three barrel structures within $|\eta| < 1.7$, and two copper/LAr hadronic endcap calorimeters covering $1.5 < \left|\eta\right| < 3.2$. The angular coverage is completed with forward copper/LAr and tungsten/LAr calorimeter modules optimized for electromagnetic and hadronic measurements, respectively. The muon spectrometer (MS) surrounds the calorimeters and is based on three large air-core toroidal superconducting magnets with eight coils each. The MS includes a system of precision tracking chambers and fast detectors for triggering.
The minimum-bias trigger scintillator, reconfigured for Run 2, detects charged particles over $2.07<|\eta|<3.86$ using two hodoscopes of 12 counters positioned at $z=\pm{3.6}{~\mathrm{m}}$.
The Zero-Degree Calorimeters (ZDCs) play a key role in identifying UPC events in heavy-ion collisions. They are located at $z=\pm\si{140}{~\mathrm{m}}$ from the interaction point, just beyond the point where the common straight-section vacuum-pipe divides back into two independent beam-pipes. The ZDC modules consist of layers of alternating quartz rods and tungsten plates that measure neutral particles at pseudorapidities $|\eta| > 8.3$.

A two-level trigger system~\cite{TRIG-2016-01} is used to select events. The first-level trigger (L1) is implemented in hardware and uses a subset of the detector information to reduce the accepted rate to at most \SI{100}{\kilo\hertz}. This is followed by the software-based high level trigger (HLT) that
reduced the accepted event rate to 1--\SI{4}{\kilo\hertz} depending on the data-taking conditions during 2018 \PbPb collisions.

A software suite~\cite{SOFT-2022-02} is used in data simulation, in the reconstruction and analysis of real and simulated data, in detector operations, and in the trigger and data acquisition systems
of the experiment.

\section{Event selection and simulations}

Photo-nuclear interactions are selected via event topologies where one of the Pb nuclei remains intact, resulting in no spectator neutrons and very sparse particle production downstream of the given nucleus. This is taken to be the photon-going direction. The datasets and event selection for photo-nuclear collisions are identical to those used in a previous measurement of two-particle azimuthal correlations~\cite{HION-2018-27}, and are briefly summarized below.

The measurements presented in this paper were performed using the $\sqrt{s_\mathrm{NN}} = 5.02$~\TeV \PbPb dataset collected with a variety of triggers in 2018, with a total integrated luminosity of 1.7~nb$^{-1}$. The 25-track high-multiplicity trigger (HMT), 15-track HMT, and the minimum-bias (MB) trigger were configured with progressively higher prescale factors, sampling 1.6~nb$^{-1}$, 0.13~nb$^{-1}$, and 1.0 ~$\mu\mathrm{b}^{-1}$ of data, respectively. Photo-nuclear candidate events were first selected by the trigger requiring one ZDC side (referred to as the Pb-going side and corresponding to $\eta<0$) to have a minimum amount of energy at L1, $E > 1$~\TeV, consistent with the presence of one or more neutrons. The other side (referred to as the photon-going side and corresponding to $\eta>0$) was required to have energy below a maximum-energy cutoff, $E < 1$~\TeV, consistent with no neutrons.
The per-nucleon energy is 2.5~\TeV, leading to a single neutron peak energy well above this 1~\TeV threshold~\cite{HION-2016-02}. The selected topology is referred to as ``0nXn''. Events were also required to satisfy an upper bound of $200$~\GeV on the total transverse energy deposited across all central calorimeters at L1, for further rejection of hadronic Pb+Pb events.

Reconstructed pseudorapidity gap quantities, constructed using charged-particle tracks and clusters of energetic calorimeter cells in each event, are used to distinguish between different physics processes such as photo-nuclear collisions, low-activity (peripheral) hadronic Pb+Pb collisions, and $\gamma\gamma\rightarrow X$ processes. The requirement of a rapidity gap above a minimum value in the photon-going direction can efficiently remove peripheral Pb+Pb events. Rather than the traditional pseudorapidity gap quantity~\cite{STDM-2011-01}, which determines the pseudorapidity difference between the edge of the detector and the closest particle, an alternative ``sum-of-gaps'' definition is used, which adds together contiguous gaps separated by particle production concentrated in a narrow pseudorapidity region. This alternative definition is used to retain a large selection efficiency for resolved photon events
where a large contiguous pseudorapidity gap may otherwise be spoiled by a hadronic fragment on the photon-going side. The quantity \sumgapgamma corresponds to the sum-of-gaps calculated in the photon-going half of the detector, and is constructed using tracks with $\pt > 0.4$~\GeV, $|\eta| < 2.5$ and clusters with $\pt > 0.2$~\GeV, $|\eta| < 4.9$  in each event.
It is calculated by first sorting the tracks and clusters in $\eta$. The differences in $\eta$ between adjacent particles, $\Delta\eta$, are included in the sum if they are larger than $0.5$. The value of $0.5$ was observed in the simulation to retain good efficiency for resolved photon events. The gap calculation is computed as always starting from mid rapidity to the edge of the detector; thus, \sumgapgamma  ranges from $0$ to $4.9$. UPC events with $\sumgap > 2.5$ are utilized in this paper following the procedure in Ref.~\cite{HION-2018-27}.

Despite the 0nXn ZDC and \sumgap selections, a residual contamination of the photo-nuclear events by \PbPb peripheral inelastic collisions remains, which is smaller than 3\%. To account for this, a small purity correction factor and associated systematic uncertainties are applied, as performed in Ref.~\cite{HION-2018-27}.
No other backgrounds, such as from $\gamma\gamma$ processes, were found to be significant after the event selection.

This paper includes comparisons with the 2016 \pPb collision data collected at $\sqrt{s_\mathrm{NN}} = 5.02$~\TeV, with an integrated luminosity of 0.1 $\mathrm{nb^{-1}}$, obtained using a minimum-bias trigger as detailed in Ref.~\cite{HION-2017-07}.
In the $p$+Pb system, positive and negative (pseudo)rapidities denote the proton- and nucleus-going directions, respectively.

\begin{figure}[!t]
\centering
\subfigure[]{\includegraphics[width=0.49\textwidth]{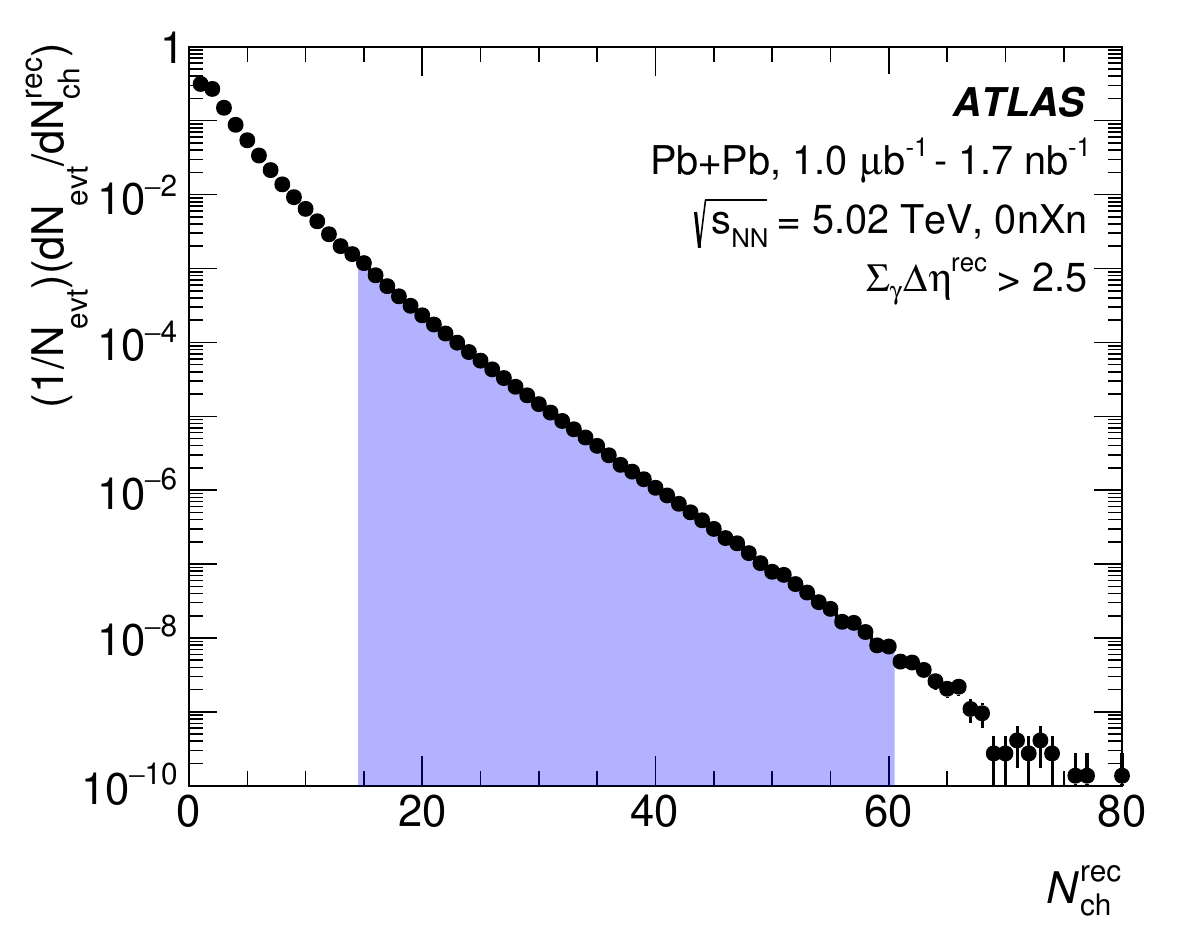}}
\subfigure[]{\includegraphics[width=0.49\textwidth]{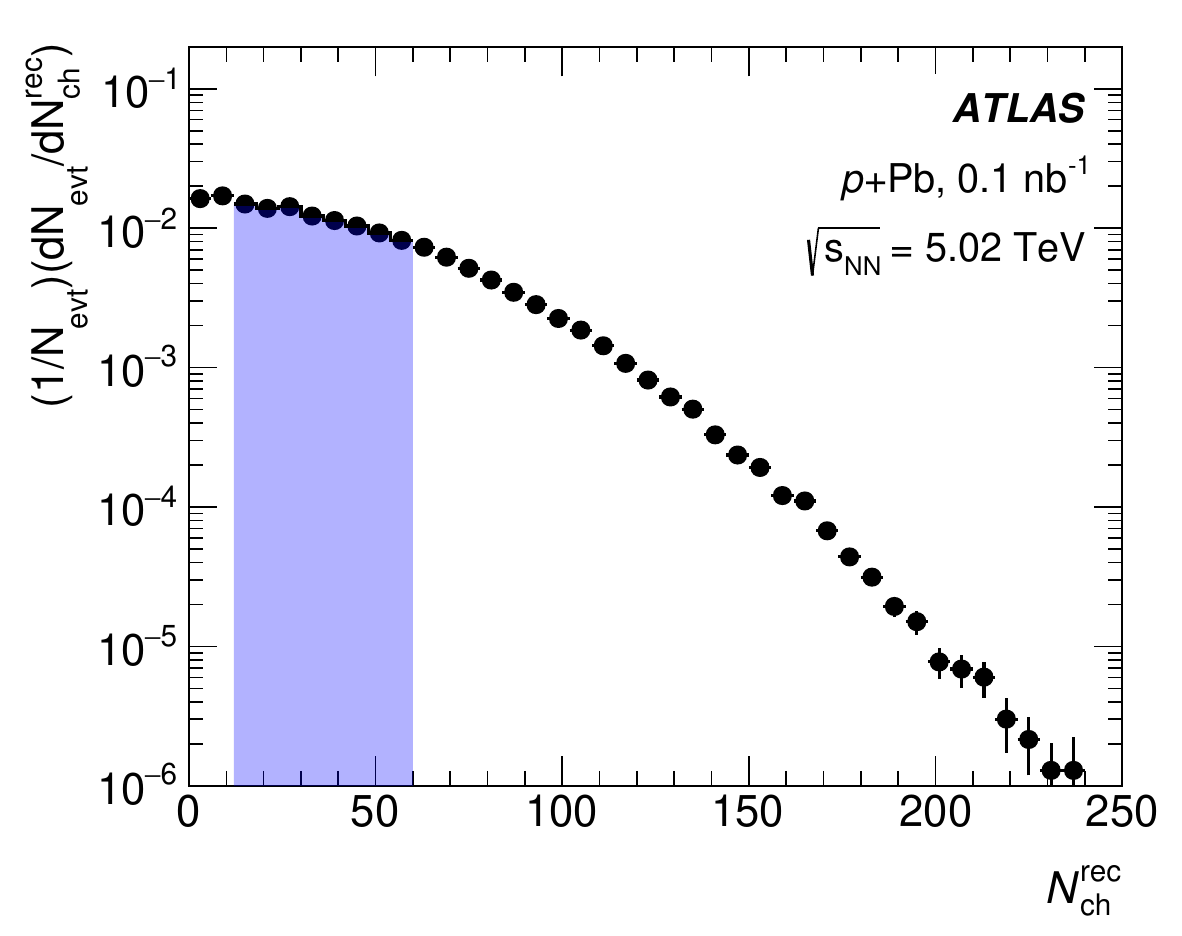}}
\caption{The multiplicity distributions (\Nch) from (a) \PbPb photo-nuclear and (b) \pPb collisions. The \Nch range utilized in this paper, $15 \leq \Nch \leq 60$, is highlighted.}
\label{fig:Nch}
\end{figure}

Each event is characterized by  the number of reconstructed tracks with $\pt > 0.4$~\GeV~and $|\Eta|<2.5$, referred to as the reconstructed charged-particle multiplicity (\Nch).
This standard ATLAS heavy-ion event class definition utilizes reconstructed tracks that are not corrected for track acceptance and efficiency, see for example Refs.~\cite{HION-2021-08,HION-2017-11,HION-2016-01,HION-2015-13}. Monte Carlo studies indicate that selections on \Nch correspond to equivalent selections on truth-level charged particles with $\pt > 0.4$~\GeV~and $|\Eta|<2.5$ as well, but with an average value of $N^{\mathrm{truth}}_{\mathrm{ch}}$ $\approx$ 1.2 $\times$ \Nch. Figure~\ref{fig:Nch}  shows the \Nch distributions in photo-nuclear \PbPb and \pPb events. The \Nch range utilized in this paper, $15 \leq \Nch \leq 60$,  is highlighted. The \Nch differential results are presented in the range $15 \leq \Nch \leq 60$ and \Nch integrated results are shown for $25 \leq \Nch \leq 60$. The \pPb events are then re-weighted to have effectively the same \Nch distribution as the UPC \PbPb events.

The simulated event sample for the photo-nuclear analysis is generated with \dpmjet + \starlight.
Events were generated with different minimum requirements on $N_\mathrm{ch}^\mathrm{rec}$ to provide a good statistical coverage over the $N_\mathrm{ch}^\mathrm{rec}$ range accessed in data.
First, the distribution of photon flux for Pb beams at the LHC was calculated using \starlight~\cite{Klein:2016yzr}. The flux distribution was passed to a multipurpose generator based on the Dual Parton Model (DPM) and referred to as \dpmjet~\cite{Engel:1994vs,Roesler:2000he}, which simulates direct and resolved photon--lead ($\gamma$+Pb) interactions at the generator level.

The DPM model is a diagrammatic way of describing particle production in hadron--hadron collisions~\cite{Wong1994}.  There are two major concepts that underlie the dual-parton model.  The first is the dual resonance model~\cite{Koba:1969rw}, which assumes that there are two alternative (or ``dual'') descriptions of hadron--hadron interactions -- the $t$-channel diagram where particles can be exchanged as a form of interaction and the $s$-channel diagram where the two incoming particles fluctuate into an intermediate state and then interact.  The second is the Veneziano scattering amplitude, which allows for a convergent calculation of the scattering amplitude for the exchange of a large set of particles.  These two concepts enable calculations in soft hadron physics through \textit{pomeron} exchange.  The pomeron is a particle with vacuum quantum numbers analogous to a closed string and can be exchanged between hadrons as a form of interaction.  Thus, through the dual resonance model there are intermediate states in elastic hadron--hadron collisions with a large number of pomerons.  These diagrams can be ``cut'' to calculate the amplitude of the inelastic process of hadrons interacting to form a large number of primarily meson final states.  The \dpmjet MC generator combines the DPM with perturbative QCD (pQCD), as well as other features, to attempt a full description of hadron--hadron, hadron--photon, and photon--photon collisions~\cite{Roesler:2000he}.
The full set of particles was then run through a full $\GEANT$~\cite{Agostinelli:2002hh} simulation of the ATLAS detector. A sample of thirteen million $\gamma$+Pb events were generated.

The simulated event sample for the \pPb analysis is generated with \hijing~\cite{Wang:1991hta}. The \hijing model combines pQCD inspired models for multiple-jet production with low-\pt multi-string phenomenology. The model thus extends the \Pythia string picture ~\cite{Bierlich:2022pfr} to include modeling of both high-energy \pp collisions, as well at $p$--nucleus and nucleus--nucleus collisions.
The geometry for multiple collisions in $p$--nucleus and nucleus--nucleus collisions is provided by MC Glauber~\cite{Miller:2007ri}. The model also includes multiple mini-jet production, nuclear shadowing of parton distribution functions, and a schematic mechanism of jet interactions in dense matter.   The phenomenological parameters are adjusted to reproduce essential features of \pp multi-particle production data for a wide energy range ($\sqrt{s_\mathrm{NN}} = 5$~\GeV~to 2 \TeV).   For the sample used here the so-called ``jet quenching'' feature is turned off.   A sample of five million \pPb \hijing events was generated.

\section{Analysis}

This paper reports charged-hadron and identified-strange-hadron yields reconstructed using tracks in the inner tracker with $\pT > 0.1$ \GeV and $|\Eta| < 2.5$.    Both utilize the same event selection criteria detailed above.   Additionally, the yields are determined in both photo-nuclear \PbPb and \pPb using identical track reconstruction and extraction methods.

\subsection{Charged hadrons}

The charged-hadron analysis utilizes tracks that originate from the collision, referred to as primaries. Primary particles are defined as charged particles with a lifetime $\tau > 3 \times 10^{-10}$~s, either directly produced in the collision or from subsequent decays of directly produced particles with $\tau < 3.0 \times 10^{-11}$~s. This definition excludes charged strange baryons that have a very small probability to actually traverse the tracker before decaying (for example the $\Xi^{-}$ with $\tau=1.6 \times 10^{-10}$~s and $\Omega^{-}$ with $\tau = 0.8 \times 10^{-10}$~s). However, it includes charged hadrons from the decay of $\Delta$ resonances and $\rho$ mesons with lifetimes shorter than $3.0 \times 10^{-8}$~s. The contribution of charged leptons is negligible and thus the tracks represent charged hadrons.
The track reconstruction follows that utilized for low pileup \pp data-taking~\cite{STDM-2010-06,ATLAS:2015AS}. The reconstructed tracks are required to satisfy quality criteria as detailed in Ref.~\cite{STDM-2015-17}. Tracks are further required to have $\pt > 0.1~\GeV$, $|\Eta| < 2.5$, and a distance of closest approach to the reconstructed vertex in both the longitudinal and transverse directions of less than 1.5~mm.

The reconstructed tracks are then used to calculate charged-hadron yields as functions of \pT in different \Eta slices:
\begin{equation}
Y_\mathrm{1}(\Eta,\pt) = \frac{1}{N_\mathrm{ev}} \frac{\mathrm{d}N^2_\mathrm{ch}}{\mathrm{d}\pt \mathrm{d}\Eta}
\label{eq:yield_eta_pt}
\end{equation}
and then yields integrated over \pt as a function of \Eta:
\begin{equation}
Y_\mathrm{2}(\Eta) = \frac{1}{N_\mathrm{ev}} \frac{\mathrm{d}N_\mathrm{ch}}{\mathrm{d}\Eta},
\label{eq:yield_eta}
\end{equation}
where $N_\mathrm{{ev}}$ is the number of selected events and $N_\mathrm{ch}$  is the number of charged particles.

The tracks entering these observables in bins of \pT and \Eta are corrected for reconstruction and selection inefficiency, as well as for contributions from tracks that are not associated with primary particles, on a per-track basis using simulation-derived correction factors.

The reconstruction efficiency is defined as the ratio of the number of truth primary charged particles whose associated reconstructed track has a truth-matched primary charged particle $N_\mathrm{truth}^\mathrm{matched}$ (as defined in Ref.~\cite{ATL-PHYS-PUB-2017-016}) to the total number of truth primary charged particles, $N_\mathrm{truth}$, as a function of both \pt and \Eta:
\begin{equation}
\varepsilon(\eta,\pt) = \frac{N_\mathrm{truth}^\mathrm{matched} (\Eta,\pt)}{N_{\mathrm{truth}}(\Eta,\pt)}.
\label{eq:recoeff_def}
\end{equation}

The contributions to reconstructed tracks that are not associated with primary particles are classified into fake tracks and secondary tracks. In order to correct for these contributions, tracks are weighted on a track-by-track basis by the ``primary fraction'', $f_\mathrm{primary}$, which is estimated as a function of reconstructed kinematics in simulated events by taking the ratio of the number of primary tracks $N_\mathrm{ch}^\mathrm{primary}$ to the number of reconstructed tracks $N_{\mathrm{ch}}$:

\begin{equation}
f\mathrm{_{primary}} (\Eta,\pt) = \frac{{N_\mathrm{{ch}}^\mathrm{primary}}(\Eta,\pt)}{N_{\mathrm{ch}} (\Eta,\pt)}.
\label{eq:prifr_def}
\end{equation}

The yields as a function of \Eta are measured for \pt $>0.1$~\GeV and extrapolated using the \dpmjet MC down to \pt $=0$~\GeV, i.e, correcting the yields by the fraction of \dpmjet charged particles with \pt~$>0.1$~\GeV to charged particles down to \pt $=0$~\GeV. These correction factors are 5--15\% in photo-nuclear \PbPb and 5--7\% in \pPb collisions.

Finally, using the yields detailed above, the \meanpt in \Eta intervals are calculated as a function of \Nch. When calculating \meanpt the extrapolation down to \pt $=0$~\GeV is performed using a Modified Hagedorn fit~\cite{Biyajima:2006mv}:

\begin{equation}
\frac{1}{N} \frac{dN}{dp_\mathrm{T}} = A_1 \frac{p_\mathrm{T}^2}{\sqrt{p_\mathrm{T}^2+m_0^2}}\left(1+\frac{p_\mathrm{T}}{p_\mathrm{1}}\right)^{-n_1},
\label{eq:hage}
\end{equation}
where $m_\mathrm{0}$ is the rest mass of considered particle, $p_\mathrm{1}$ and $n_\mathrm{1}$ are the free parameters, and $A_\mathrm{1}$ is the normalization constant. The \meanpt in each \Eta bin and in each \Nch bin is calculated by finding the mean value of the fit results for $\pt > 0$~\GeV.

\subsection{Identified strange hadrons}

Only the \ks, \lmda and \csd originating from the primary vertex are considered. The contribution of secondary \lmda from $\Sigma^0$ decay is included, while contributions from the decay products of heavy baryons (\csd, $\Omega^-$) are excluded from the definition of primary \lmda. For \lmda and \csd, the definition includes only the baryon state and does not represent an average of baryons and anti-baryons. Anti-baryons have an additional acceptance correction due to annihilation processes, and \GEANT is known to not model this correctly~\cite{ALICE:2010vtz,ALICE:2010hjm}.

Identified strange hadrons are reconstructed using oppositely-charged tracks with $\pT > 0.1$ \GeV and $|\Eta| < 2.5$, which are fitted to a common secondary vertex using a Kalman filter~\cite{Fruhwirth:1987fm}. The \ks candidates in the $\pi^+ + \pi^-$ decay mode (branching ratio of 69.2\%) are required to satisfy the following criteria:
\begin{itemize}
\item The \Chi of the two-track vertex fit is required to be less than 15 (with one degree of freedom).
\item The cosine of the pointing angle in the transverse plane (\costheta) between the \ks momentum vector and the \ks flight direction, defined as the line connecting the reconstructed primary vertex to the decay direction, is required to fulfill the requirements:
\begin{itemize}
\item $\costheta > 0.999$ for \PbPb photo-nuclear collisions, except in the most backward rapidity bin,
\y:~$[-2.5,-1.6]$, $\costheta > 0.9999$.
\item $\costheta > 0.995$ for \pPb, except in the most backward and forward rapidity bins of \pPb, \y:~$[-2.5,-1.6]$ and \y:~$[1.6,2.5]$, $\costheta > 0.999$
\end{itemize}
\item Requirements on the minimum values of the variables \lxysigma and \ptsigma, in bins of \pT and \y. These are optimized using the Toolkit for Multivariate Data Analysis (TMVA) package within the ROOT framework~\cite{hoecker2009tmvatoolkitmultivariate}, where $L_\mathrm{xy}$ is the distance from the reconstructed primary vertex to the reconstructed secondary vertex (decay vertex of the \ks candidate) in the transverse plane, $\sigma_{L_\mathrm{xy}}$ is the uncertainty on $L_\mathrm{xy}$ reconstruction, \pt is the reconstructed momentum of the \ks candidate and $\sigma_{\pt}$ is the uncertainty in the \pt reconstruction.
\end{itemize}

The \lmda candidates in the $p + \pi^-$ decay mode (branching ratio of 63.9\%) are required to satisfy the following criteria:
\begin{itemize}
\item The \pT of the \lmda candidate is greater than 0.5 \GeV.
\item The \Chi of the two-track vertex fit is required to be less than 15 (with one degree of freedom).
\item The cosine of the pointing angle (\costheta) is required to be greater than 0.999.
\item Requirements on the minimum values of the variables \lxysigma and \ptsigma, in bins of \pT and \y, determined using the TMVA package as above.
\end{itemize}

The \csd candidates in the $\lmda + \pi^-$ decay mode (branching ratio of 99.5\%) are required to satisfy the following criteria:
\begin{itemize}
\item The \pT of the \csd candidate is greater than 1 \GeV, and $|y| < 1.6$.
\item The \Chi of the two-track vertex fit for reconstructing both \lmda and \csd is required to be less than 15 (with one degree of freedom).
\item The cosine of the pointing angle (\costheta) associated with \csd vertex is required to be greater than 0.9992.
\item The cosine of the pointing angle associated with \lmda vertex is required to be greater than 0.99999.
\item Requirements on the minimum values of the variables \lxysigma associated with vertex fit for reconstructing \lmda and \csd, in bins of \pT and \y, determined using the TMVA package as above
\end{itemize}

Below the minimum \pT value for \lmda and \csd, the efficiency is very low due to the slow pions. In contrast, the \ks can be measured down to \pt = 0 \GeV.   The tight selection criteria for the pointing angle and the minimum value of \lxysigma significantly improve the signal significance in all kinematic regions.

\begin{figure}[!t]
\centering
\subfigure[]{\includegraphics[width=0.49\textwidth]{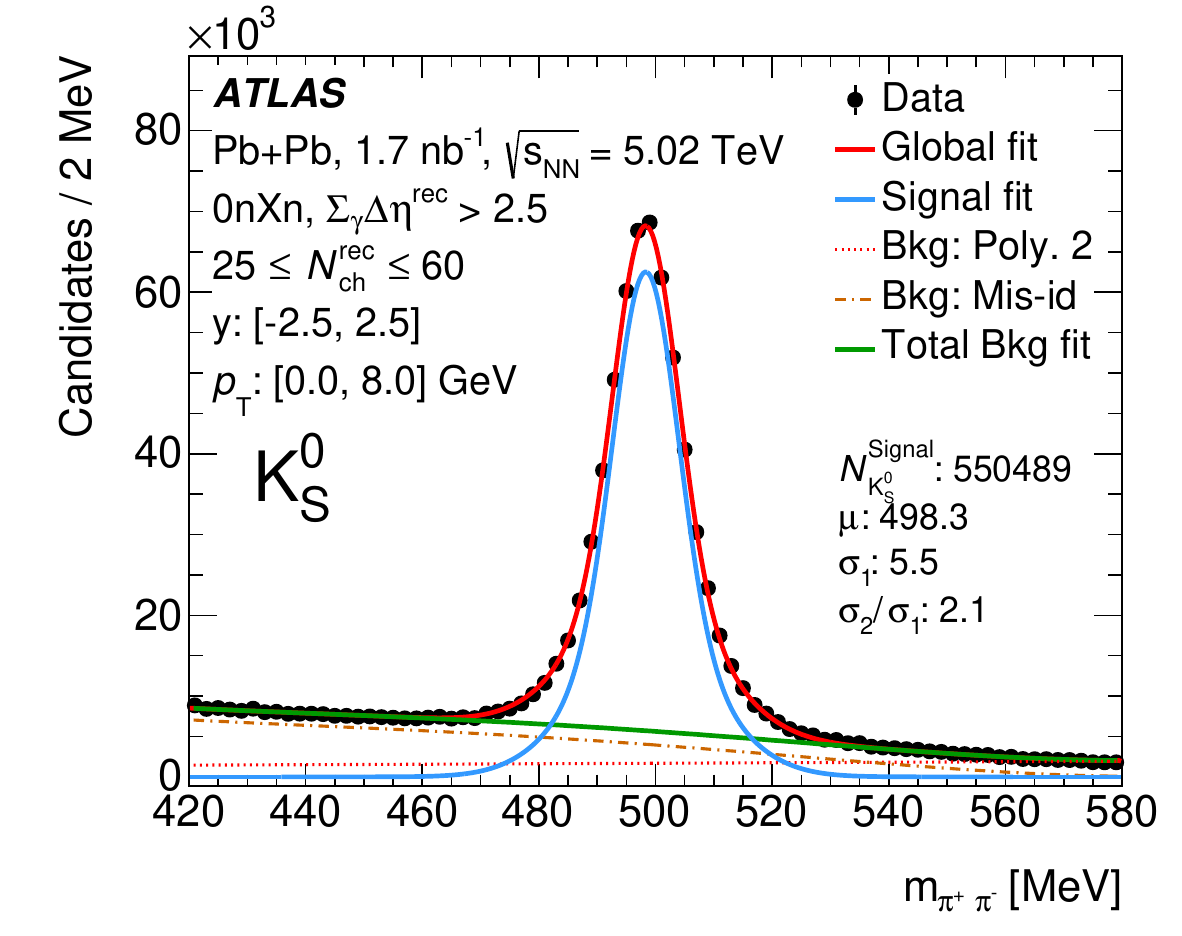}}
\subfigure[]{\includegraphics[width=0.49\textwidth]{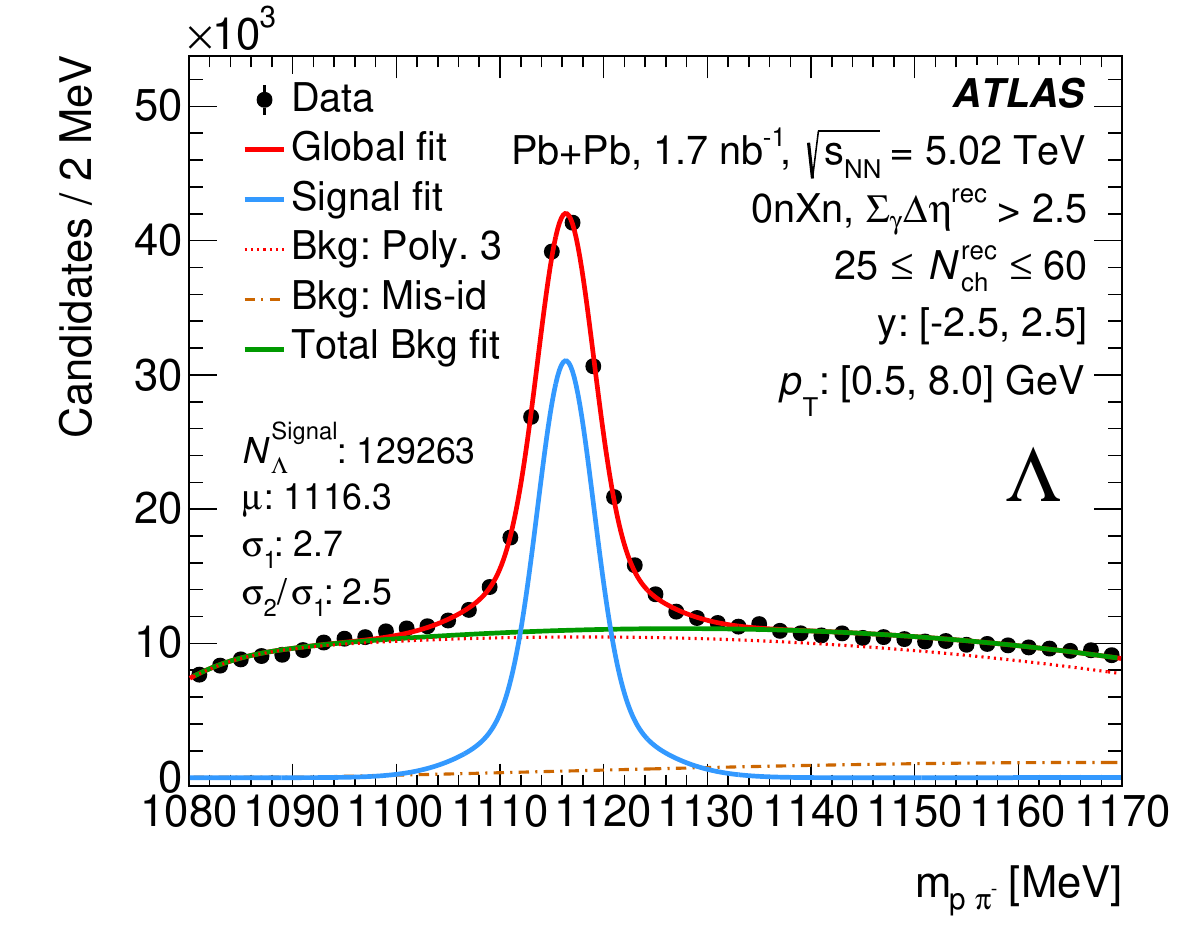}}
\subfigure[]{\includegraphics[width=0.49\textwidth]{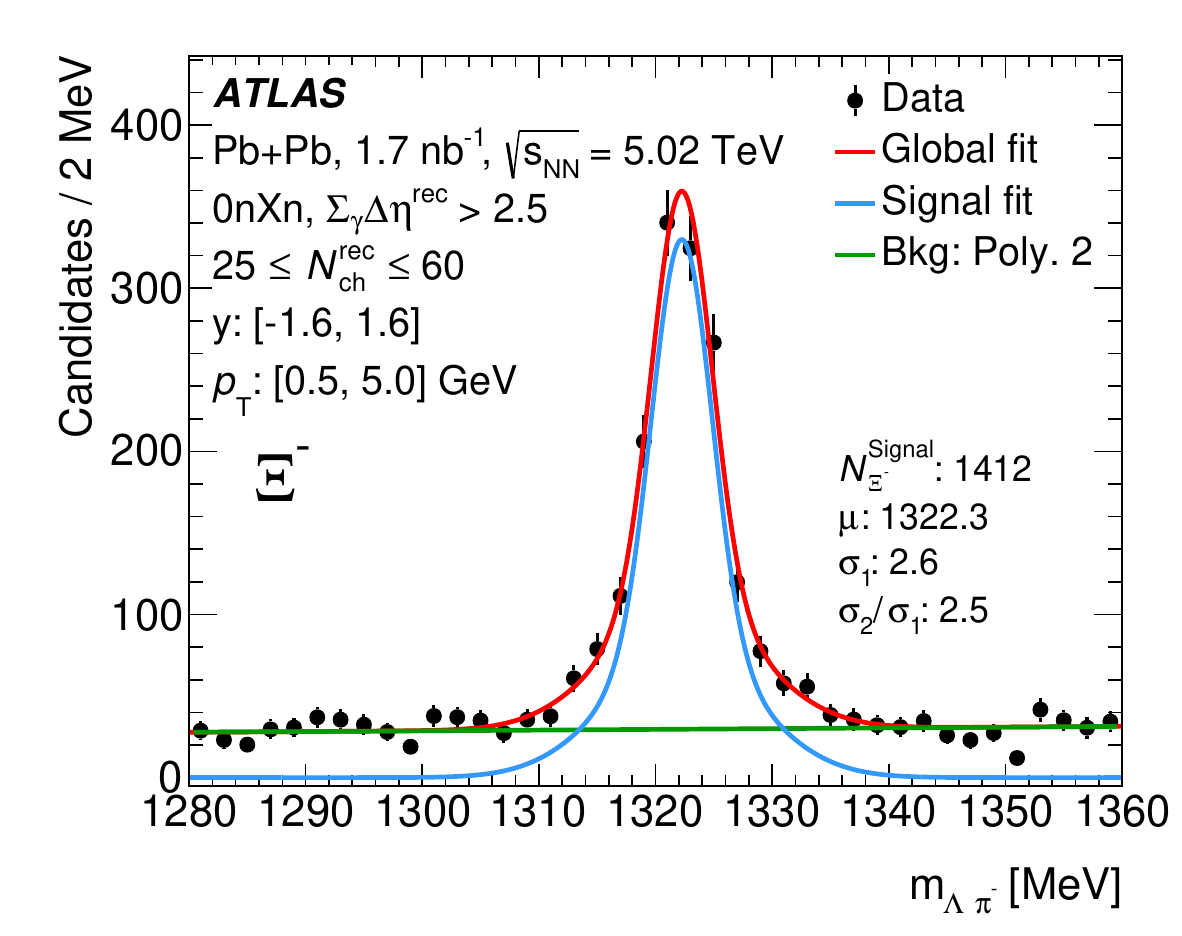}}
\caption{The invariant-mass distributions of (a) \ks, (b) \lmda, and (c) \csd in \PbPb photo-nuclear collisions. The data are fitted to a signal (double Gaussian) and a background component (detailed in the text). The signal counts, and fit parameters for the mean $\mu$ and widths $\sigma_1$ and $\sigma_2$ of the double Gaussian are also shown in units of \MeV.}
\label{fig:mass_upc}
\end{figure}

Figure~\ref{fig:mass_upc} shows the resulting invariant-mass distributions for \ks, \lmda, and \csd in \PbPb photo-nuclear collisions.
The number of signal candidates in a given \pt and \y bin is determined by fitting the invariant-mass distribution of the strange hadron candidates in that bin. The fit utilizes a double Gaussian for the signal peak and a second-order (third-order for \lmda) polynomial for the combinatorial background. The background component also includes a functional form modeling the cases where the particles are mis-identified, e.g., a truth \ks that has its decay products fit under the assumption of a parent \lmda particle.     These functional forms are determined from \dpmjet MC.
The ratio of the widths and amplitudes of the two Gaussian distributions is constrained based on the MC truth-matched invariant-mass distributions within the corresponding \pt-y bin. The fit ranges utilized in this analysis are [420, 580] \MeV for \ks, [1080, 1170] \MeV for \lmda, and [1280, 1360] \MeV for \csd. The quality of the fit is evaluated based on reasonable $\chi^2$ values.

\begin{figure}[!t]
\centering
\subfigure[]{\includegraphics[width=0.49\textwidth]{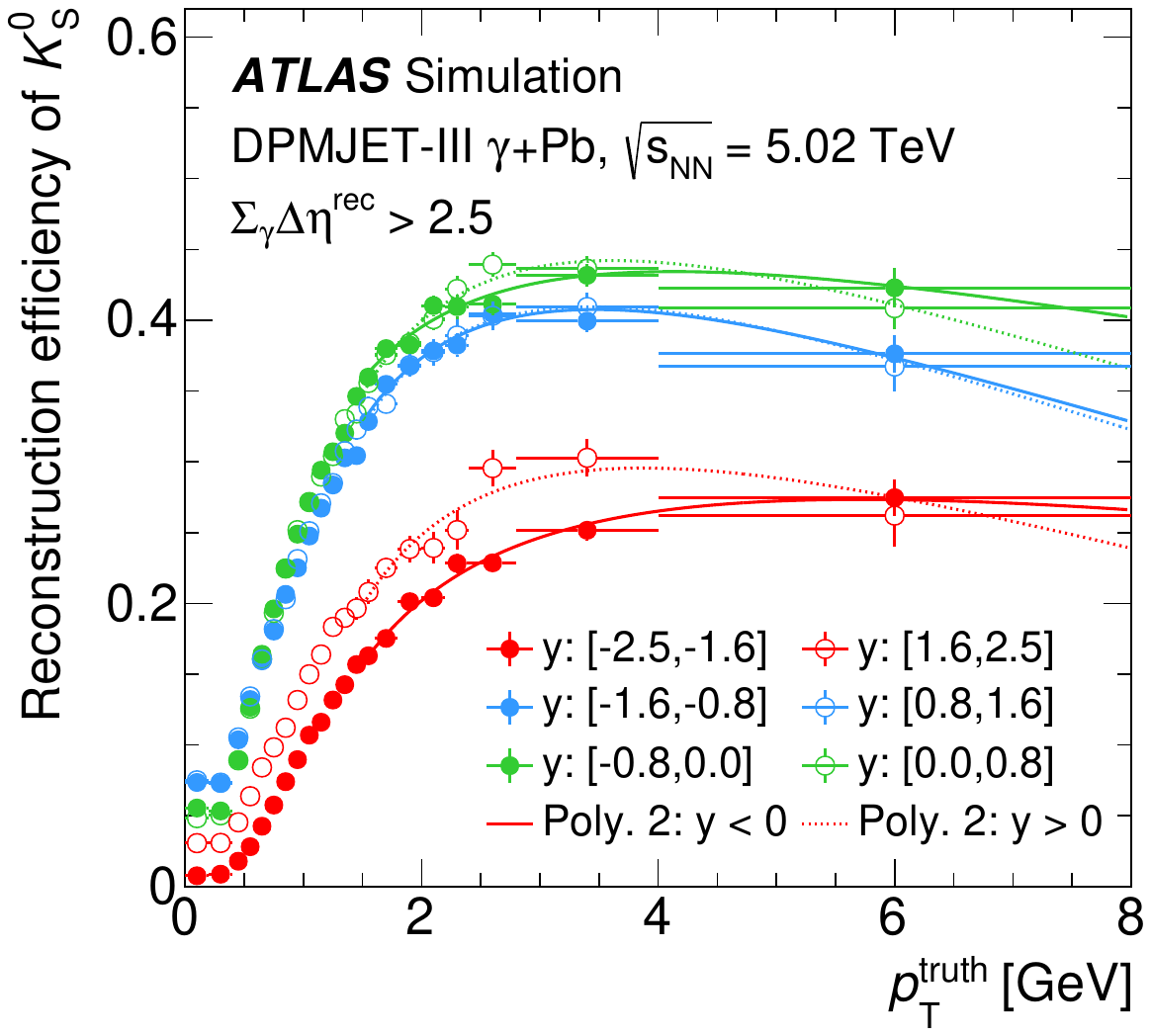}}
\subfigure[]{\includegraphics[width=0.49\textwidth]{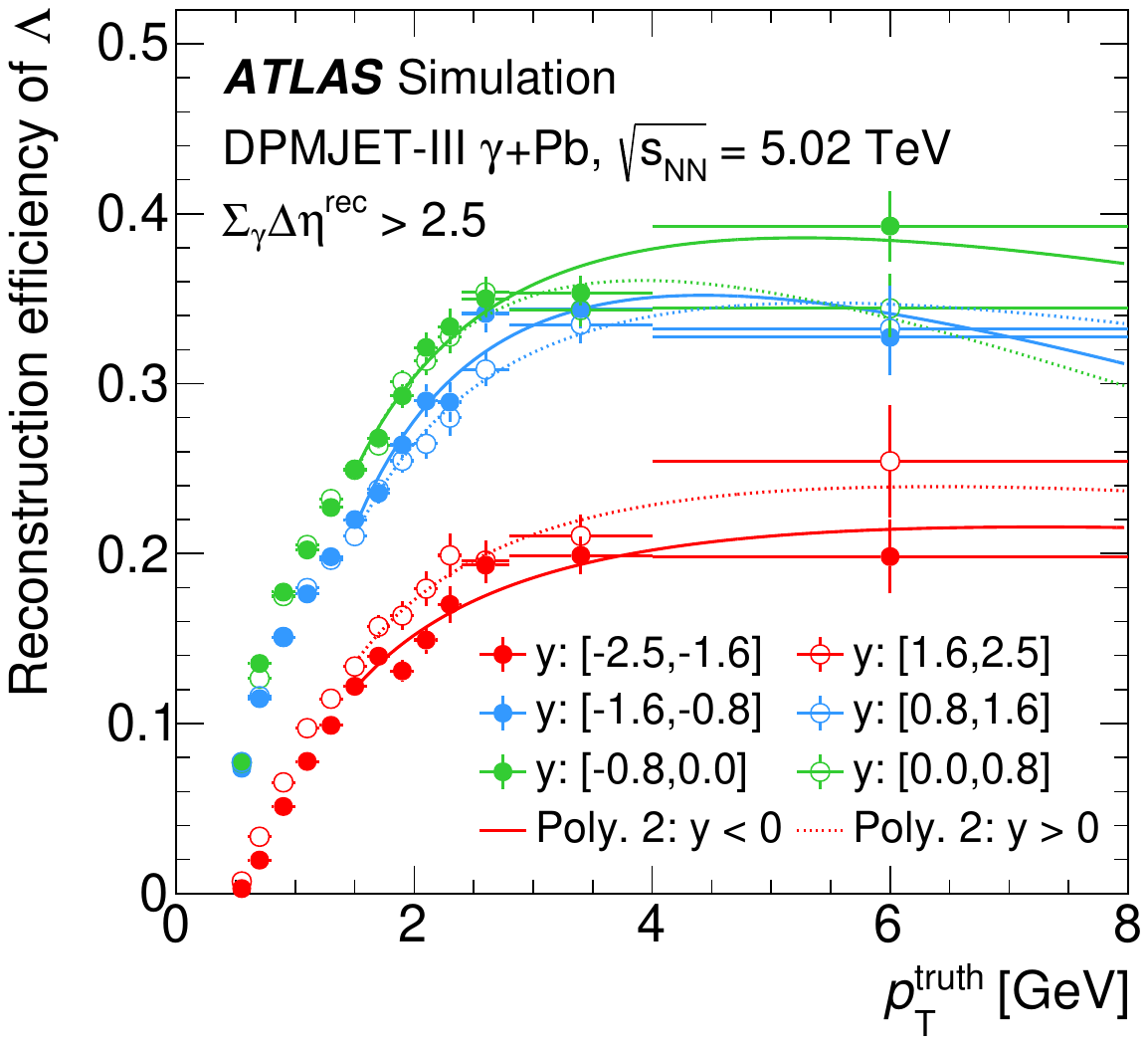}}
\subfigure[]{\includegraphics[width=0.49\textwidth]{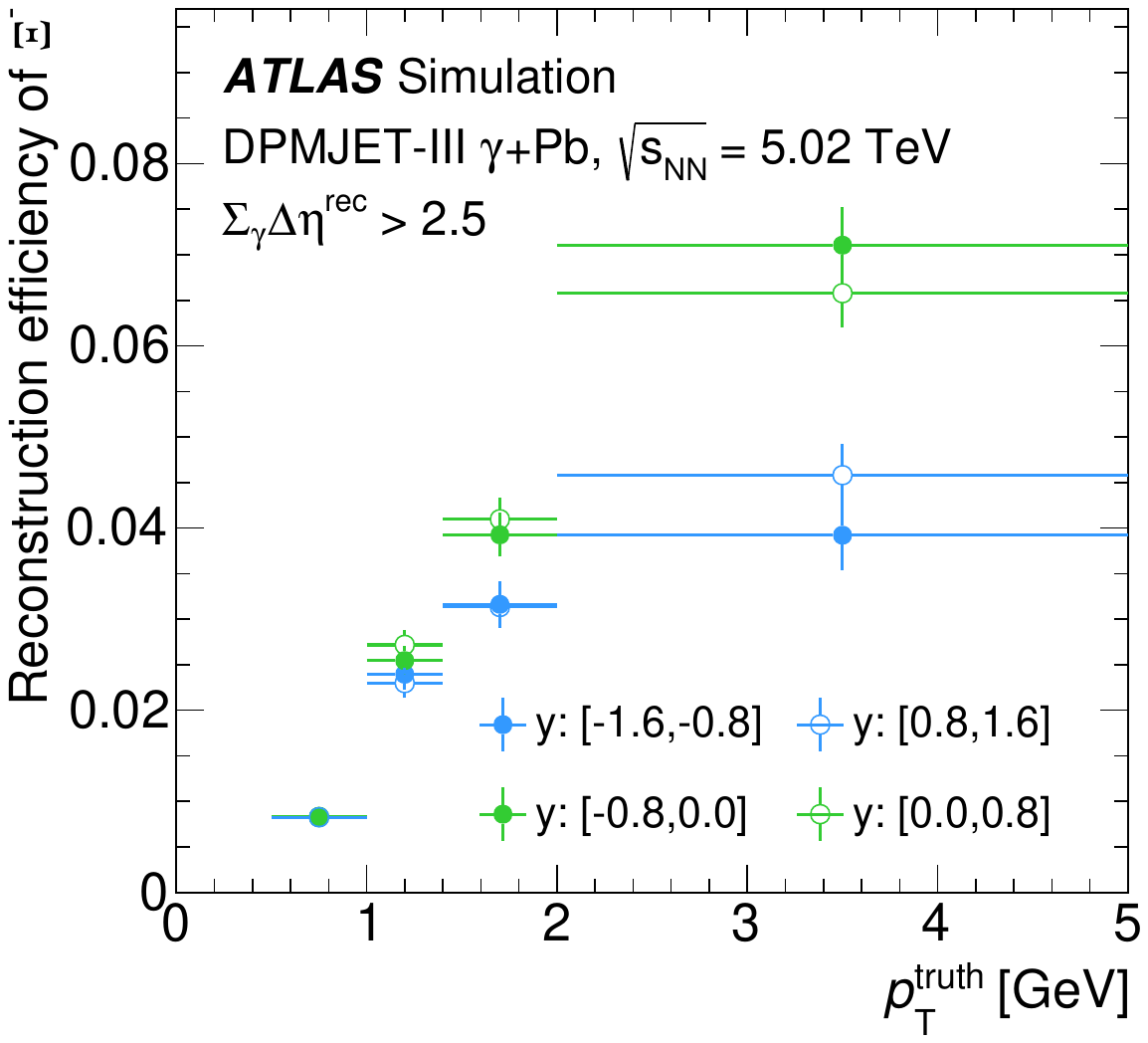}}
\caption{The reconstruction efficiencies of (a) \ks, (b) \lmda, and (c) \csd as a function of $\pt^{\mathrm{truth}}$ in intervals of $\y^{\mathrm{truth}}$ as determined using \dpmjet. The efficiencies are fitted using polynomial functions for \ks and \lmda.   Statistical uncertainties are shown as vertical lines.}
\label{fig:recoeff_ks_lmda_csd}
\end{figure}

The signal candidates, obtained in bins of \pt and \y, are corrected for reconstruction and selection inefficiencies, using Eq.~\ref{eq:recoeff_def} on the secondary vertex candidates, and for signal inefficiency, which accounts for the missed fraction of reconstructed candidates due to the TMVA-optimized selection requirements. Furthermore, signal candidates are corrected for the contributions of secondaries, as defined by Eq.~\ref{eq:prifr_def}. The largest sources of secondaries are from hadronic interactions of particles with the detector material and the decay products of heavier strange baryons, which contribute to the \lmda yield at the 10\% level~\cite{STDM-2010-09}. Figure~\ref{fig:recoeff_ks_lmda_csd} shows the reconstruction efficiencies for \ks, \lmda, and \csd calculated using \dpmjet $\gamma$+Pb simulations. High-\pt neutral strange particles start to have a sufficiently large relativistic boost that the decay occurs after some of the silicon-detector components, leading to missing hits and an efficiency drop.

The corrected number of signal candidates is used to calculate the identified-hadron yield as a function of \pt in different rapidity bins using Eq.~\ref{eq:yield_eta_pt}, and as a function of \y using Eq.~\ref{eq:yield_eta}.

\begin{figure}[!ht]
\centering
\subfigure[]{\includegraphics[width=0.49\textwidth]{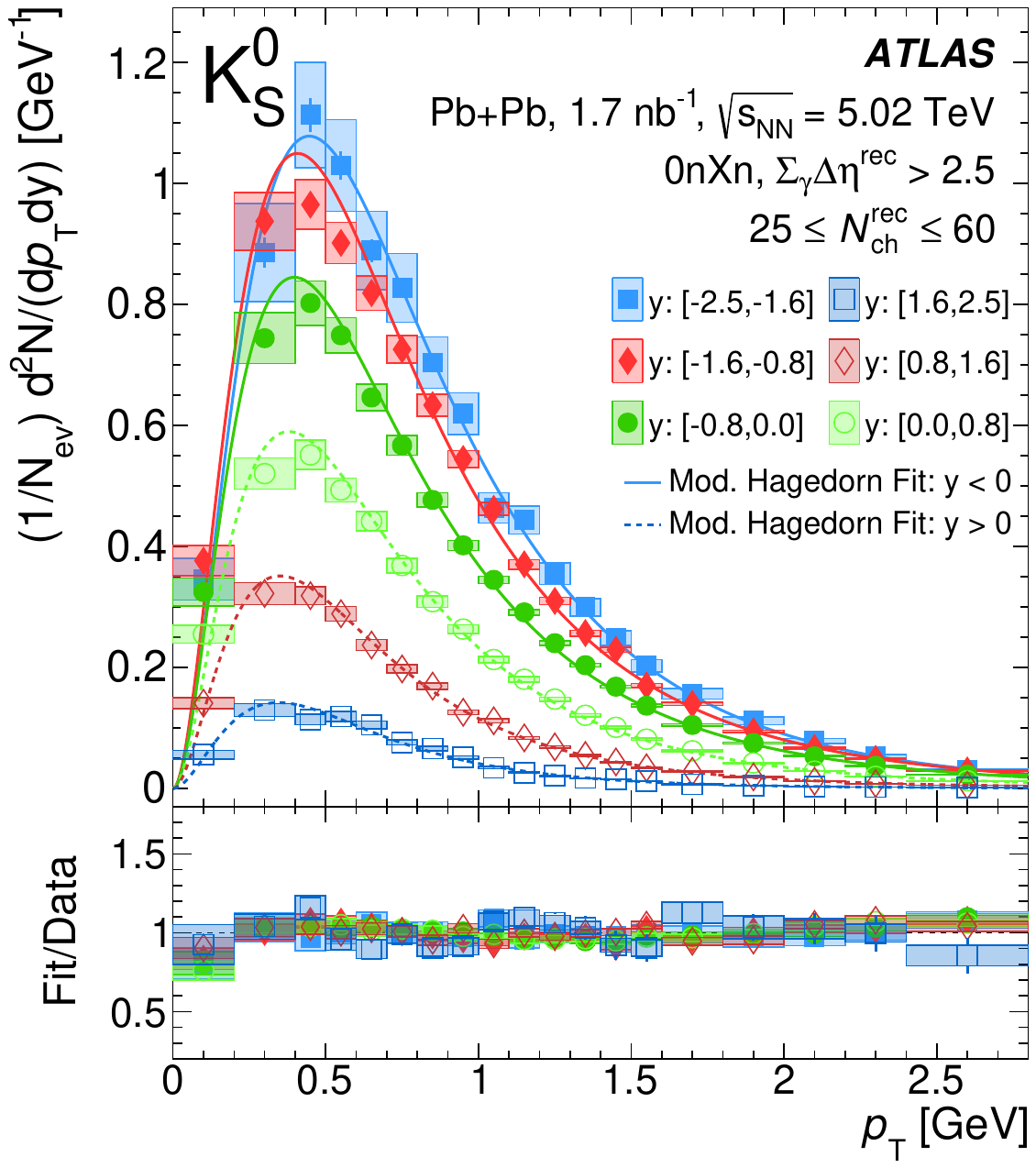}}
\subfigure[]{\includegraphics[width=0.49\textwidth]{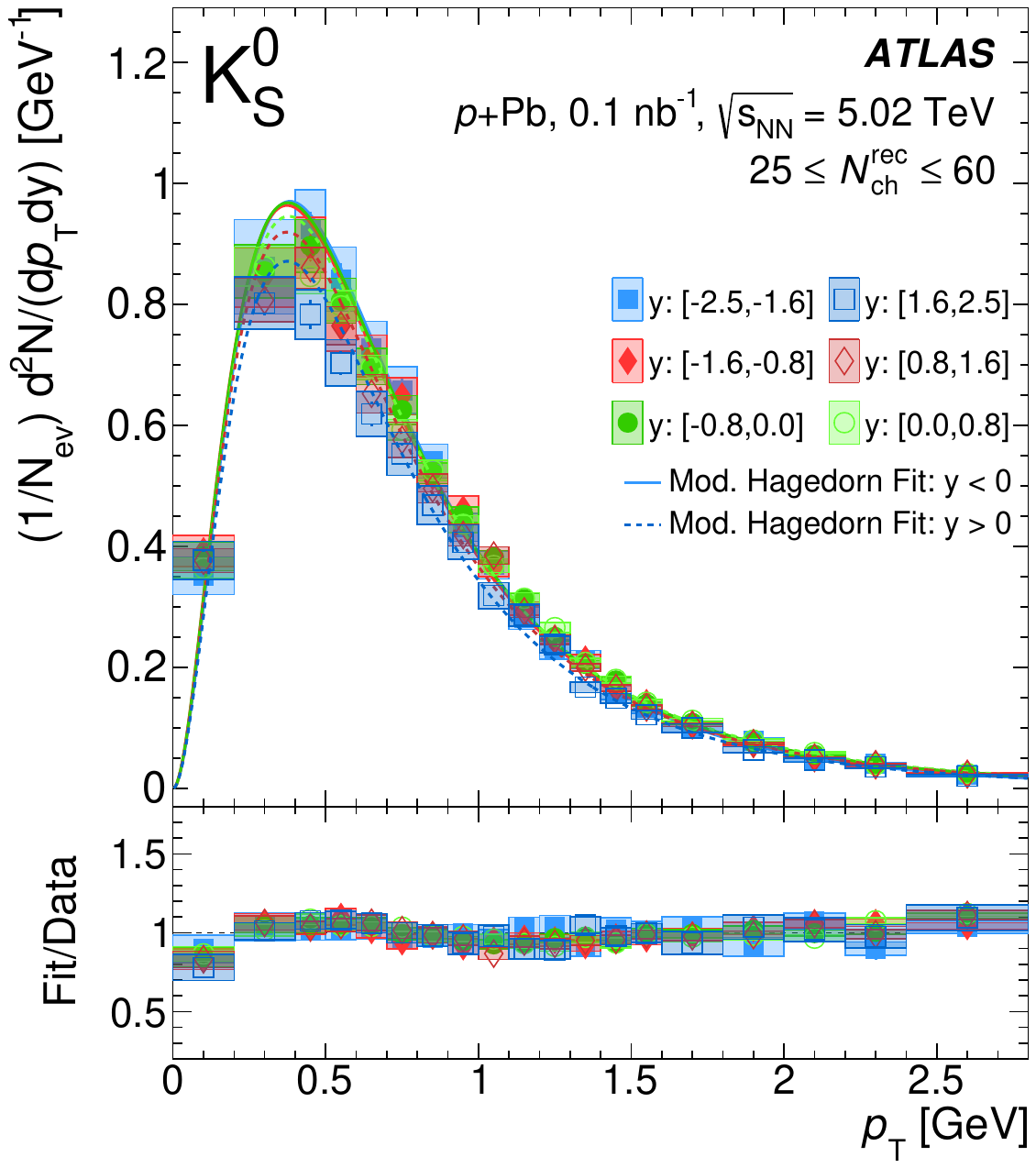}}
\caption{The \ks yields as a function of \pt in different \y selections in (a) \PbPb photo-nuclear and (b) \pPb collisions. The bottom panels show the ratio of the Modified Hagedorn fit results to the data. Statistical uncertainties are shown as vertical lines and systematic uncertainties are shown as colored boxes.}
\label{fig:pTspectra_extrap_k0s}
\end{figure}

\begin{figure}[!ht]
\centering
\subfigure[]{\includegraphics[width=0.49\textwidth]{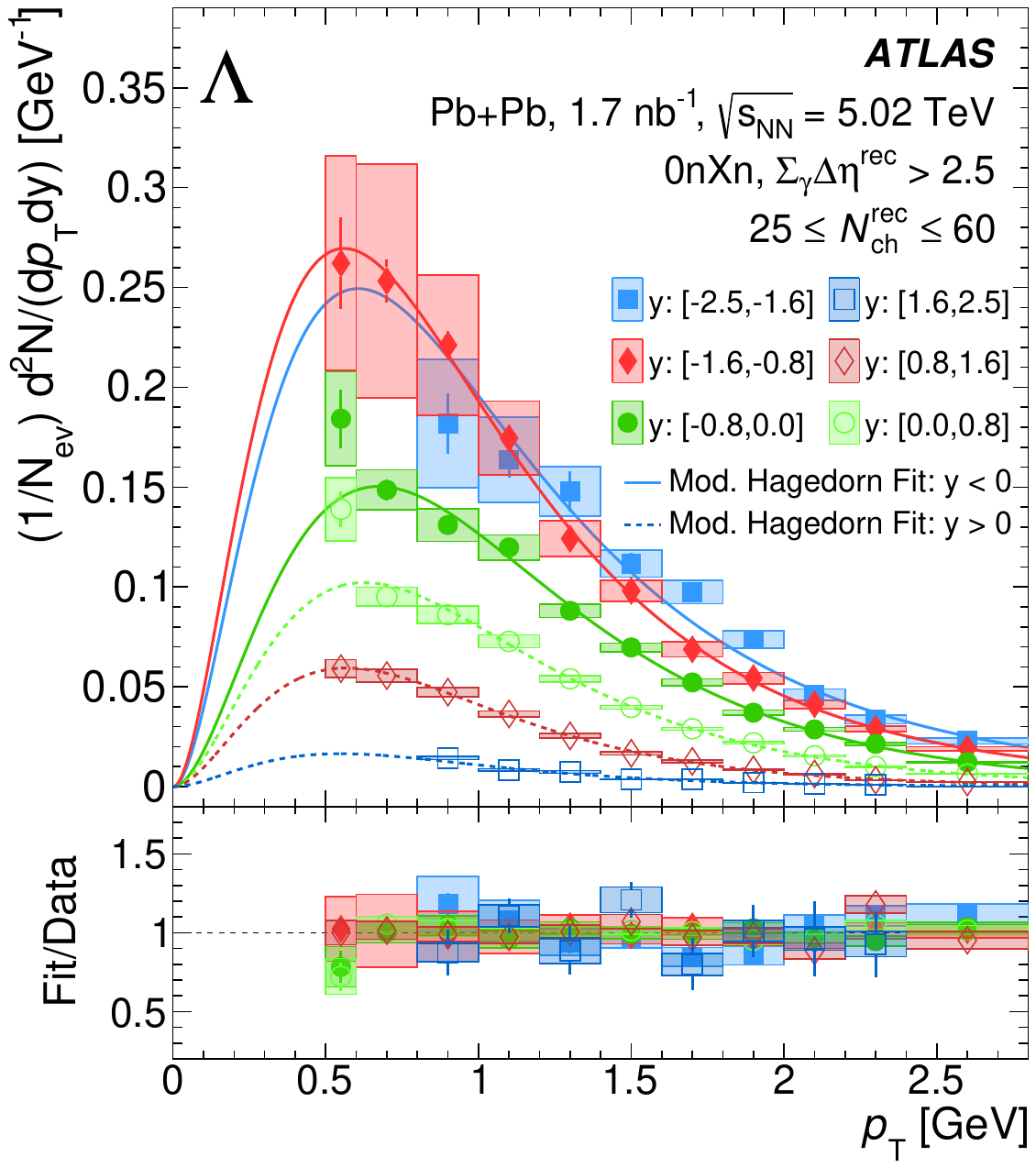}}
\subfigure[]{\includegraphics[width=0.49\textwidth]{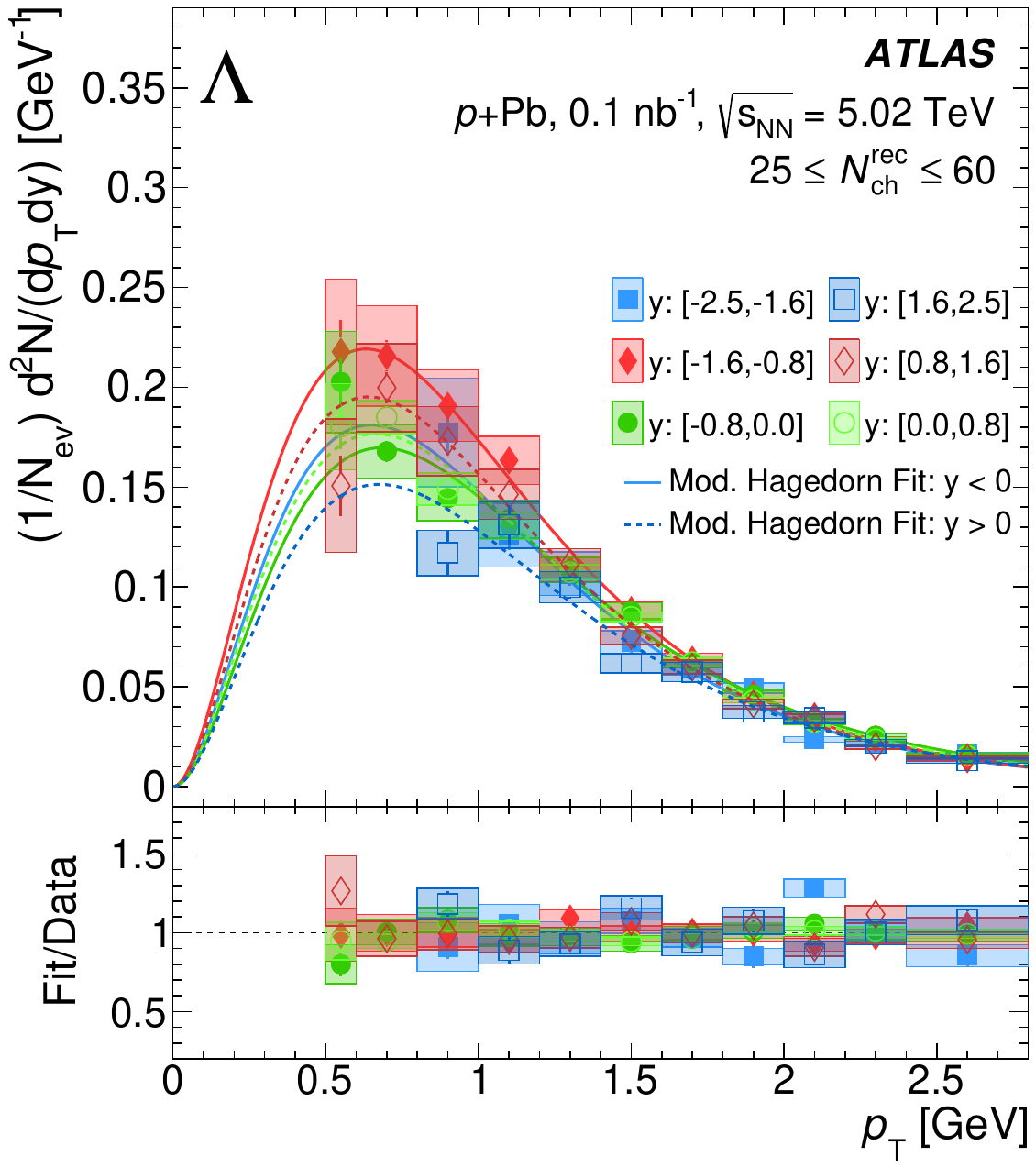}}
\caption{The \lmda yields as a function of \pt in different \y selections in (a) \PbPb photo-nuclear and (b) \pPb collisions. The bottom panels show the ratio of the Modified Hagedorn fit results to the data. Statistical uncertainties are shown as vertical lines and systematic uncertainties are shown as colored boxes.}
\label{fig:pTspectra_extrap_lmda}
\end{figure}

The minimum \pt values are 0~\GeV~for \ks, and 0.5~\GeV~for \csd. For \lmda, the minimum \pt is 0.5~\GeV~for~$|\y| <$~1.6 and 0.8~\GeV~for $1.6 < |\y| < 2.5$. Thus the \pt-integrated yields for \ks are simply calculated by summing the yields as a function of \pt, whereas for \lmda and \csd the yield below the minimum \pt must be accounted for. Hence, an extrapolation procedure is performed for \lmda and \csd, using the Modified Hagedorn functional fit, given by Eq.~\ref{eq:hage}. Figures~\ref{fig:pTspectra_extrap_k0s} and~\ref{fig:pTspectra_extrap_lmda} show the \ks and \lmda yields in \PbPb photo-nuclear and \pPb collisions as a function of \pt across six \y selections, with the Modified Hagedorn fit results. The fit to the \ks yield, where measurements extend down to \pt = 0~\GeV, shows that the function provides a good description of the data. This confirms that the same fit function can be used to extrapolate the \lmda and \csd yields down to \pt = 0~\GeV.
The fit is performed using statistical uncertainties only and is then repeated for each systematic uncertainty variation. Approximately 20\% of the total yield lies in the region of extrapolation ($\pt < 0.5$~\GeV) for \lmda, and is larger for the most forward/backward rapidities (where $\pt < 0.8$~\GeV). Approximately 20--30\% of the total yield lies in the region of extrapolation ($\pt < 0.5$~\GeV) for \csd. Systematic uncertainties on these extrapolations, and from other sources, are discussed below.

Finally, using the \pt-integrated yields determined above, the \meanpt and the ratio of strange-hadron to charged-hadron yields are calculated as a function of \Nch. These values are calculated after extrapolating down to \pt $=0$~\GeV.

\clearpage
\section{Systematic uncertainties}

The sources of systematic uncertainties in this measurement are described in the following. For the event selection criteria in \PbPb photo-nuclear events, the primary sources contributing to both the charged-hadron and identified-hadron yields include uncertainties associated with the purity estimation and pseudorapidity gap selection. In both \PbPb photo-nuclear and \pPb collisions, uncertainties are assigned to track selection by relaxing specific hit requirements in turn, to contributions from fake and secondary tracks by varying $f_\mathrm{primary}$ in Eq.~\ref{eq:prifr_def} by 50\% of the rate, and to the mis-modeling of the detector material~\cite{STDM-2015-02,STDM-2010-06}.   An uncertainty is also assigned for bin migration due to track momentum resolution.    This bin migration effect is quite small and the uncertainty is set by turning off the bin migration entirely.   Furthermore, uncertainties on the fit values of the track reconstruction efficiency are included as systematic uncertainties.

For the identified-hadron analysis, specific uncertainties are assigned to the selection requirements by varying the specific requirements from the nominal values.   The uncertainty in the signal extraction is further quantified by varying the fit range of invariant-mass distributions. Finally, to assess the uncertainty in the extrapolation to \pt = 0~\GeV, a varied fit functional form for the \pt distribution based on Tsallis statistics~\cite{Biro:2003vz} is utilized. The fitting procedure is re-done for all other systematic variations.

In all cases, the ratio of the varied to nominal result is either smoothened via fit or directly used, and the values are assigned as the systematic uncertainty.
All uncertainty contributions are added in quadrature and then symmetrized by taking the maximum variation at each point to determine the total systematic uncertainty. Other potential sources of uncertainty, such as those related to the trigger efficiency~\cite{HION-2018-27}, were found to be negligible compared to the ones described above.

\begin{figure}[!ht]
\centering
\subfigure[]{\includegraphics[width=0.49\textwidth]{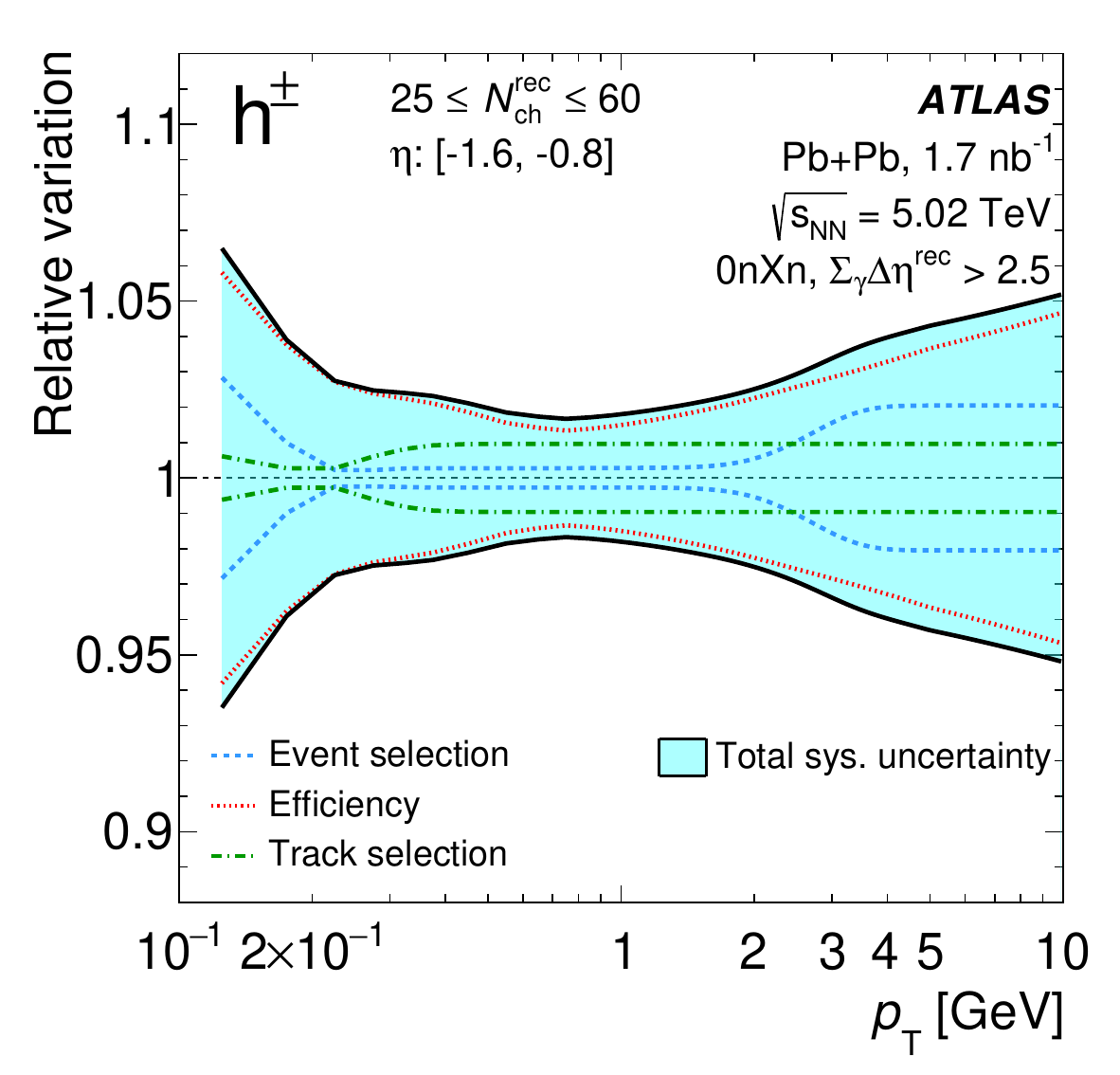}}
\subfigure[]{\includegraphics[width=0.49\textwidth]{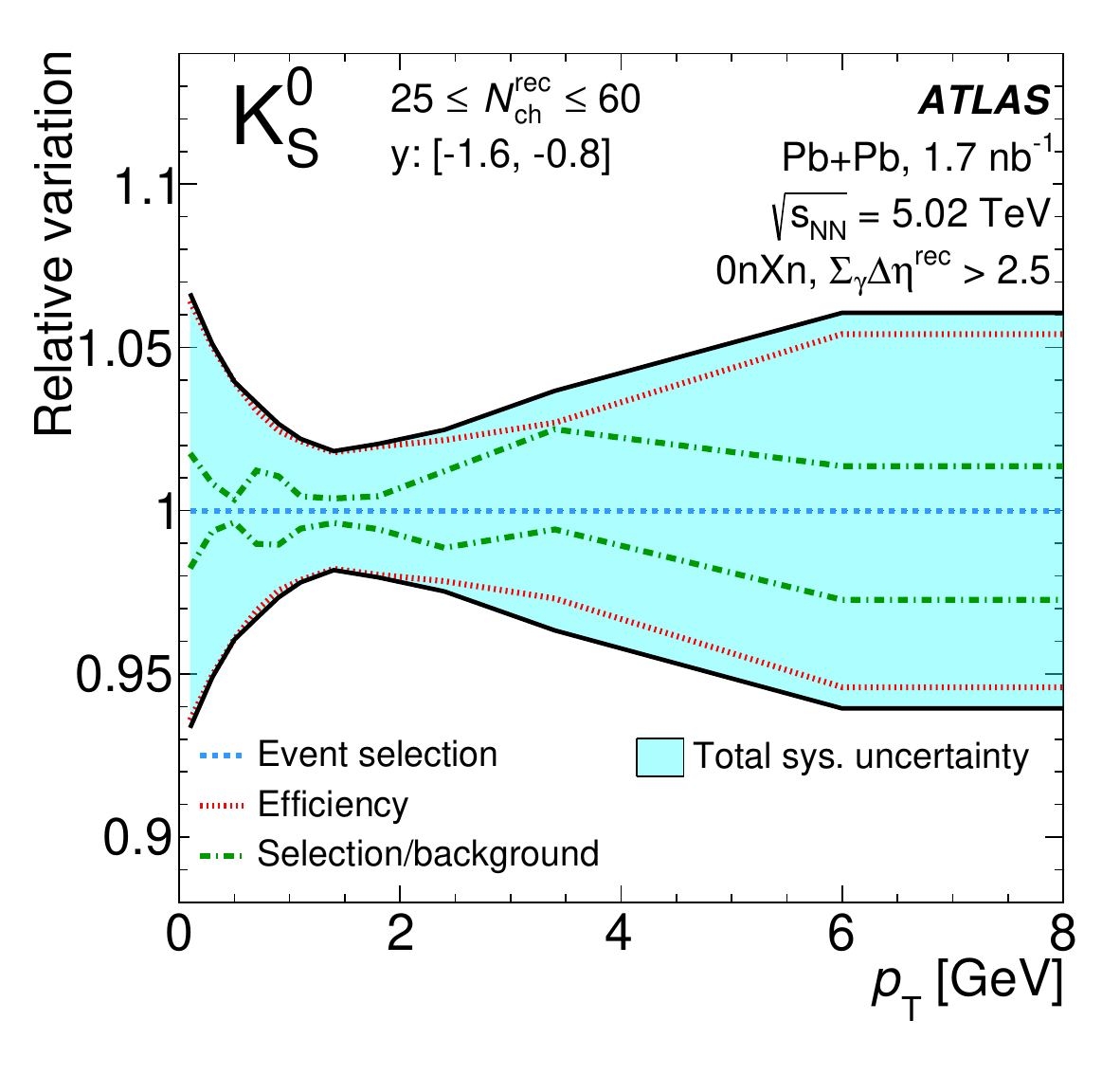}}
\subfigure[]{\includegraphics[width=0.49\textwidth]{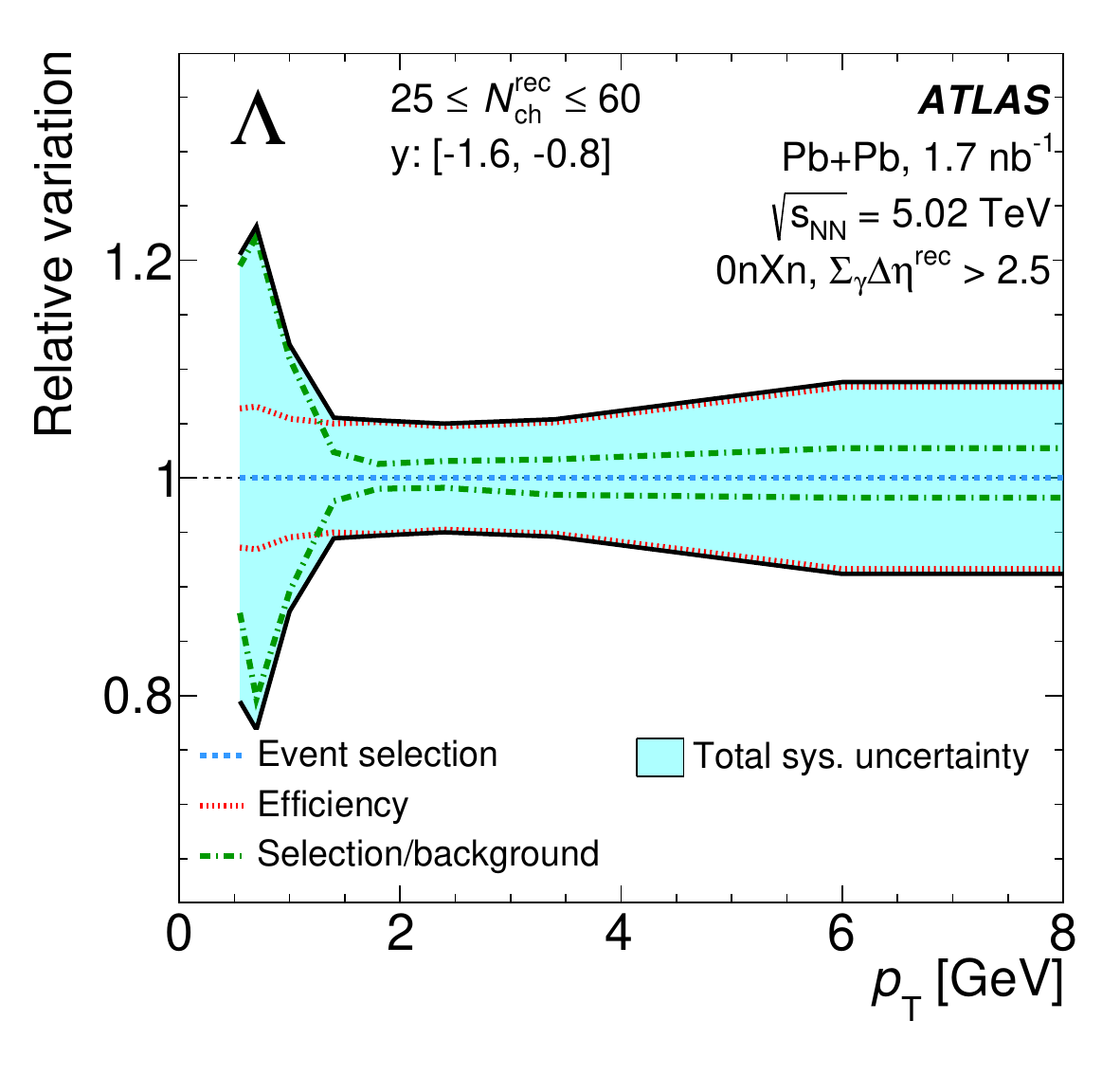}}
\subfigure[]{\includegraphics[width=0.49\textwidth]{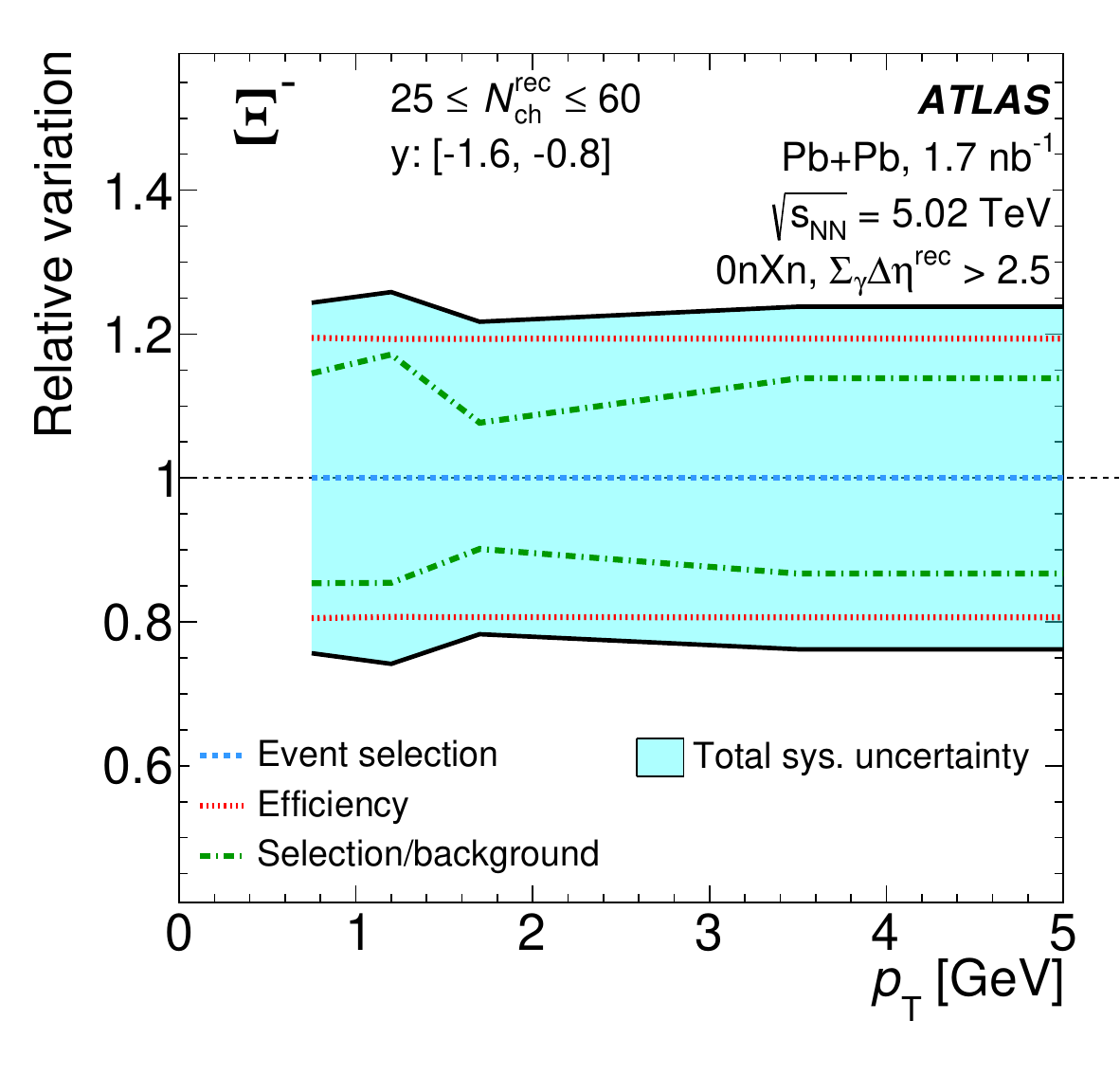}}
\caption{The relative systematic uncertainties in (a) charged hadrons, (b) \ks, (c) \lmda, and (d) \csd yields in photo-nuclear collisions as a function of \pt. All uncertainty contributions are added in quadrature and the result is symmetrized to obtain the full systematic uncertainty.}
\label{fig:systematic_uncertainty_pT}
\end{figure}

\begin{figure}[!ht]
\centering
\subfigure[]{\includegraphics[width=0.49\textwidth]{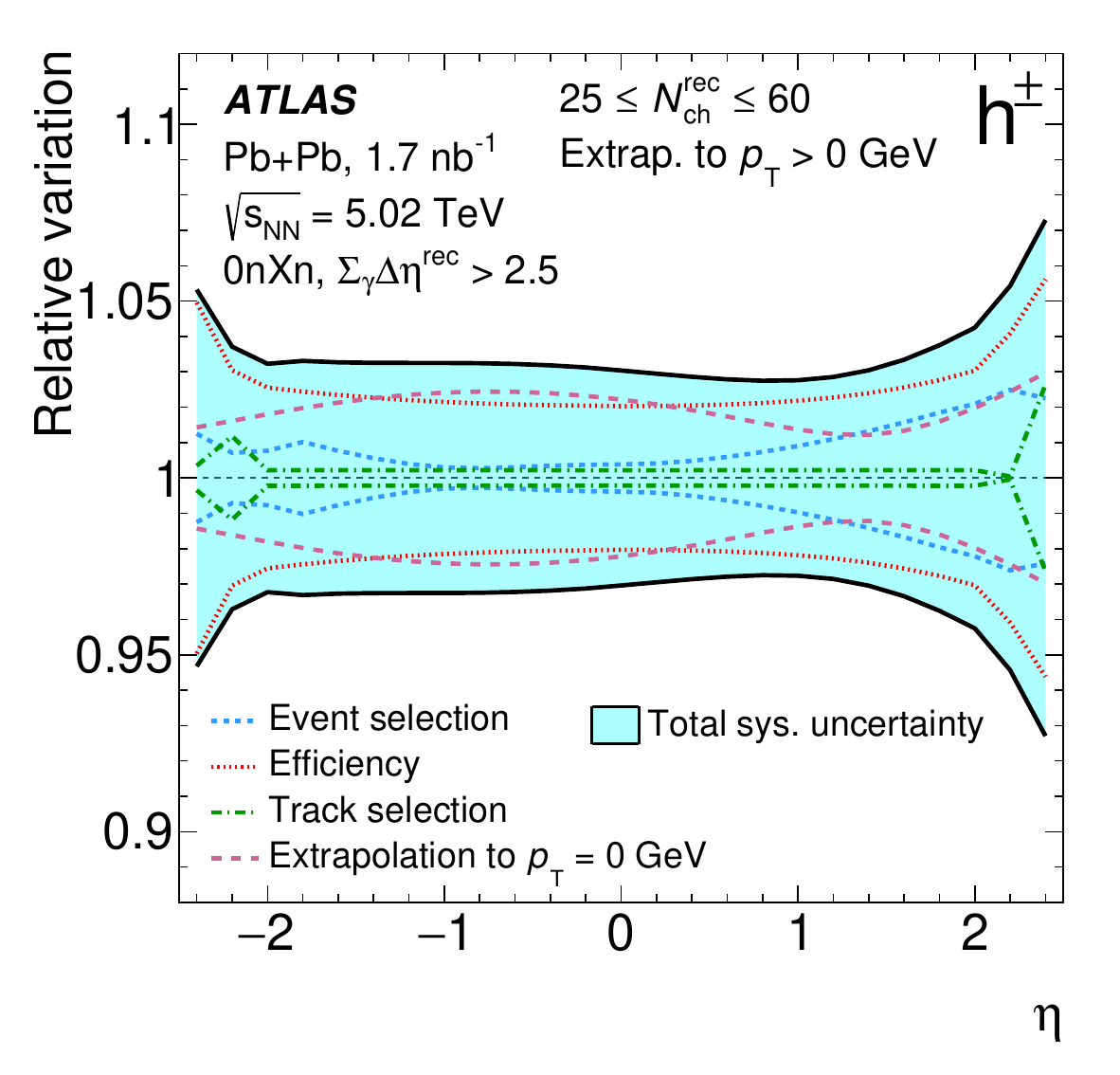}}
\subfigure[]{\includegraphics[width=0.49\textwidth]{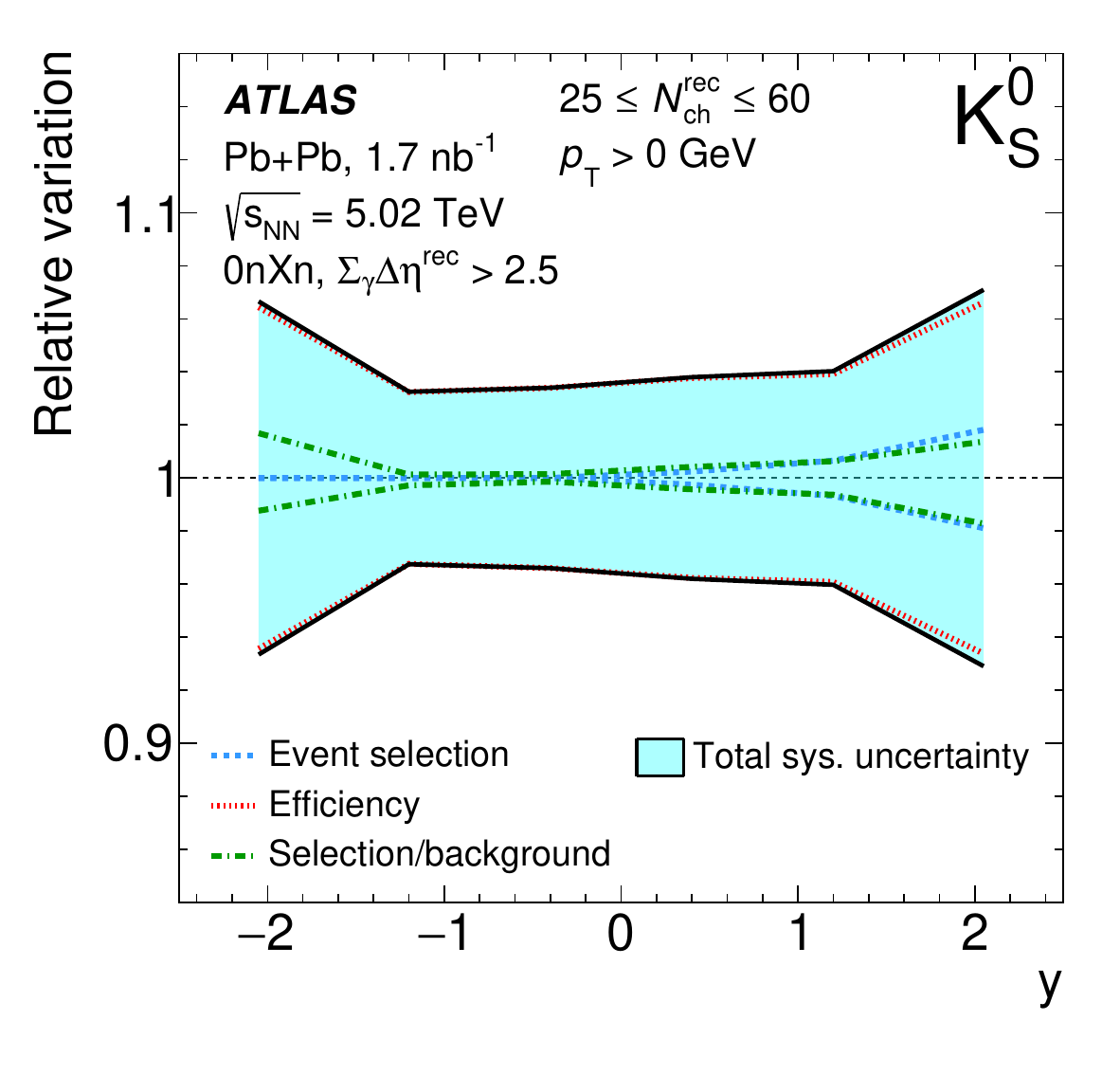}}
\subfigure[]{\includegraphics[width=0.49\textwidth]{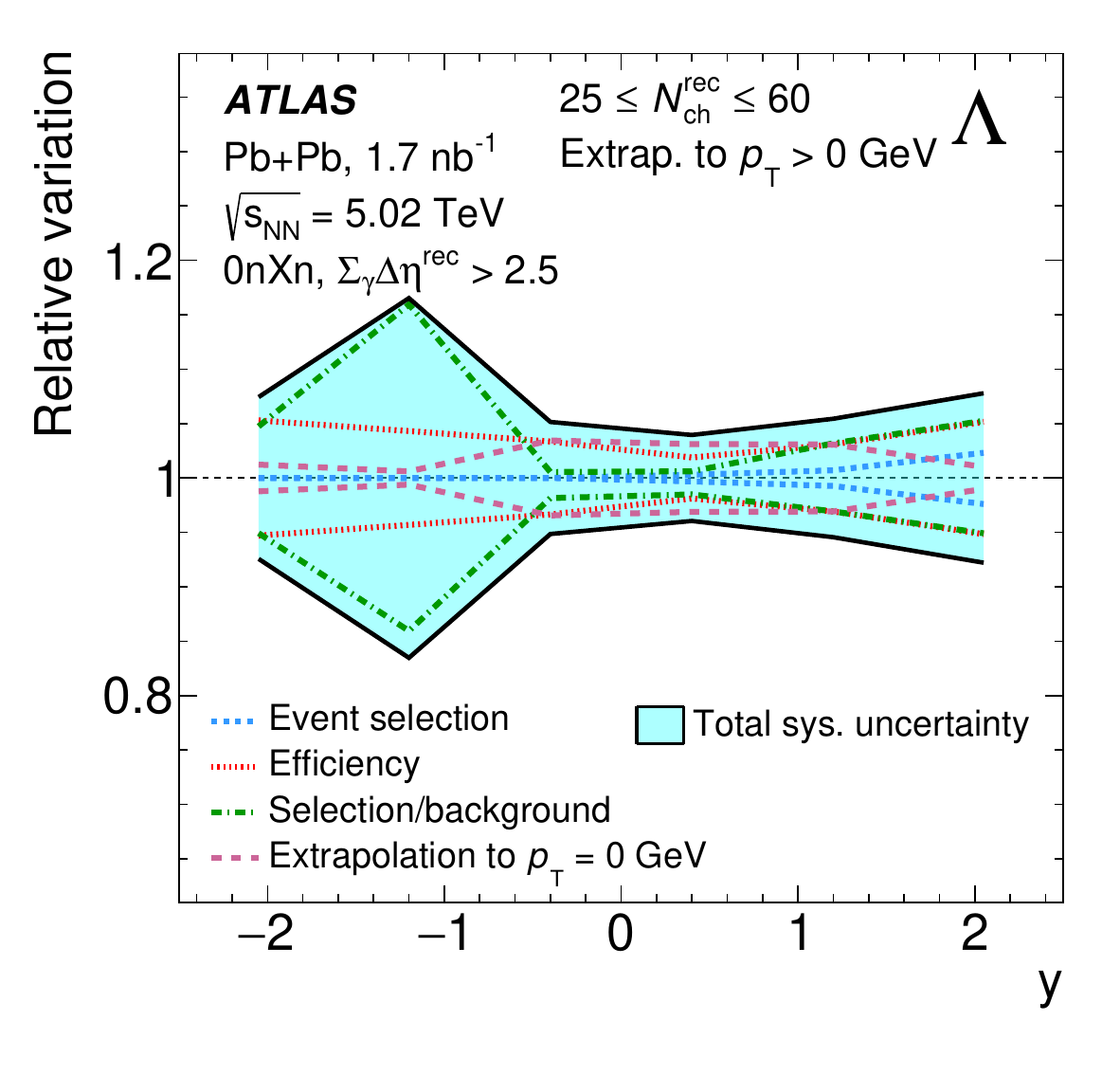}}
\subfigure[]{\includegraphics[width=0.49\textwidth]{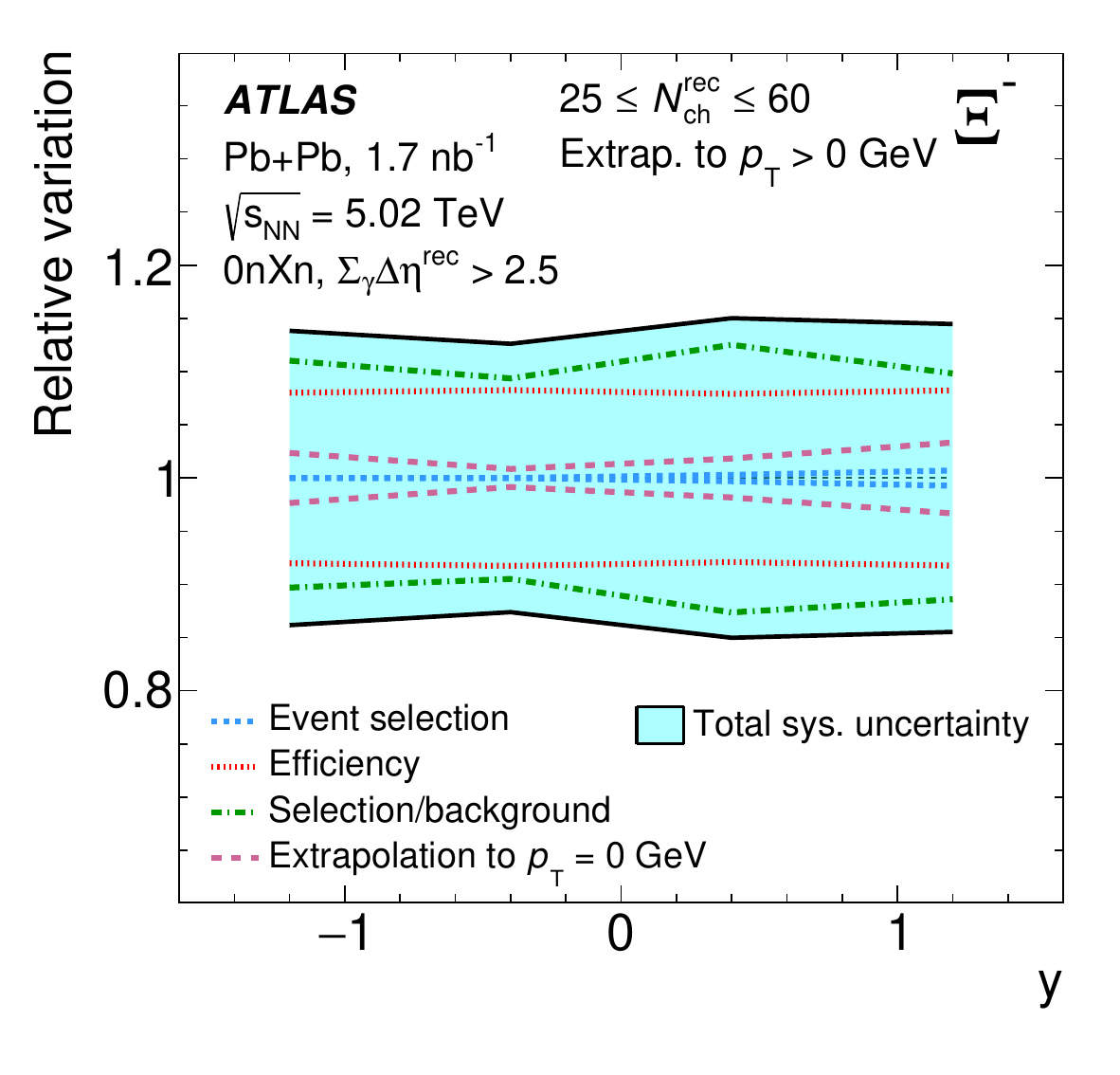}}
\caption{The relative systematic uncertainties in (a) charged hadrons, (b) \ks, (c) \lmda, and (d) \csd yields in photo-nuclear collisions as a function of (pseudo) rapidity. The yield of \ks is measured down to \pT~=~0~\GeV. For charged hadrons, \lmda and \csd, the measured yield is extrapolated to \pT = 0~\GeV. All uncertainty contributions are added in quadrature and the result is symmetrized to obtain the full systematic uncertainty.}
\label{fig:systematic_uncertainty_y}
\end{figure}

Figures~\ref{fig:systematic_uncertainty_pT} and~\ref{fig:systematic_uncertainty_y} present the relative systematic uncertainties on charged hadrons, \ks, \lmda, and \csd yields in photo-nuclear collisions as a function of \pt and (pseudo) rapidity, respectively. The dominant uncertainty in the low \pt region ($\pt < 1$~\GeV) arises from variations in the detector materials and bin-migration effects. For identified hadrons, in addition to these uncertainties, the uncertainty in the background correction contributes significantly. For the \meanpt calculation, the total uncertainties are on the order of 3–-5\%. As a function of rapidity, a significant uncertainty for charged hadrons and \lmda, \csd particles comes from the extrapolation of the yield down to \pT~=~0~\GeV.  For the \ks there is no such uncertainty as the yield is measured down to \pT = 0~\GeV. The uncertainty is dominated by systematic rather than statistical effects.

\clearpage
\newpage

\section{Results}

\begin{figure}[!t]
\centering
\subfigure[]{\includegraphics[width=0.49\textwidth]{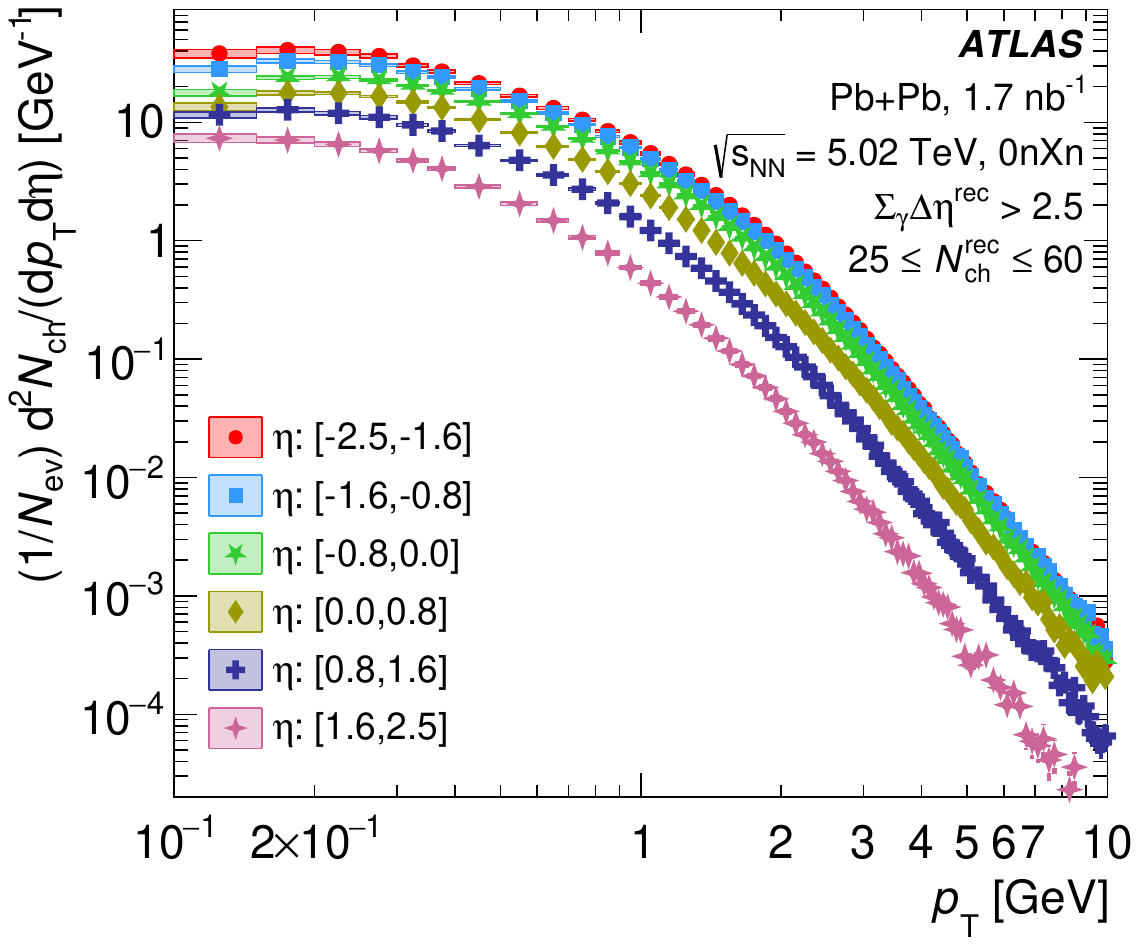}}
\subfigure[]{\includegraphics[width=0.49\textwidth]{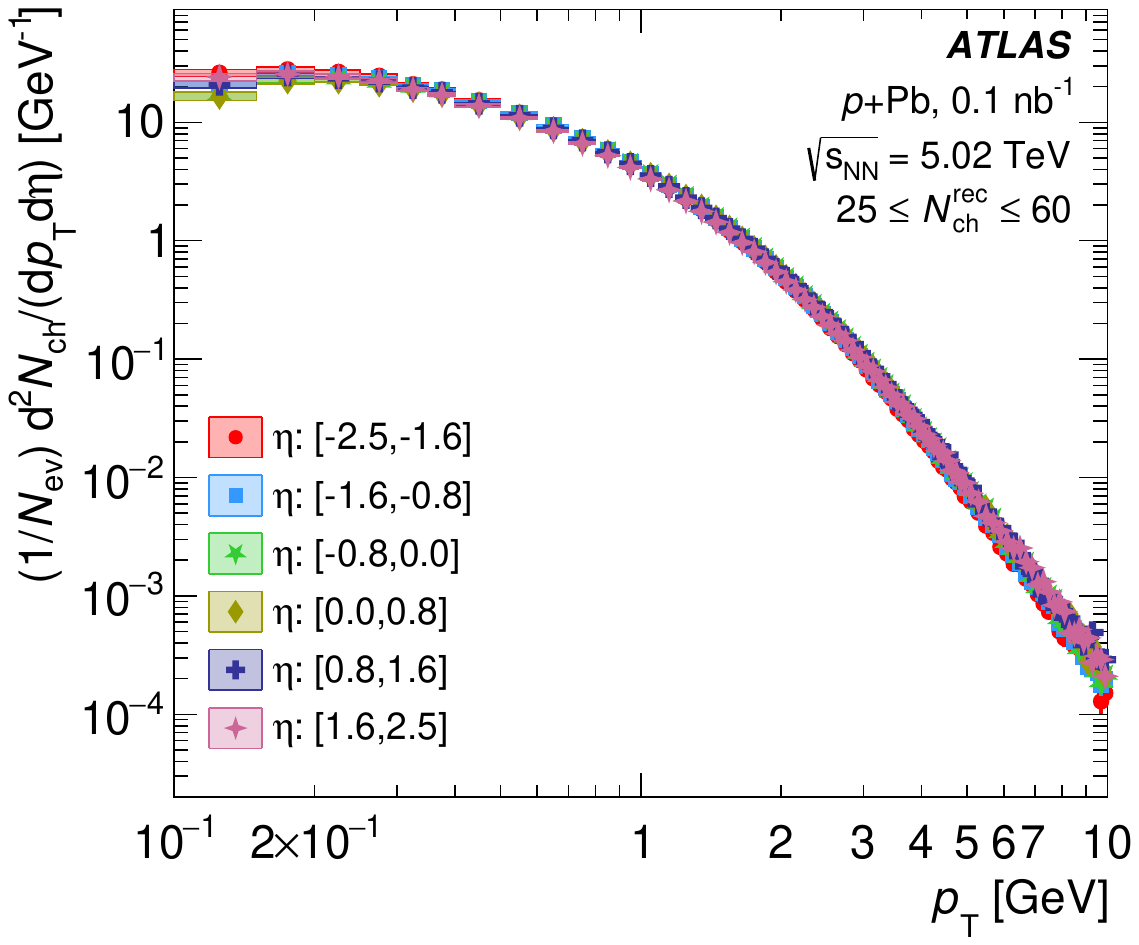}}
\caption{The charged-hadron yields as a function of \pt in six $\eta$ selections for (a) \PbPb photo-nuclear and (b) \pPb collisions. Statistical uncertainties are shown as vertical lines and systematic uncertainties are shown as colored boxes.}
\label{fig:yieldch}
\end{figure}

The resulting yields as a function of \pt for charged hadrons are shown in Figure~\ref{fig:yieldch} in six pseudorapidity selections spanning $-2.5 < \eta < 2.5$ in \PbPb photo-nuclear and \pPb collisions. The yields are calculated for collisions with $25 \le \Nch \le 60$, with the \pPb \Nch distribution re-weighted to match that of the photo-nuclear collision sample. The photo-nuclear collision yields have a strong pseudorapidity dependence, with much lower multiplicity in the photon-going direction (positive $\eta$). In contrast, the \pPb collision yields are nearly $\eta$-symmetric. Unlike in more central \pPb collisions~\cite{HION-2012-14}, the yields in the low-multiplicity selection ($25 \leq \Nch \leq 60$) are symmetric in rapidity; hence, the $\eta$ distribution is more \pp-like.

\begin{figure}[!p]
\centering
\subfigure[]{\includegraphics[width=0.32\textwidth]{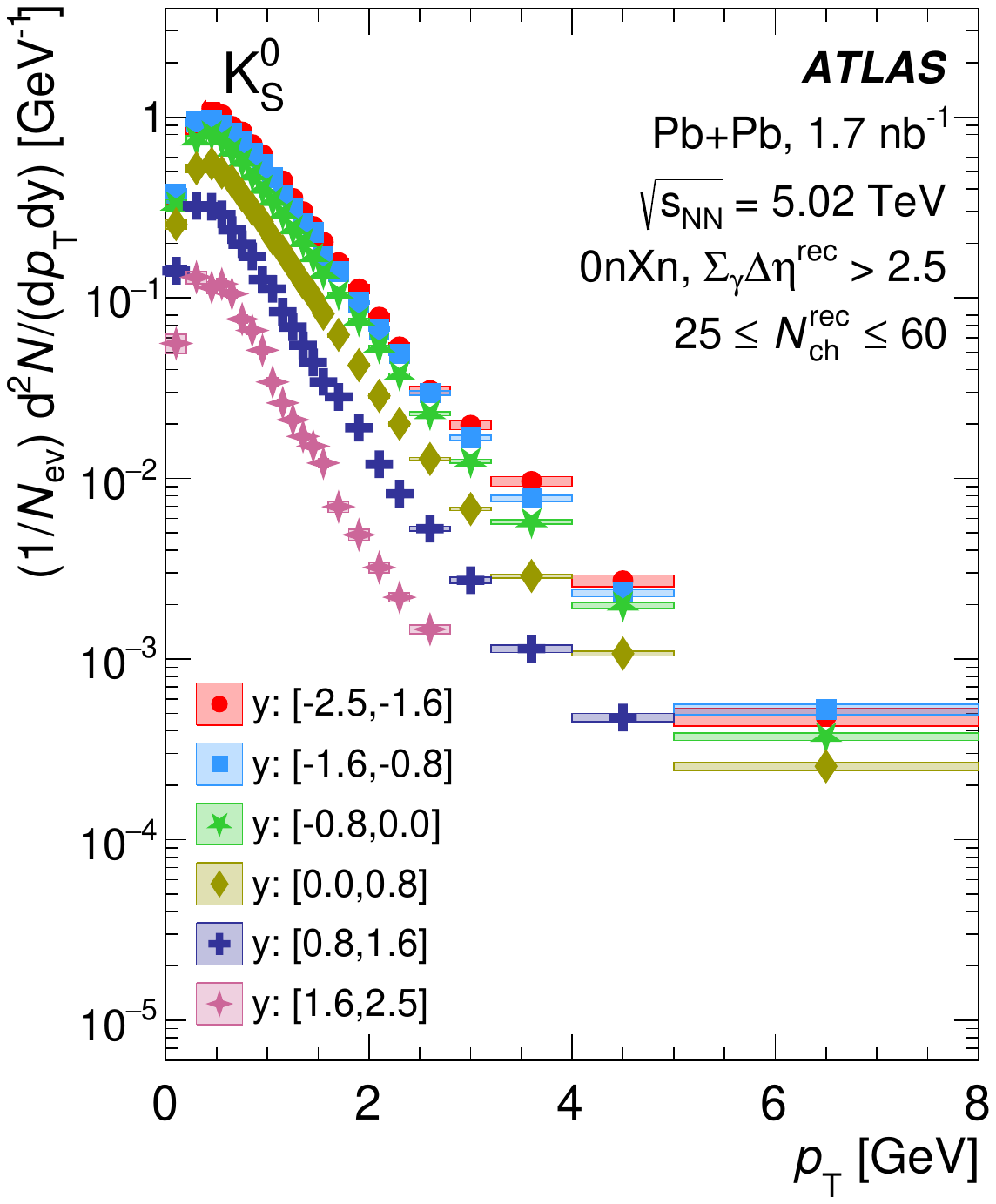}}
\subfigure[]{\includegraphics[width=0.32\textwidth]{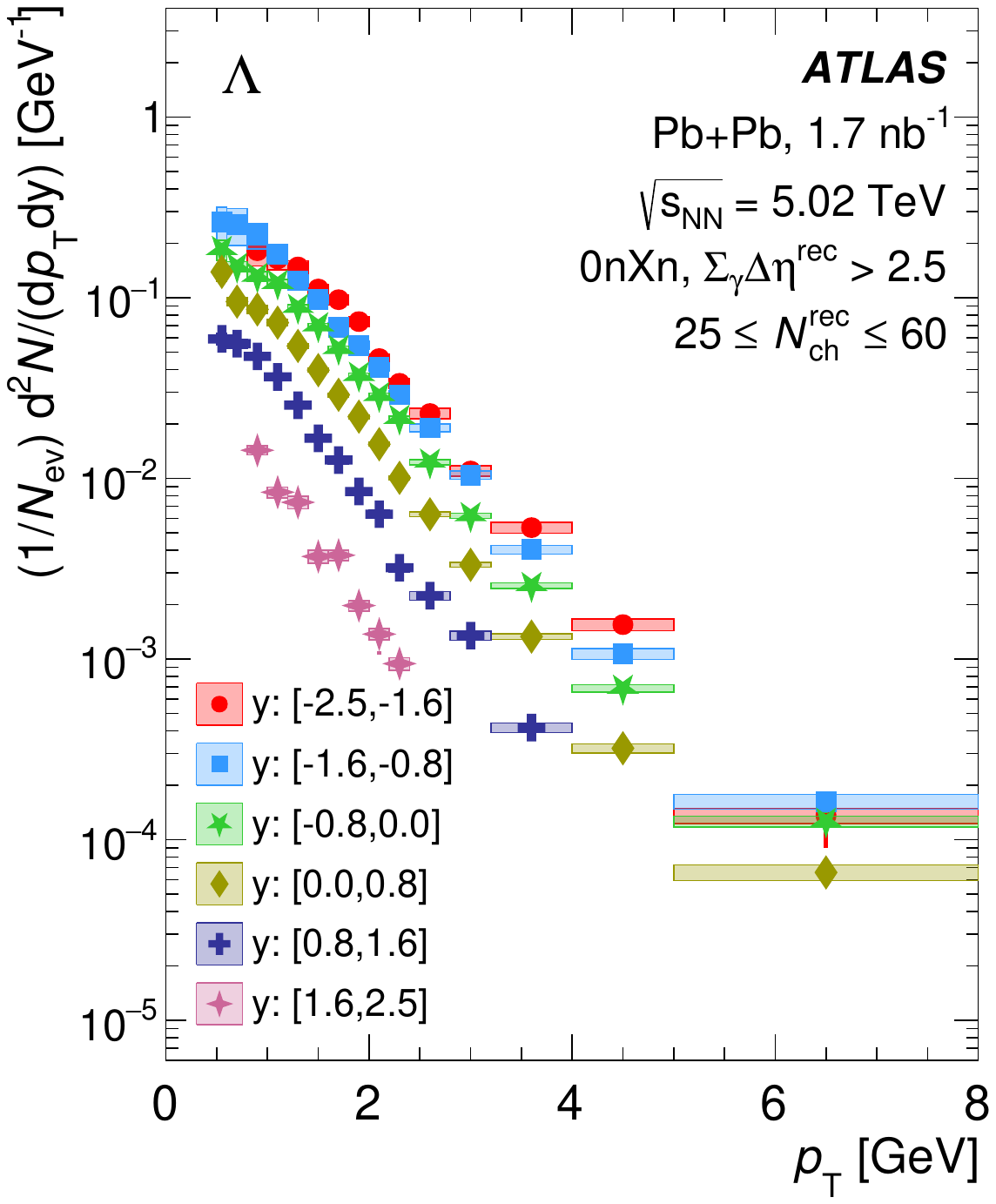}}
\subfigure[]{\includegraphics[width=0.32\textwidth]{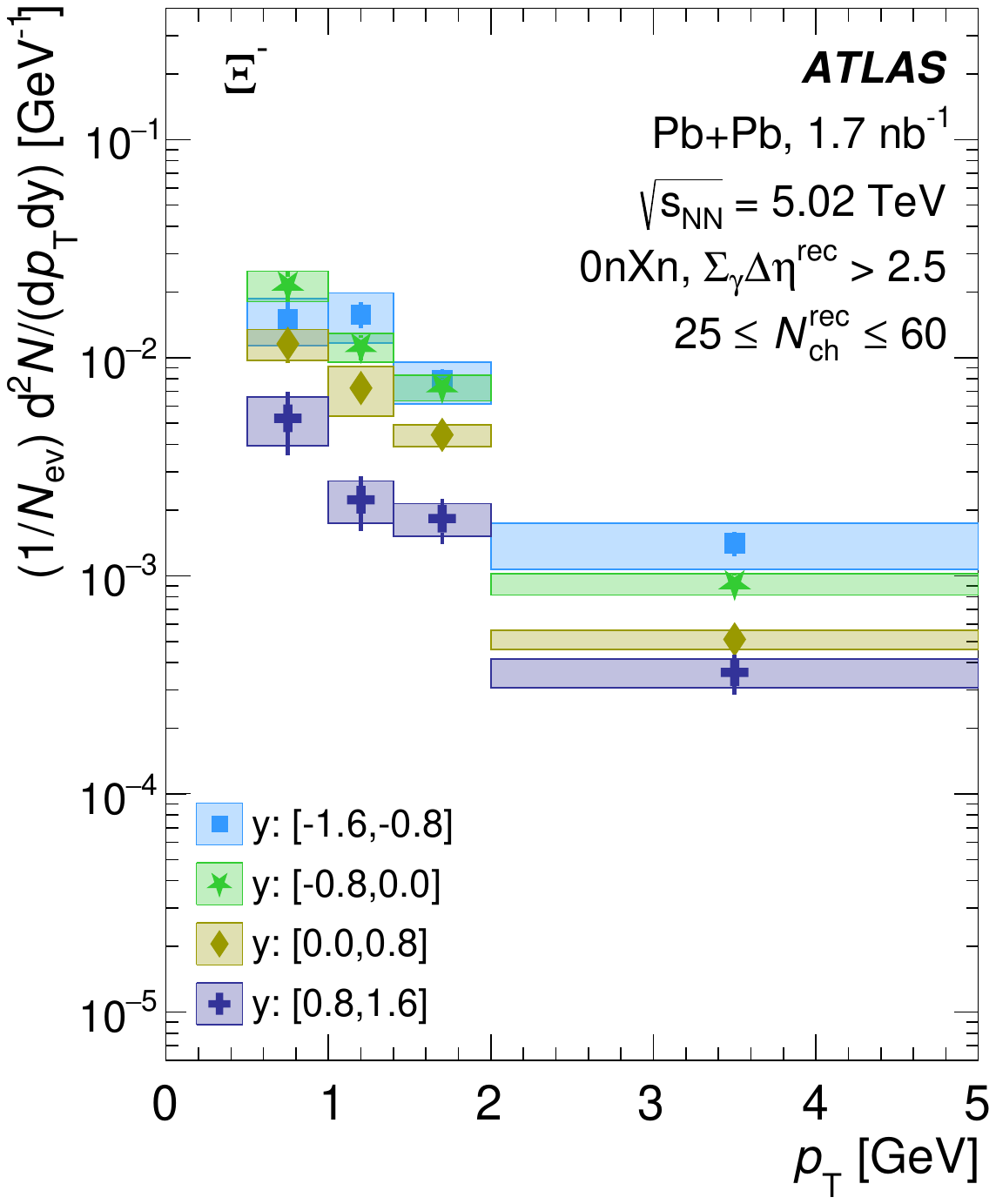}}
\subfigure[]{\includegraphics[width=0.32\textwidth]{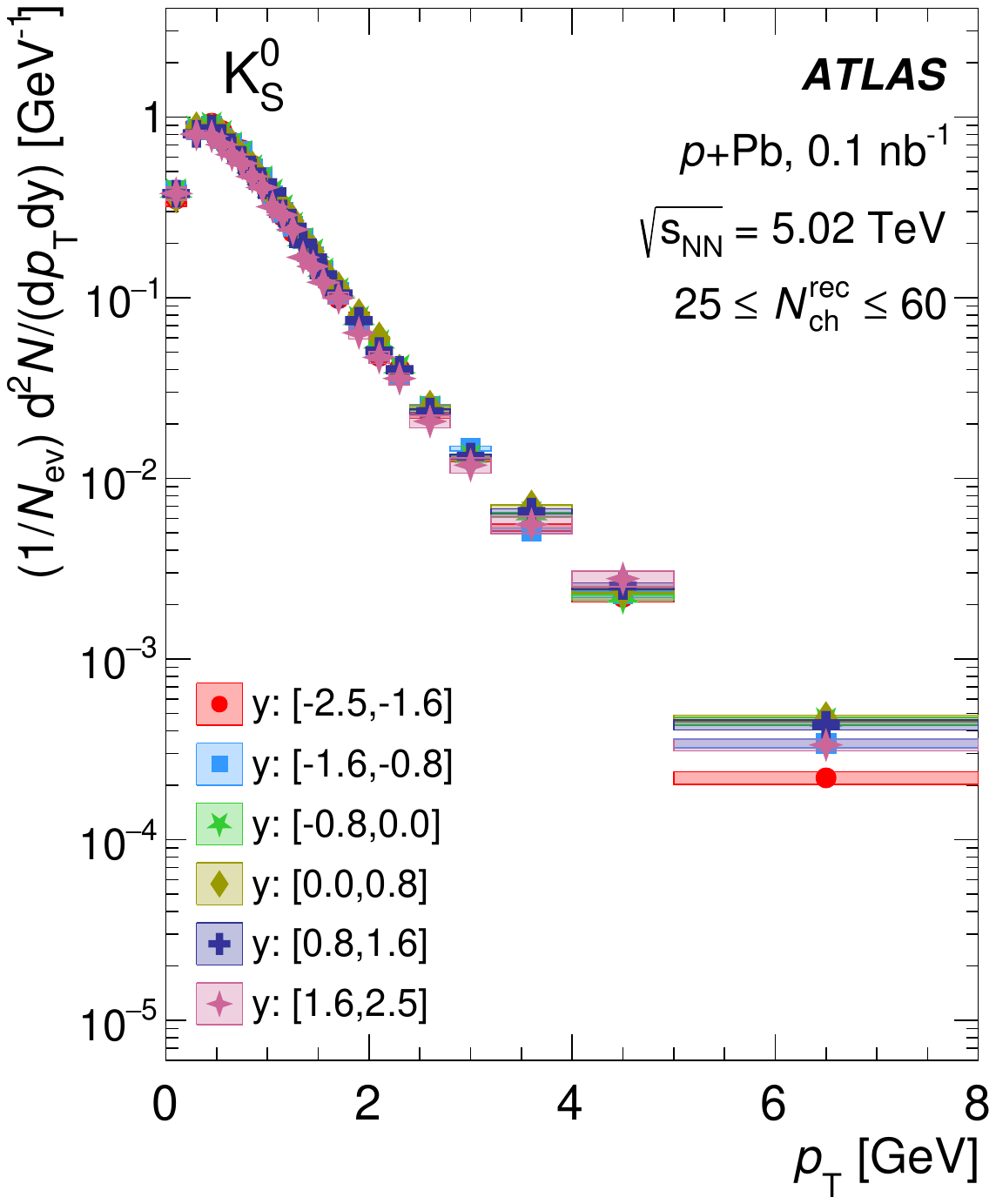}}
\subfigure[]{\includegraphics[width=0.32\textwidth]{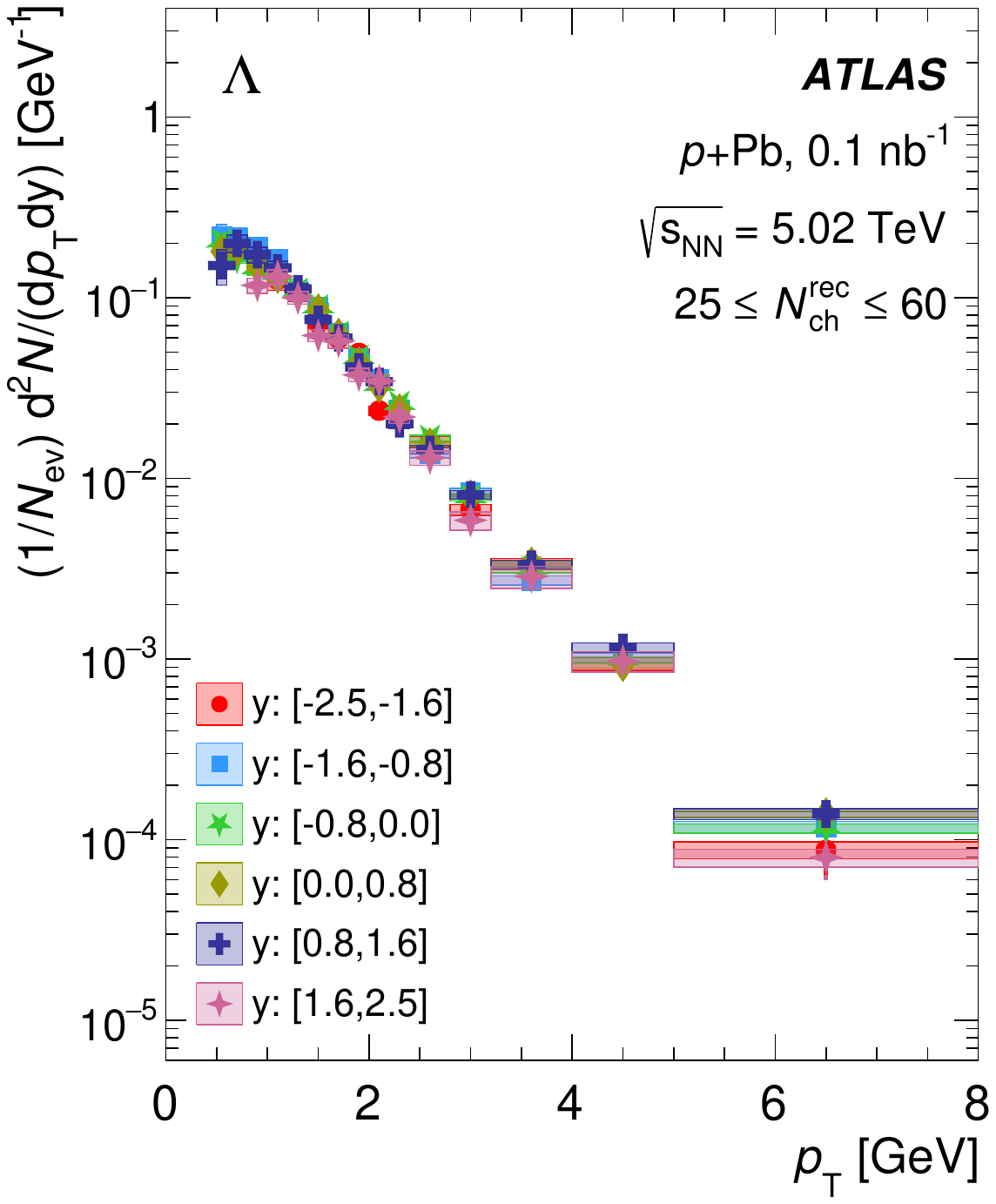}}
\subfigure[]{\includegraphics[width=0.32\textwidth]{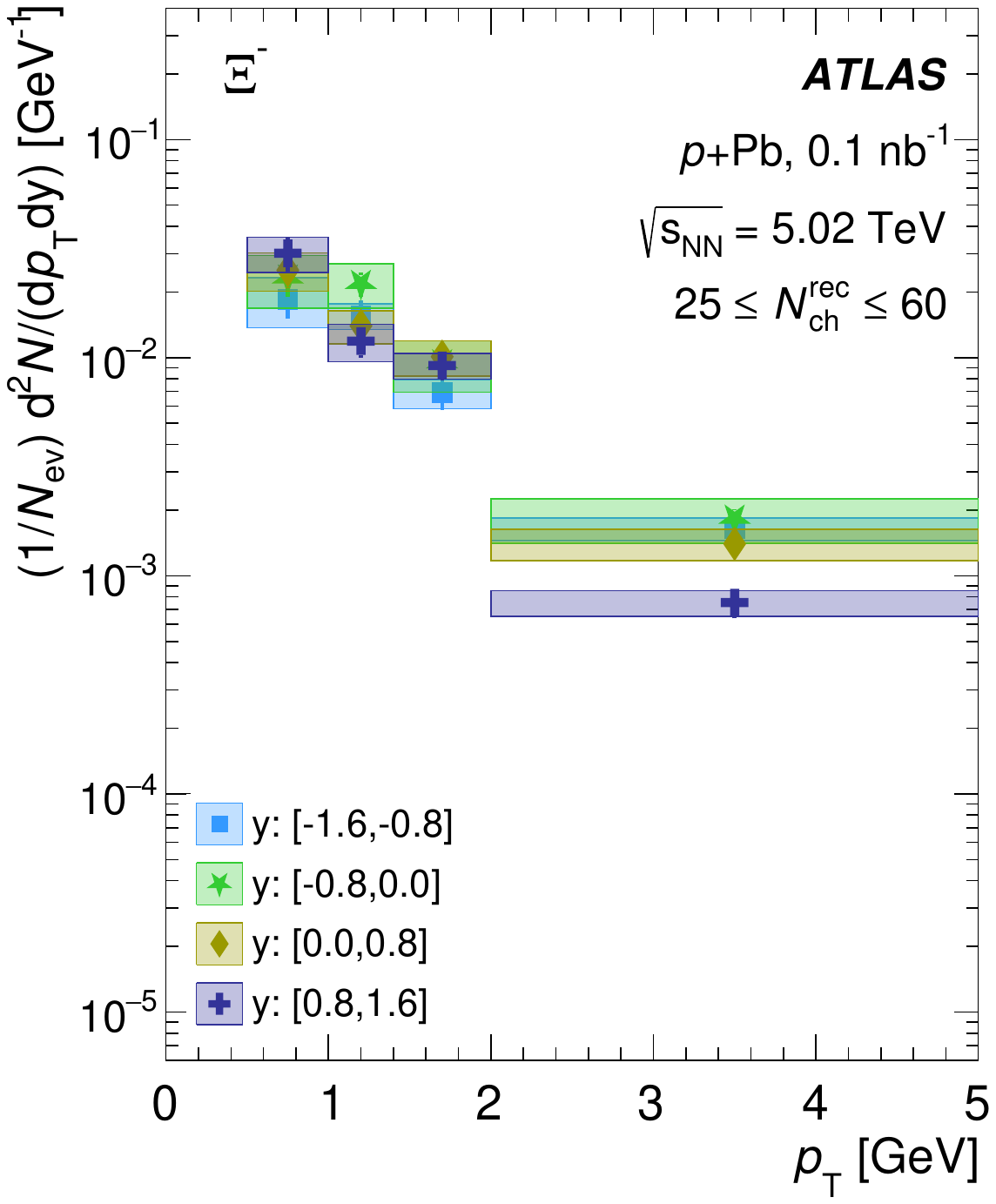}}
\caption{The (a) \ks, (b) \lmda, and (c) \csd yields are  shown as a function of \pt in six \y selections in \PbPb photo-nuclear collisions.   The (d) \ks, (e) \lmda, and (f) \csd yields are  shown as a function of \pt in six \y selections in \pPb collisions.   Statistical uncertainties are shown as vertical lines and systematic uncertainties are shown as colored boxes.}
\label{fig:pTspectra}
\end{figure}

Figure~\ref{fig:pTspectra} shows the yields as a function of \pt for identified strange hadrons \ks, \lmda, and \csd in six rapidity selections spanning $-2.5 < \y < 2.5$  in \PbPb photo-nuclear and \pPb collisions for the same event selection as the charged-hadron yields. The \lmda and \csd yields are for baryons only and are not the average of baryons and anti-baryons. The \lmda yields include decays from $\Sigma_0$, but otherwise are not inclusive of other baryon feed-down contributions, e.g., from \csd and $\Omega$. As in the charged-hadron case, the yields of all strange hadrons have a strong rapidity dependence in photo-nuclear collisions.

\clearpage

\begin{figure}[!t]
\centering
\subfigure[]{\includegraphics[width=0.49\textwidth]{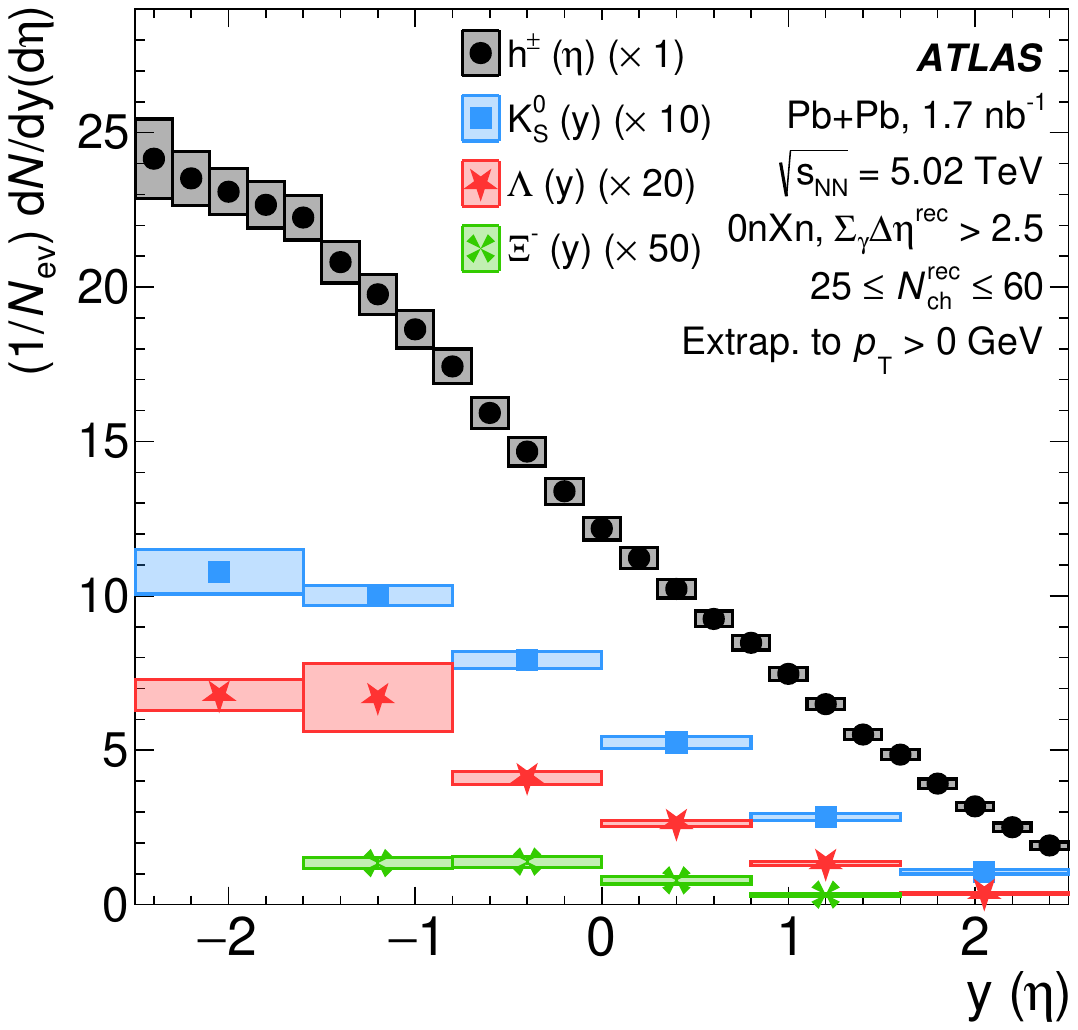}}
\subfigure[]{\includegraphics[width=0.49\textwidth]{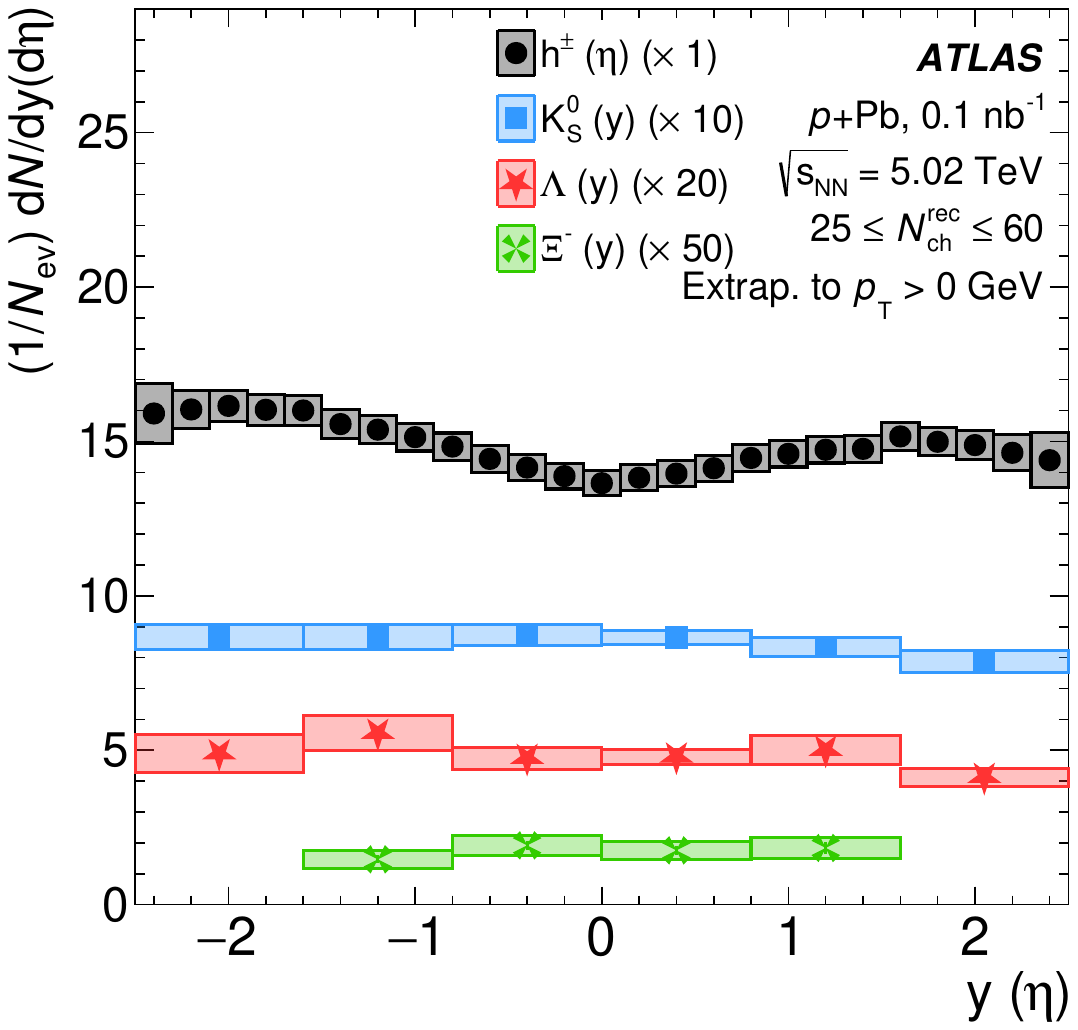}}
\caption{The charged-hadron yields as a function of $\eta$ and the \ks, \lmda, and \csd yields as a function of \y for (a) \PbPb photo-nuclear and (b) \pPb collisions. Statistical uncertainties are shown as vertical lines and systematic uncertainties are shown as colored boxes.}
\label{fig:dNdy}
\end{figure}

The yields as a function of \pt are then integrated over the \pt range of the measurements, and extrapolated to be inclusive over all \pt, i.e., $\pt>0$. The resulting \pt-integrated yields as a function of pseudorapidity, for charged hadrons, and rapidity, for identified strange hadrons are shown in Figure~\ref{fig:dNdy}. The \pt-integrated yields again show a strong rapidity asymmetry in photo-nuclear collisions and are nearly rapidity symmetric in \pPb collisions.

\begin{figure}[!p]
\centering
\subfigure[]{\includegraphics[width=0.9\textwidth]{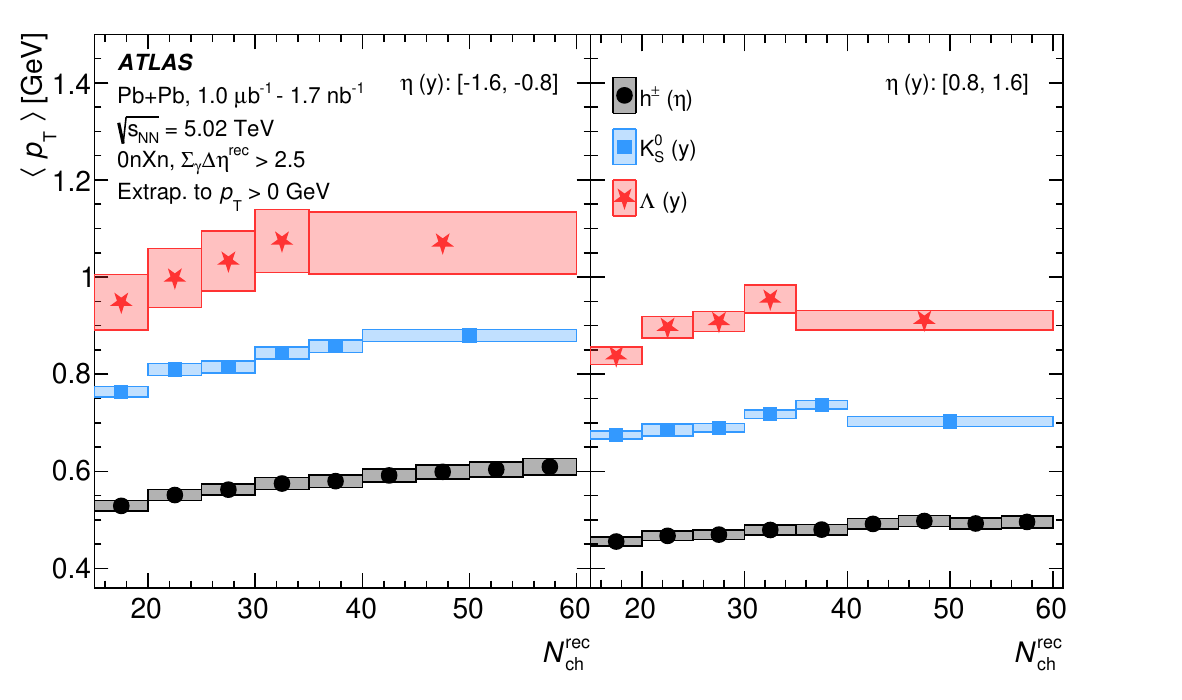}}
\subfigure[]{\includegraphics[width=0.9\textwidth]{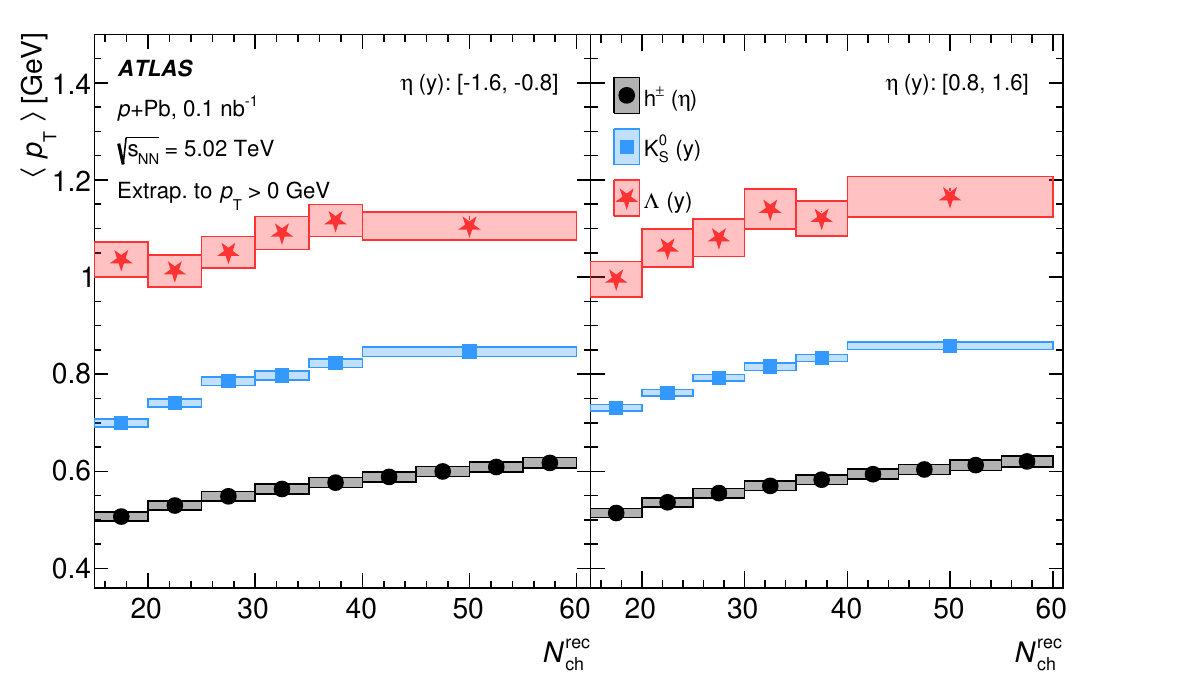}}
\caption{The $\left< \pt \right>$ for charged hadrons, \ks, and \lmda in (a) \PbPb photo-nuclear and (b) \pPb collisions as a function of \Nch. The left (right) panels are for a backward (forward) rapidity interval. Statistical uncertainties are shown as vertical lines and systematic uncertainties are shown as colored boxes.}
\label{fig:meanpT}
\end{figure}

Next, the \pt distributions are characterized in terms of the \meanpt, calculated to correspond to the mean value for $\pt>0$. These \meanpt values for charged hadrons, \ks, and \lmda particles, in two rapidity intervals, as a function of finer intervals of \Nch are shown in Figure~\ref{fig:meanpT}.  The top (bottom) sub-figures correspond to photo-nuclear \PbPb (\pPb) collisions, with the left (right) panels corresponding to backward (forward) rapidity. In both photo-nuclear and \pPb collisions, the \meanpt increases with increasing \Nch and there is a distinct ordering with \meanpt (charged hadrons) $<$ \meanpt (\ks) $<$ \meanpt (\lmda). Under the assumption that charged hadrons are dominated by pions, the pattern follows a distinct mass ordering. In photo-nuclear collisions, the \meanpt show a large rapidity asymmetry, with much lower values for all particles at forward rapidity. In contrast, the \pPb results are consistent with being rapidity symmetric.

For the \csd, the \meanpt can only be calculated in one selection of \Nch, $25 \le \Nch  \le 60$, and correspond to values of $1.60 \pm 0.02 (\mathrm{stat}) \pm 0.24 (\mathrm{sys})~\GeV $
($1.04 \pm 0.01 (\mathrm{stat}) \pm 0.12 (\mathrm{sys})~\GeV$)
for photo-nuclear collisions and $1.33 \pm 0.03 (\mathrm{stat}) \pm 0.18 (\mathrm{sys})~\GeV $  $(1.06 \pm 0.02 (\mathrm{stat}) \pm 0.25 (\mathrm{sys})~\GeV)$ for \pPb collisions, at backward (forward) rapidity. These values suggest that the \csd may have a slightly higher \meanpt than \lmda particles, although  consistent within large uncertainties.

\clearpage

\begin{figure}[!t]
\centering
\subfigure[]{\includegraphics[width=0.49\textwidth]{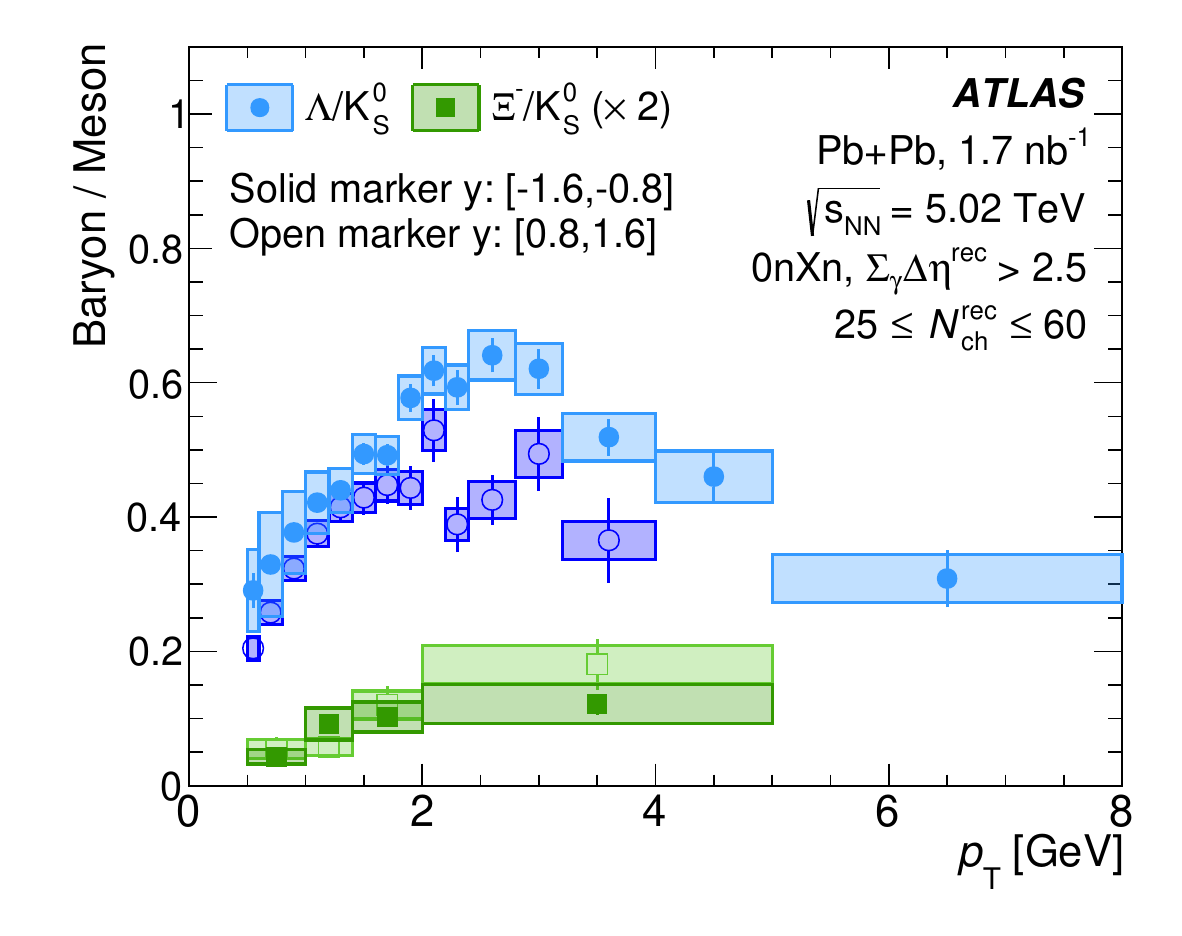}}
\subfigure[]{\includegraphics[width=0.49\textwidth]{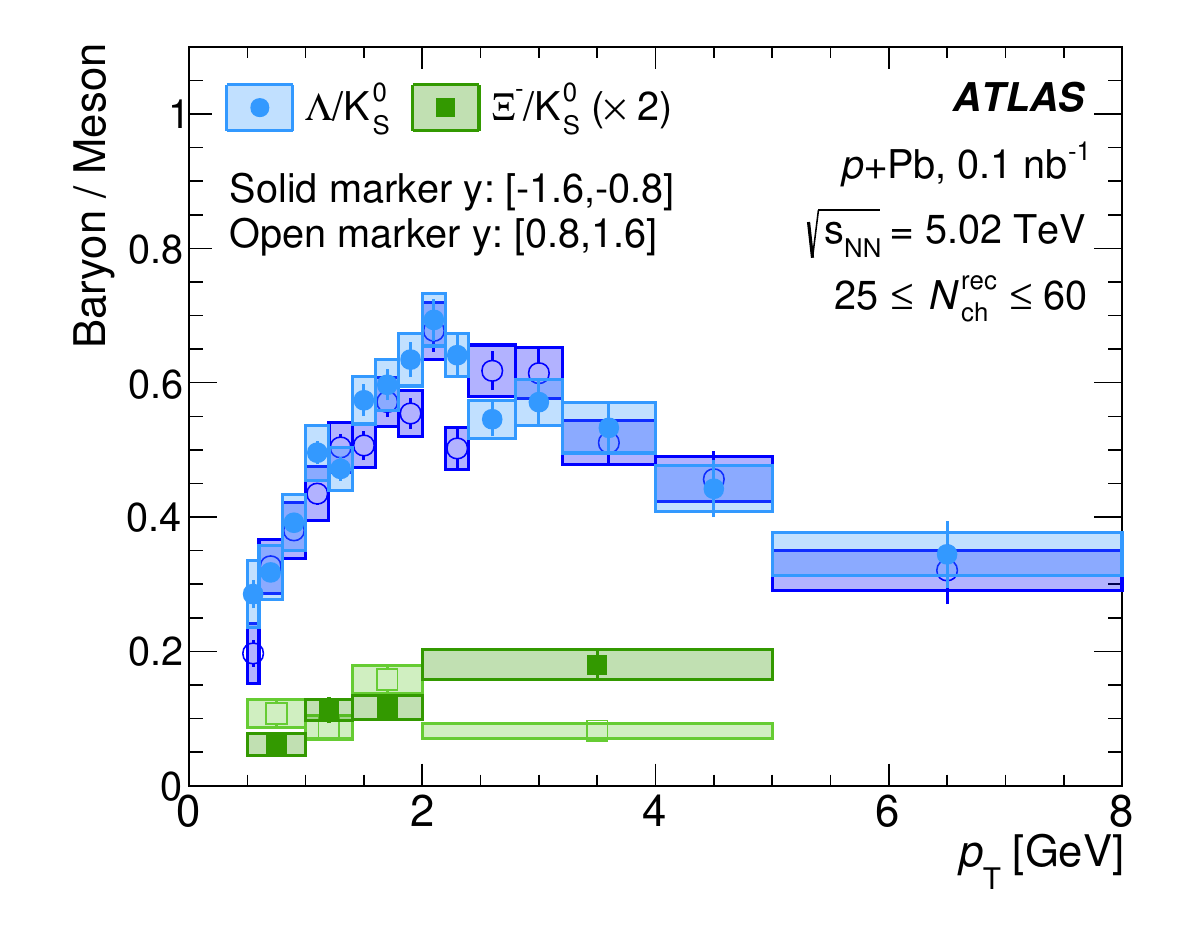}}
\caption{The ratio of \lmda/\ks and \csd/\ks yields as a function of \pt in two rapidity intervals for (a) \PbPb photo-nuclear and (b) \pPb collisions. Statistical uncertainties are shown as vertical lines and systematic uncertainties are shown as colored boxes.}
\label{fig:BaryonToMeson_ratios}
\end{figure}

To further elucidate the data trends, the ratios of \lmda/\ks and \csd/\ks as a function of \pt, for two rapidity intervals, in \PbPb photo-nuclear and \pPb collisions are shown in Figure~\ref{fig:BaryonToMeson_ratios}. The most striking observation is the larger \lmda/\ks ratio at intermediate \pt $\approx 1.5-4.0$~\GeV when measured at backward rapidity compared to forward rapidity in photo-nuclear collisions. This baryon enhancement at intermediate \pt is reminiscent of the ``baryon anomaly'' observed in \pPb and \PbPb collisions~\cite{ALICE:2013cdo} and is  often associated with quark recombination as the dominant hadronization mechanism~\cite{Fries:2008hs}. The ratios in \pPb in both rapidity intervals are comparable to the backward rapidity photo-nuclear values. For the \csd, the uncertainties preclude any strong conclusions.

\begin{figure}[!p]
\centering
\subfigure[]{\includegraphics[width=0.9\textwidth]{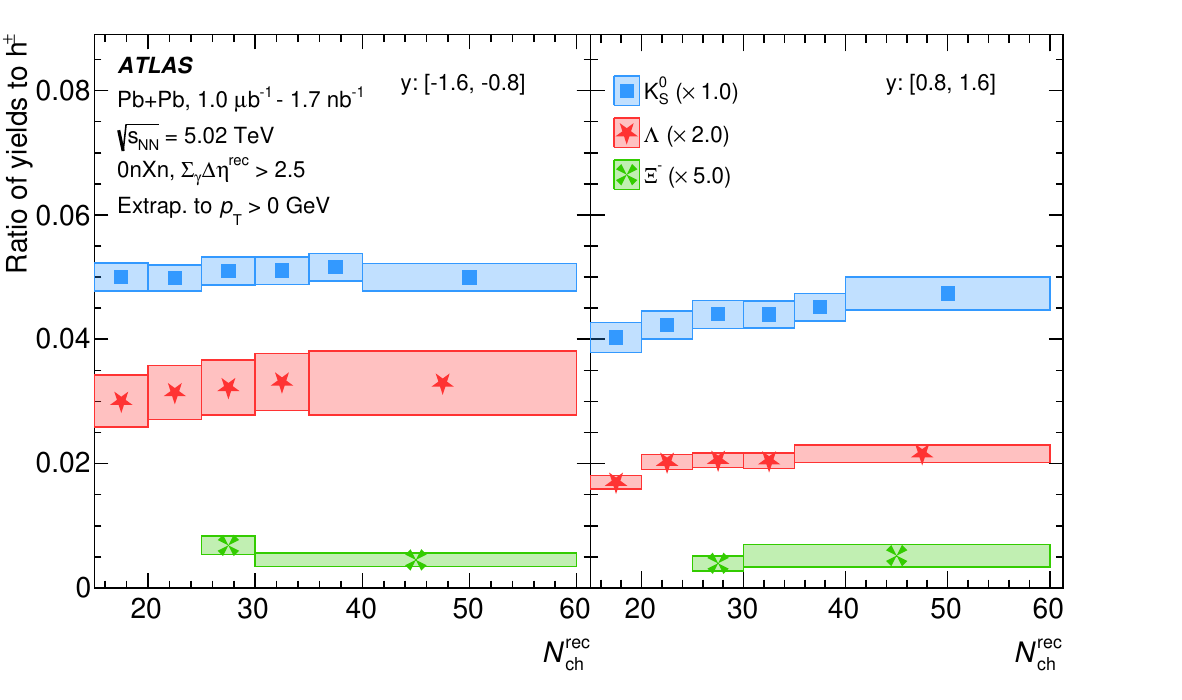}}
\subfigure[]{\includegraphics[width=0.9\textwidth]{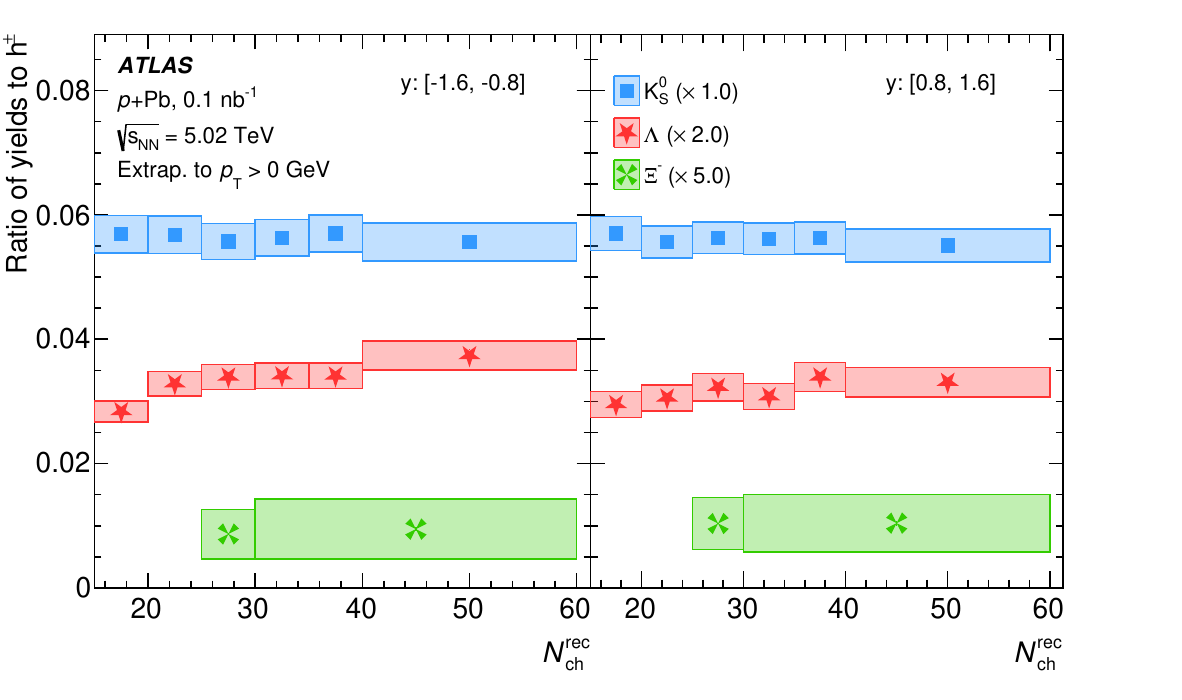}}
\caption{The ratios of identified-strange-hadron yields to charged-hadron yields as a function of \Nch for (a) \PbPb photo-nuclear collisions and (b) \pPb collisions. The left (right) panels are for a backward (forward) rapidity interval. Statistical uncertainties are shown as vertical lines and systematic uncertainties are shown as colored boxes.}
\label{fig:V0sToCH_ratios}
\end{figure}

Finally, the ratio of identified-strange-hadron yields relative to charged-hadron yields is calculated as a function of \Nch. The resulting ratios at backward and forward rapidities are shown as a function of \Nch in Figure~\ref{fig:V0sToCH_ratios} for photo-nuclear and \pPb collisions. In photo-nuclear collisions, there is a clear increase in strange hadron yields relative to charged hadrons between backward and forward rapidity.     If there is strangeness enhancement due to final-state interactions, this would be consistent with the larger ratios in the Pb-going direction.  The ratios are similar between backward rapidity photo-nuclear yields and the \pPb yields in both rapidity intervals. Overall, the ratios are generally consistent with being flat, i.e., not exhibiting an \Nch dependence over the given range, with a hint of a rise in the \lmda to charged hadron ratios. It is notable that the large strangeness enhancement observed in \pp, \pPb, and \PbPb collisions in Ref.~\cite{ALICE:2016fzo} is actually quite small for \ks and \lmda in the multiplicity range corresponding to the measurements presented here.

\clearpage
\newpage

\section{Discussion}

These experimental results are now compared with \dpmjet MC for photo-nuclear collisions and \hijing for \pPb collisions, respectively, neither of which includes any final-state interactions or QGP formation. Comparisons are also made to the so-called ``hybrid'' model that explicitly includes a hydrodynamic modeling of QGP formation in both photo-nuclear and \pPb collisions~\cite{Zhao:2022ayk}. The hybrid model incorporates initial conditions by an extension of MC Glauber to three-dimensions, time evolution via viscous hydrodynamics using the publicly available package $\textsc{music}$~\cite{Schenke:2010nt}, and finally hadronic scattering via the publicly available package $\textsc{urqmd}$~\cite{Bleicher:1999xi}. For the photo-nuclear case, following Ref.~\cite{Baltz:2007kq}, a parametrization is used for the photon energy and hence the center-of-mass energy distribution. The $\gamma$+Pb collisions are treated via the vector meson dominance picture, i.e., the virtual photon state may be decomposed into a set of vector meson states, like $\rho$, $\omega$, and $\phi$, in a low virtuality regime, $Q^2 = 0.0625~\GeV^{2}$. The virtual photon is treated as a vector meson with two ``partonic participants'' in the MC Glauber calculation (in contrast to the three ``partonic participants'' for the proton projectile in the \pPb case).

When making such comparisons to the data, the event selection criteria used here are important to incorporate. In particular, the yields presented here are characterized by specific \Nch intervals, e.g., $25 \le \Nch \le 60$. As discussed earlier, MC studies indicate that the selections on \Nch correspond to equivalent selections on truth-level charged particles with $\pt > 0.4$~\GeV and $|\Eta|<2.5$ as well, but with an average value of $N^{\mathrm{truth}}_{\mathrm{ch}}$ $\approx$ 1.2 $\times$ \Nch. Each event is additionally characterized by the sum-of-gaps using reconstructed tracks and energy clusters.
Monte Carlo studies indicate that this selection corresponds to equivalent selections on truth-level particles with $\pT > 0.45$~\GeV and $|\eta| < 4.9$.  Differences between the reconstruction  and truth level selections are less than 2--3\%. For the \dpmjet and \hijing simulations, the exact event selection criteria used in this measurement are applied through a full \GEANT simulation of the detector response and reconstruction, i.e., to match the experimental event selection of \Nch and \sumgap. Once an event satisfies these criteria, its truth-level particles are included in the yield calculation. In the hybrid model case, the model framework was calibrated with \pPb measurements at a center-of-mass energy of 5.02 \TeV and then made predictions for $\gamma$+Pb collisions.

\begin{figure}[!t]
\centering
\subfigure[]{\includegraphics[width=0.49\textwidth]{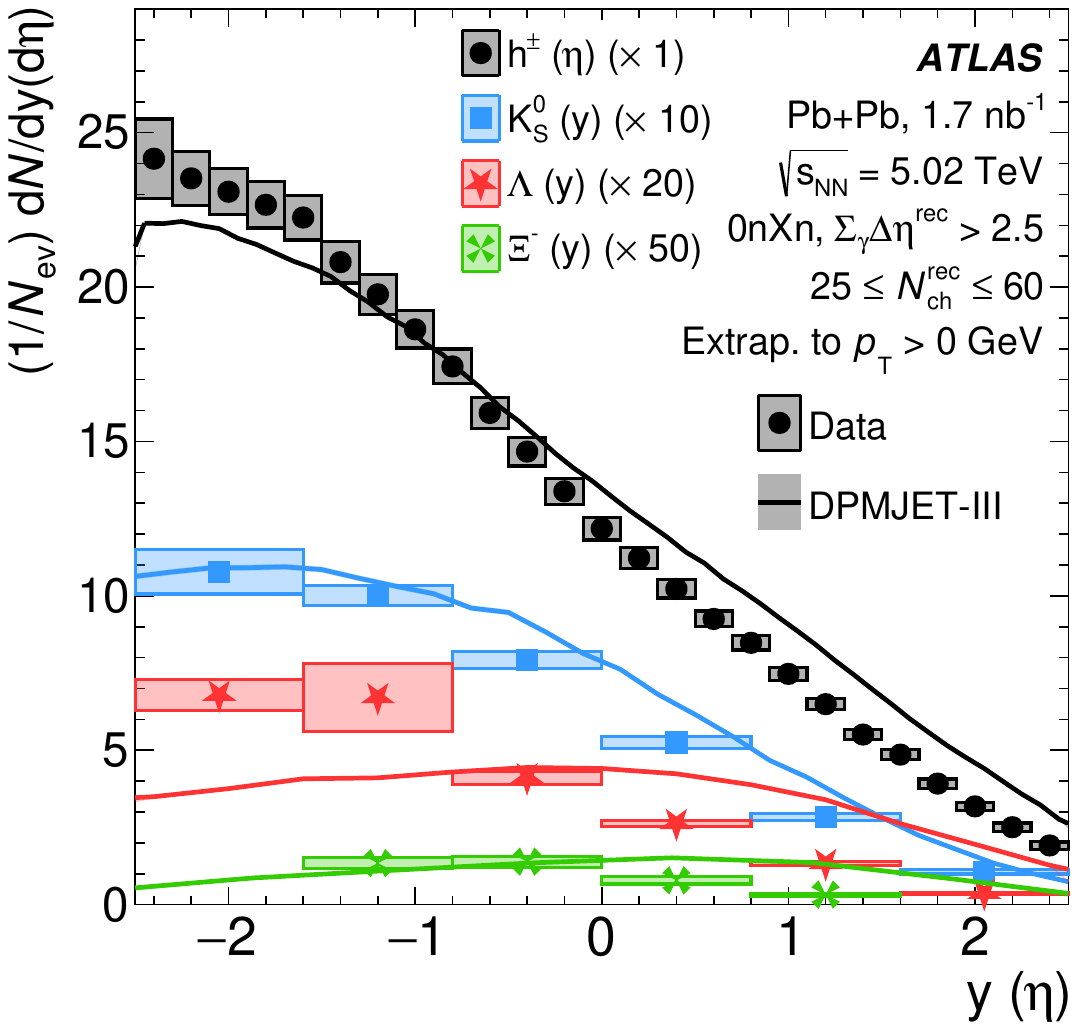}}
\subfigure[]{\includegraphics[width=0.49\textwidth]{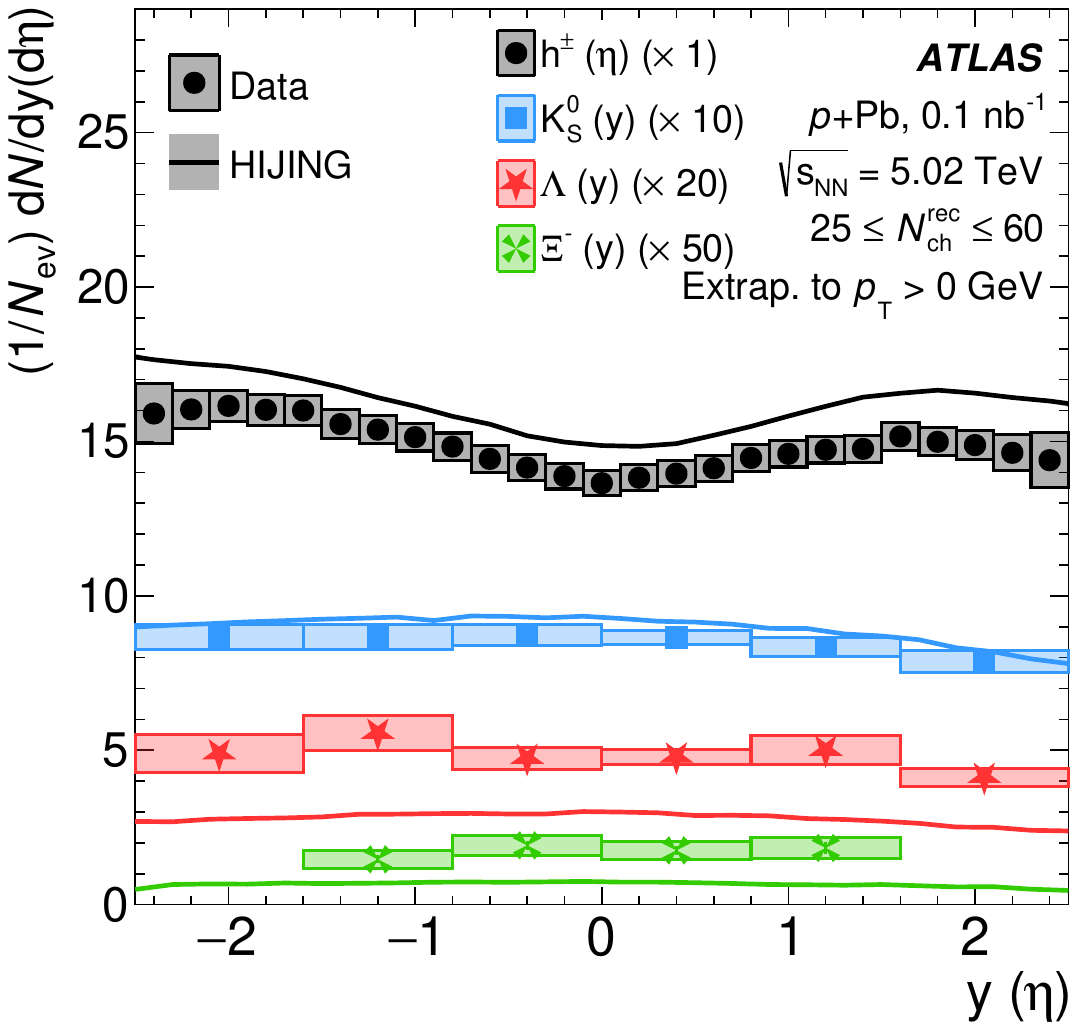}}
\caption{The charged-hadron yields as a function of $\eta$ and the \ks, \lmda, and \csd yields as a function of \y for (a) \PbPb photo-nuclear collisions, with comparisons to the MC model \dpmjet, and (b) \pPb collisions, with comparisons to \hijing. Statistical uncertainties are shown as vertical lines and systematic uncertainties are shown as colored boxes. The statistical uncertainties in the MC simulations are represented by colored bands, though they are negligible.}
\label{fig:dNdy_dataMC}
\end{figure}

Figure~\ref{fig:dNdy_dataMC} shows the measured charged hadron, \ks, \lmda, and \csd yields as a function of rapidity, compared to MC results from \dpmjet for photo-nuclear collisions and \hijing for \pPb collisions.
Both calculations describe the charged hadron and \ks rapidity dependence and the overall normalization at the 15--25\% level. In contrast, the \lmda and \csd yields are poorly described with \hijing under-predicting the strange baryon yields by almost a factor of two and \dpmjet over-predicting the yields at forward rapidity and under-predicting at backward rapidity.

\begin{figure}[!t]
\centering
\subfigure[]{\includegraphics[width=0.49\textwidth]{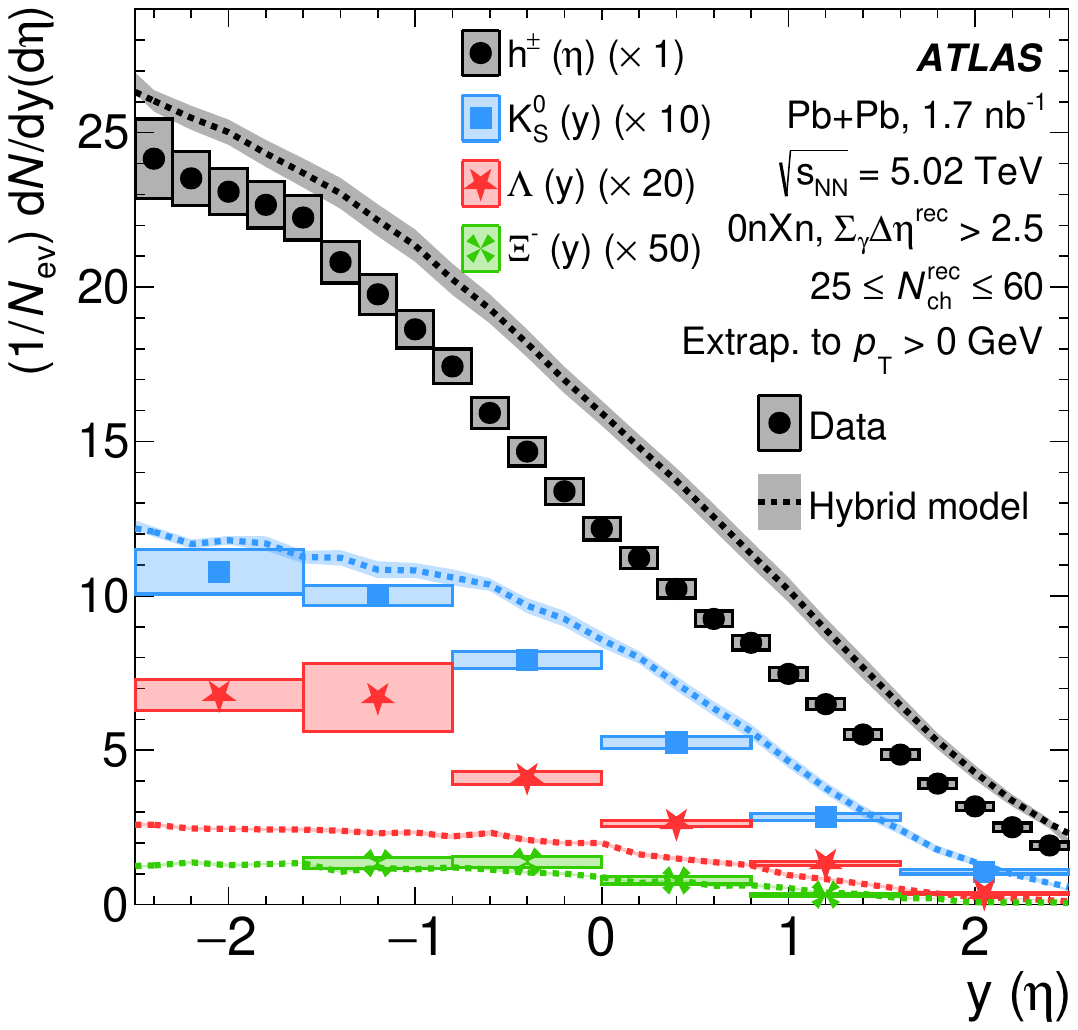}}
\subfigure[]{\includegraphics[width=0.49\textwidth]{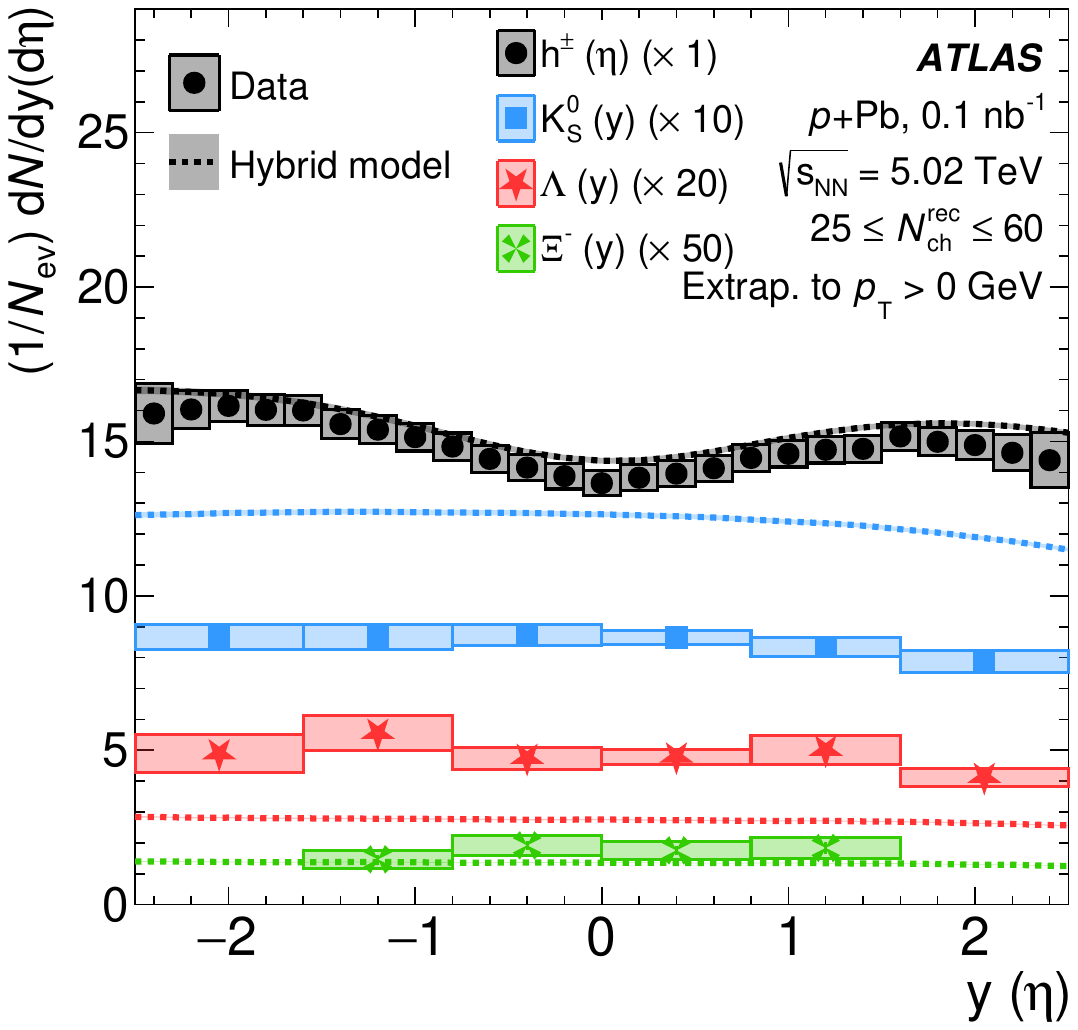}}
\caption{The charged-hadron yields as a function of $\eta$ and the \ks, \lmda, and \csd yields as a function of \y for (a) \PbPb photo-nuclear collisions and (b) \pPb collisions. Both are compared to the hybrid model calculations. Statistical uncertainties are shown as vertical lines and systematic uncertainties are shown as colored boxes. The statistical uncertainties of the model calculations are shown by colored bands.}
\label{fig:dNdy_datahydro}
\end{figure}

Figure~\ref{fig:dNdy_datahydro} shows the same ATLAS results now compared with calculations from the hybrid model. The charged hadron yields are well described in \pPb collisions, and only qualitatively described in photo-nuclear collisions. For both collision systems, the \ks are over-predicted, the \lmda under-predicted, and the \csd well described.  Since the level of disagreement for strange hadrons (\ks and \lmda) is similar in both \pPb and photo-nuclear collisions, it is likely a generic failing of the modeling for strangeness and/or baryons in general in the hadronization of the hydrodynamic QGP.

\clearpage

\begin{figure}[!p]
\centering
\subfigure[]{\includegraphics[width=0.9\textwidth]{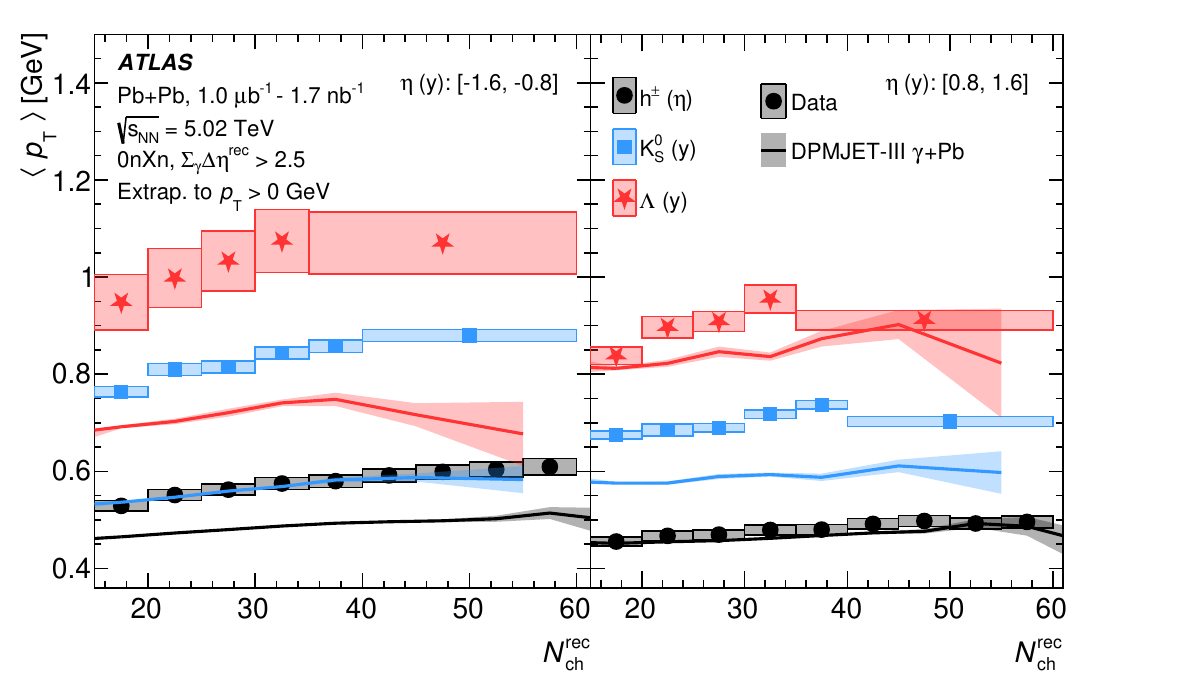}}
\subfigure[]{\includegraphics[width=0.9\textwidth]{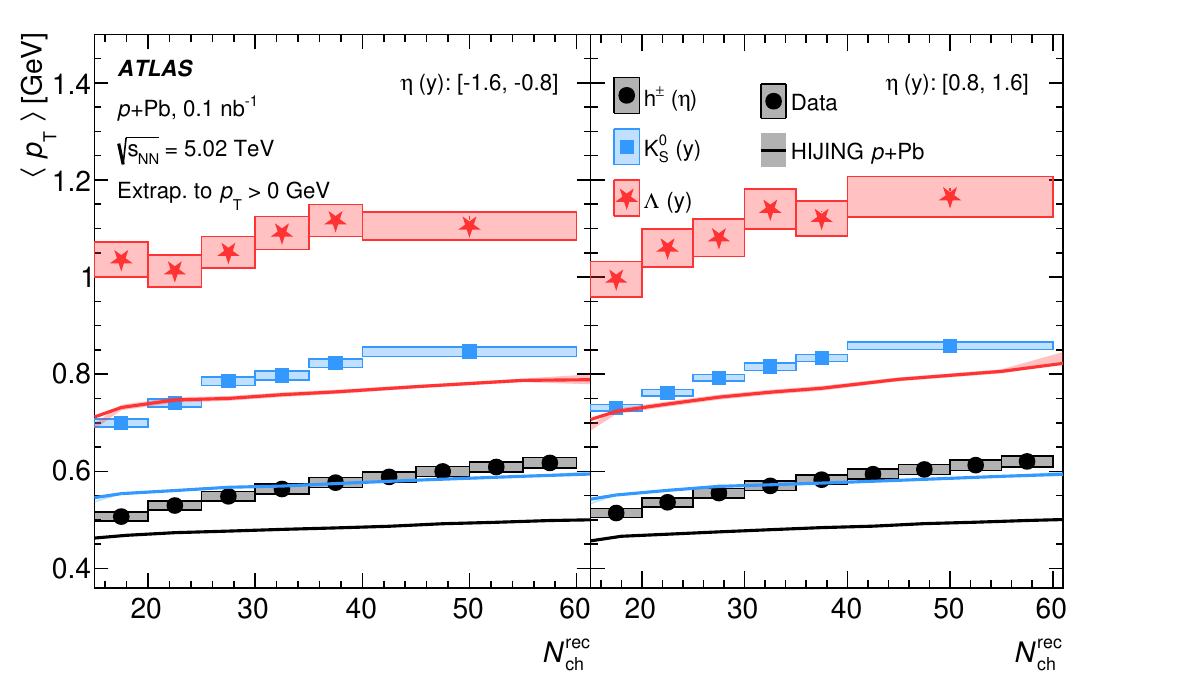}}
\caption{The \meanpt for charged hadrons, \ks, and \lmda in (a) \PbPb photo-nuclear and (b) \pPb collisions as a function of \Nch. The left (right) panels are for a backward (forward) rapidity selection. Statistical uncertainties are shown as vertical lines and systematic uncertainties are shown as colored boxes. Also shown are comparisons to the MC model \dpmjet and \hijing for \PbPb photo-nuclear and \pPb collisions, respectively, with the shaded bands indicating the statistical uncertainties.}
\label{fig:meanpT_dataMC}
\end{figure}

\begin{figure}[!p]
\centering
\subfigure[]{\includegraphics[width=0.9\textwidth]{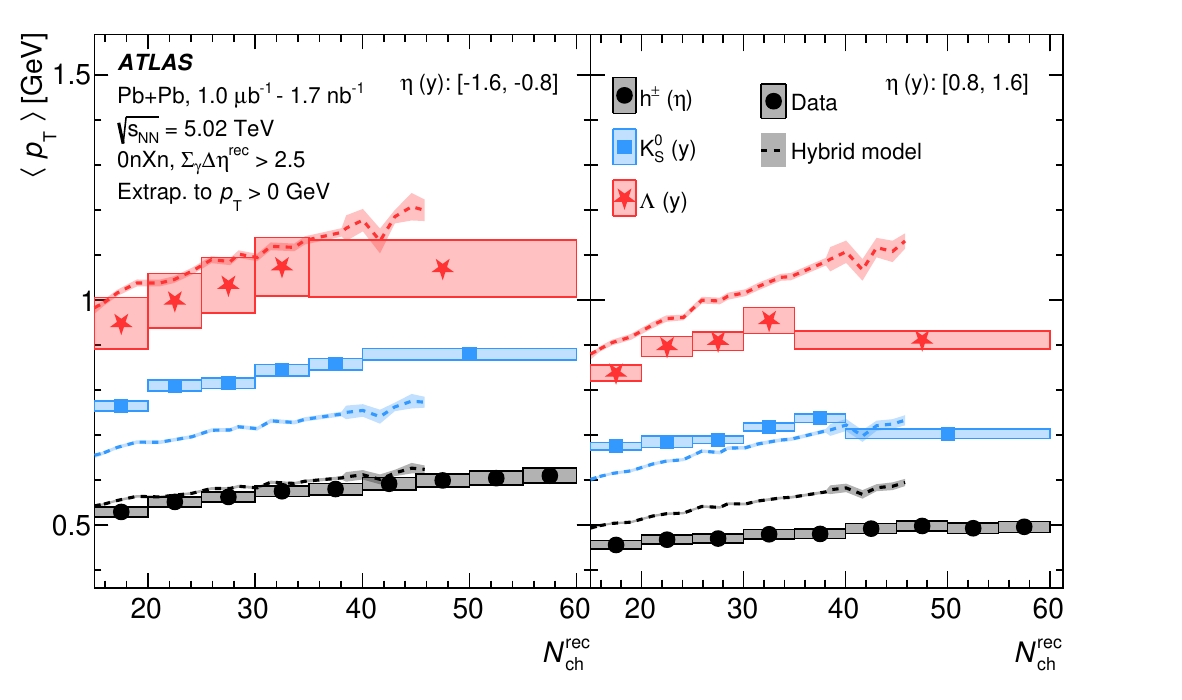}}
\subfigure[]{\includegraphics[width=0.9\textwidth]{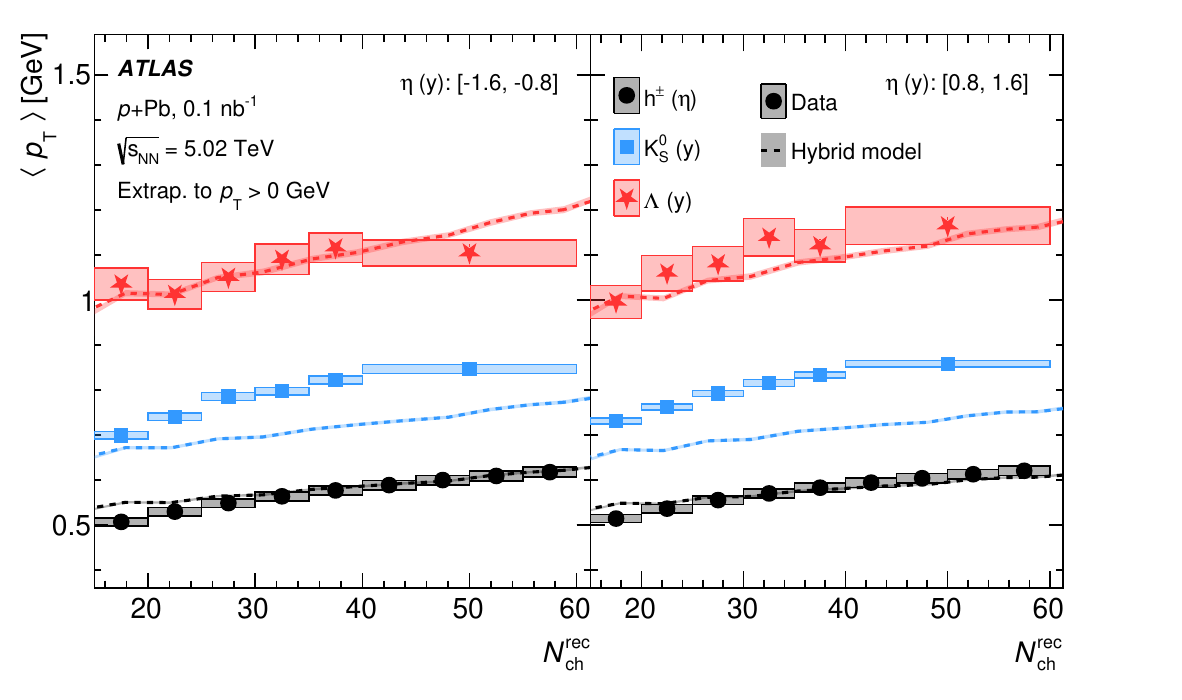}}
\caption{The \meanpt for charged hadrons, \ks, and \lmda in (a) \PbPb photo-nuclear and (b) \pPb collisions as a function of \Nch. The left (right) panels are for a backward (forward) rapidity selection. Statistical uncertainties are shown as vertical lines and systematic uncertainties are shown as colored boxes. Also shown are comparisons to the hybrid hydrodynamic model for both collision types.}
\label{fig:meanpT_datahydro}
\end{figure}

It is also instructive to compare the experimental results for \meanpt as a function of \Nch as shown in Figure~\ref{fig:meanpT_dataMC} with the MC models \dpmjet and \hijing and Figure~\ref{fig:meanpT_datahydro} with the hybrid model. Both MC models substantially under-predict the \meanpt for all particles, and also under-predict the  difference in \meanpt between hadrons of different masses. Strikingly, \dpmjet predicts a higher \meanpt for \lmda particles at forward rapidity compared to backward rapidity, exactly opposite to the trend in data.

In contrast, the hybrid model provides a reasonable description of the \meanpt of charged hadrons and \lmda particles in both photo-nuclear and \pPb collisions, including the higher \meanpt values at backward rapidity compared to forward rapidity in photo-nuclear collisions. However, the \meanpt values for \ks are under-predicted in all cases. Until these deficiencies in the hybrid model are resolved, first in the \pPb case, stronger conclusions regarding QGP formation in photo-nuclear events remains premature.

\FloatBarrier

\clearpage
\section{Conclusion}

This paper reports a measurement of the yields of charged hadrons and identified \ks, \lmda, \csd in high-energy photo-nuclear collisions. Events are selected from 1.7 nb$^{-1}$ of $\sqrt{s_\mathrm{NN}} = 5.02$~\TeV \PbPb data collected by the ATLAS detector at the LHC in 2018. The yields are measured as a function of \pt and rapidity for different \Nch intervals. The results are compared with \pPb collision data at comparable \Nch intervals. These photo-nuclear events reveal a strong rapidity asymmetry in all particle yields, with fewer particles in the photon-going direction as expected. In the Pb-going direction, the particles exhibit larger average transverse momentum that increases for higher mass particles. There is also a significant enhancement of the \lmda/\ks (baryon/meson) ratio, and a hint of an enhancement of overall strange particle production. These observations in the Pb-going direction are generally consistent with what is observed in the \pPb Pb-going direction, giving credence to the hypothesis that the photo-nuclear collisions are dominated by vector meson dominance, i.e., hadronic $\rho$ meson - Pb collisions. Comparisons with MC models reveal a very incomplete modeling of the physics processes. The hybrid model, based on the assumption of QGP formation and hydrodynamic expansion, captures some features such as the larger mean \pt at backward rapidity in photo-nuclear events, but fails to quantitatively describe the yields of identified strange particles.

\section*{Acknowledgments}
%
%

%
%

%
%

We thank CERN for the very successful operation of the LHC and its injectors, as well as the support staff at
CERN and at our institutions worldwide without whom ATLAS could not be operated efficiently.

The crucial computing support from all WLCG partners is acknowledged gratefully, in particular from CERN, the ATLAS Tier-1 facilities at TRIUMF/SFU (Canada), NDGF (Denmark, Norway, Sweden), CC-IN2P3 (France), KIT/GridKA (Germany), INFN-CNAF (Italy), NL-T1 (Netherlands), PIC (Spain), RAL (UK) and BNL (USA), the Tier-2 facilities worldwide and large non-WLCG resource providers. Major contributors of computing resources are listed in Ref.~\cite{ATL-SOFT-PUB-2025-001}.

We gratefully acknowledge the support of ANPCyT, Argentina; YerPhI, Armenia; ARC, Australia; BMWFW and FWF, Austria; ANAS, Azerbaijan; CNPq and FAPESP, Brazil; NSERC, NRC and CFI, Canada; CERN; ANID, Chile; CAS, MOST and NSFC, China; Minciencias, Colombia; MEYS CR, Czech Republic; DNRF and DNSRC, Denmark; IN2P3-CNRS and CEA-DRF/IRFU, France; SRNSFG, Georgia; BMBF, HGF and MPG, Germany; GSRI, Greece; RGC and Hong Kong SAR, China; ICHEP and Academy of Sciences and Humanities, Israel; INFN, Italy; MEXT and JSPS, Japan; CNRST, Morocco; NWO, Netherlands; RCN, Norway; MNiSW, Poland; FCT, Portugal; MNE/IFA, Romania; MSTDI, Serbia; MSSR, Slovakia; ARIS and MVZI, Slovenia; DSI/NRF, South Africa; MICIU/AEI, Spain; SRC and Wallenberg Foundation, Sweden; SERI, SNSF and Cantons of Bern and Geneva, Switzerland; NSTC, Taipei; TENMAK, T\"urkiye; STFC/UKRI, United Kingdom; DOE and NSF, United States of America.

Individual groups and members have received support from BCKDF, CANARIE, CRC and DRAC, Canada; CERN-CZ, FORTE and PRIMUS, Czech Republic; COST, ERC, ERDF, Horizon 2020, ICSC-NextGenerationEU and Marie Sk{\l}odowska-Curie Actions, European Union; Investissements d'Avenir Labex, Investissements d'Avenir Idex and ANR, France; DFG and AvH Foundation, Germany; Herakleitos, Thales and Aristeia programmes co-financed by EU-ESF and the Greek NSRF, Greece; BSF-NSF and MINERVA, Israel; NCN and NAWA, Poland; La Caixa Banking Foundation, CERCA Programme Generalitat de Catalunya and PROMETEO and GenT Programmes Generalitat Valenciana, Spain; G\"{o}ran Gustafssons Stiftelse, Sweden; The Royal Society and Leverhulme Trust, United Kingdom.

In addition, individual members wish to acknowledge support from Armenia: Yerevan Physics Institute (FAPERJ); CERN: European Organization for Nuclear Research (CERN DOCT); Chile: Agencia Nacional de Investigaci\'on y Desarrollo (FONDECYT 1230812, FONDECYT 1230987, FONDECYT 1240864); China: Chinese Ministry of Science and Technology (MOST-2023YFA1605700, MOST-2023YFA1609300), National Natural Science Foundation of China (NSFC - 12175119, NSFC 12275265); Czech Republic: Czech Science Foundation (GACR - 24-11373S), Ministry of Education Youth and Sports (ERC-CZ-LL2327, FORTE CZ.02.01.01/00/22\_008/0004632), PRIMUS Research Programme (PRIMUS/21/SCI/017); EU: H2020 European Research Council (ERC - 101002463); European Union: European Research Council (ERC - 948254, ERC 101089007, ERC, BARD, 101116429), European Regional Development Fund (SMASH COFUND 101081355, SLO ERDF), Horizon 2020 Framework Programme (MUCCA - CHIST-ERA-19-XAI-00), European Union, Future Artificial Intelligence Research (FAIR-NextGenerationEU PE00000013), Horizon 2020 (EuroHPC - EHPC-DEV-2024D11-051), Italian Center for High Performance Computing, Big Data and Quantum Computing (ICSC, NextGenerationEU); France: Agence Nationale de la Recherche (ANR-21-CE31-0013, ANR-21-CE31-0022, ANR-22-EDIR-0002); Germany: Baden-Württemberg Stiftung (BW Stiftung-Postdoc Eliteprogramme), Deutsche Forschungsgemeinschaft (DFG - 469666862, DFG - CR 312/5-2); China: Research Grants Council (GRF); Italy: Istituto Nazionale di Fisica Nucleare (ICSC, NextGenerationEU), Ministero dell'Università e della Ricerca (NextGenEU PRIN20223N7F8K M4C2.1.1); Japan: Japan Society for the Promotion of Science (JSPS KAKENHI JP22H01227, JSPS KAKENHI JP22H04944, JSPS KAKENHI JP22KK0227, JSPS KAKENHI JP23KK0245); Norway: Research Council of Norway (RCN-314472); Poland: Ministry of Science and Higher Education (IDUB AGH, POB8, D4 no 9722), Polish National Science Centre (NCN 2021/42/E/ST2/00350, NCN OPUS 2023/51/B/ST2/02507, NCN OPUS nr 2022/47/B/ST2/03059, NCN UMO-2019/34/E/ST2/00393, UMO-2020/37/B/ST2/01043, UMO-2022/47/O/ST2/00148, UMO-2023/49/B/ST2/04085, UMO-2023/51/B/ST2/00920, UMO-2024/53/N/ST2/00869); Portugal: Foundation for Science and Technology (FCT); Spain: Generalitat Valenciana (Artemisa, FEDER, IDIFEDER/2018/048), Ministry of Science and Innovation (MCIN \& NextGenEU PCI2022-135018-2, MICIN \& FEDER PID2021-125273NB, RYC2019-028510-I, RYC2020-030254-I, RYC2021-031273-I, RYC2022-038164-I); Sweden: Carl Trygger Foundation (Carl Trygger Foundation CTS 22:2312), Swedish Research Council (Swedish Research Council 2023-04654, VR 2021-03651, VR 2022-03845, VR 2022-04683, VR 2023-03403, VR 2024-05451), Knut and Alice Wallenberg Foundation (KAW 2018.0458, KAW 2022.0358, KAW 2023.0366); Switzerland: Swiss National Science Foundation (SNSF - PCEFP2\_194658); United Kingdom: Leverhulme Trust (Leverhulme Trust RPG-2020-004), Royal Society (NIF-R1-231091); United States of America: U.S. Department of Energy (ECA DE-AC02-76SF00515), Neubauer Family Foundation.

%
%


%

%
\clearpage
\printbibliography
\clearpage
 
\begin{flushleft}
\hypersetup{urlcolor=black}
{\Large The ATLAS Collaboration}

\bigskip

\AtlasOrcid[0000-0002-6665-4934]{G.~Aad}$^\textrm{\scriptsize 104}$,
\AtlasOrcid[0000-0001-7616-1554]{E.~Aakvaag}$^\textrm{\scriptsize 17}$,
\AtlasOrcid[0000-0002-5888-2734]{B.~Abbott}$^\textrm{\scriptsize 123}$,
\AtlasOrcid[0000-0002-0287-5869]{S.~Abdelhameed}$^\textrm{\scriptsize 119a}$,
\AtlasOrcid[0000-0002-1002-1652]{K.~Abeling}$^\textrm{\scriptsize 55}$,
\AtlasOrcid[0000-0001-5763-2760]{N.J.~Abicht}$^\textrm{\scriptsize 49}$,
\AtlasOrcid[0000-0002-8496-9294]{S.H.~Abidi}$^\textrm{\scriptsize 27b}$,
\AtlasOrcid[0009-0003-6578-220X]{M.~Aboelela}$^\textrm{\scriptsize 45}$,
\AtlasOrcid[0000-0002-9987-2292]{A.~Aboulhorma}$^\textrm{\scriptsize 36e}$,
\AtlasOrcid[0000-0001-5329-6640]{H.~Abramowicz}$^\textrm{\scriptsize 157}$
% ...
\bigskip
\\

% ...
$^{17}$Department for Physics and Technology, University of Bergen, Bergen; Norway.\\
% ...
$^{27}$$^{(a)}$University of Colorado Boulder, Department of Physics, Colorado;$^{(b)}$Physics Department, Brookhaven National Laboratory, Upton NY; United States of America.\\
% ...
$^{36}$$^{(a)}$Facult\'e des Sciences Ain Chock, Universit\'e Hassan II de Casablanca;$^{(b)}$Facult\'{e} des Sciences, Universit\'{e} Ibn-Tofail, K\'{e}nitra;$^{(c)}$Facult\'e des Sciences Semlalia, Universit\'e Cadi Ayyad, LPHEA-Marrakech;$^{(d)}$LPMR, Facult\'e des Sciences, Universit\'e Mohamed Premier, Oujda;$^{(e)}$Facult\'e des sciences, Universit\'e Mohammed V, Rabat;$^{(f)}$Institute of Applied Physics, Mohammed VI Polytechnic University, Ben Guerir; Morocco.\\
% ...
$^{45}$Physics Department, Southern Methodist University, Dallas TX; United States of America.\\
% ...
$^{49}$Fakult\"{a}t Physik , Technische Universit{\"a}t Dortmund, Dortmund; Germany.\\
% ...
$^{55}$II. Physikalisches Institut, Georg-August-Universit\"{a}t G\"ottingen, G\"ottingen; Germany.\\
% ...
$^{104}$CPPM, Aix-Marseille Universit\'e, CNRS/IN2P3, Marseille; France.\\
% ...
$^{119}$$^{(a)}$New York University Abu Dhabi, Abu Dhabi;$^{(b)}$United Arab Emirates University, Al Ain; United Arab Emirates.\\
% ...
$^{123}$Homer L. Dodge Department of Physics and Astronomy, University of Oklahoma, Norman OK; United States of America.\\
% ...
$^{157}$Raymond and Beverly Sackler School of Physics and Astronomy, Tel Aviv University, Tel Aviv; Israel.\\
% ...
$^{a}$ Also at Affiliated with an institute formerly covered by a cooperation agreement with CERN.\\
$^{b}$ Also at An-Najah National University, Nablus; Palestine.\\
$^{c}$ Also at Borough of Manhattan Community College, City University of New York, New York NY; United States of America.\\
$^{d}$ Also at Center for Interdisciplinary Research and Innovation (CIRI-AUTH), Thessaloniki; Greece.\\
$^{e}$ Also at Centre of Physics of the Universities of Minho and Porto (CF-UM-UP); Portugal.\\
$^{f}$ Also at CERN, Geneva; Switzerland.\\
% ...

\end{flushleft}


%

%
%

%

%
%
%
%

\end{document}